\documentclass[%
longbibliography,
 amsmath,amssymb,
 aps,
prb,
floatfix,
]{revtex4-1}

\usepackage{graphicx}
\usepackage{dcolumn}
\usepackage{bm}

\begin{document}

\title{Perpendicular space accounting of localized states in a quasicrystal}

\author{Murod Mirzhalilov}
 \affiliation{Department of Physics , Bilkent University, Ankara, 06800, TURKEY}
\author{M.\"O. Oktel}
\email{oktel@bilkent.edu.tr}
\affiliation{Department of Physics , Bilkent University, Ankara, 06800, TURKEY}

\date{\today}

\begin{abstract}
Quasicrystals can be described as projections of sections of higher dimensional periodic lattices into real space. The image of the lattice points in the projected out dimensions, called the perpendicular space,  carries valuable information about the local structure of the real space lattice. In this paper, we use perpendicular space projections to analyze the elementary excitations of a quasicrystal. In particular, we consider the vertex tight binding model on the two dimensional Penrose lattice and investigate the properties of strictly localized states using their perpendicular space images. Our method reproduces the previously reported frequencies for the six types of localized states in this model. We also calculate the overlaps between different localized states and show that the number of type-5 and type-6 localized states which are independent from the four other types is a factor of golden ratio $\tau=(1+\sqrt{5})/2$ higher than previously reported values. Two orientations of the same type-5 or type-6 which are supported around the same site are shown to be linearly dependent with the addition of other types. We also show through exhaustion of all lattice sites in perpendicular space that any point in the Penrose lattice is either in the support of at least one localized state or is forbidden by local geometry to host a strictly localized state.

\end{abstract}

\maketitle

\section{\label{sec:Introduction} Introduction}

Since their discovery in the 1980s \cite{she84} quasicrystals have provided unique problems which challenge the connection between order and periodicity in solids. While there has been immense progress in understanding the structural properties of quasicrystalline systems\cite{els85,kal85,lev86,soc86}, elementary excitations in quasicrystals are much less studied \cite{ste18}. The absence of periodicity, thus quasi-momentum,  makes it hard to label the eigenstates of a quasiperiodic system. While some fundamental issues have been overcome in one dimensional models\cite{koh83,ost83,koh87}, the results in two and three dimensions rely mostly on large scale computations of spectra \cite{tsu91}. 

There has been a resurgence of interest in quasiperiodic systems due to experiments which can create quasiperiodic order in engineered systems such as photonic crystals \cite{var13}, polaritonic cavities\cite{tan14} and most recently ultracold atoms\cite{sin15,vie19}. While the system sizes in these experiments are much smaller compared to quasicrystalline solids, they are free of impurities and provide an opportunity to measure quantities which are not accessible for solids. Similarly, quasicrystalline order may lead to novel phases in models which are well understood for periodic systems\cite{kam18,sak17,fli20}.  Particularly cold atom realizations of quasicrystals may soon explore the effects of local quasicrystal structure in strongly interacting systems. 

A quasicrystal can be defined as the projection of a segment of a periodic lattice in a higher dimension \cite{bru81}. While other methods for defining quasiperiodic order exist\cite{lev86}, the projection method has a geometric appeal which provides insight about the quasiperiodic order.  The projection property has been successfully applied to calculate the x-ray diffraction spectra\cite{els85,kal85} and other observables related to structure of quasicrystalline systems \cite{htr92}. However, the investigation of eigenstates have mostly relied on numerical calculations in periodic approximants to quasicrystals \cite{tsu91} or on the self-similar properties obtained through inflation-deflation transformations \cite{koh83}. The perpendicular space images of eigenstates in the one dimensional Fibonacci chain \cite{mjp16}, and spin models on quasicrystals \cite{sja08} have been investigated. Perpendicular space images have been calculated for particular eigenstates satisfying an ansatz \cite{kka14}  for two dimensional tight binding models\cite{mjk17}. 

The perpendicular space images of a lattice point carries valuable information about the local connectivity of that point in real space\cite{kdu85}. In general, points which are close to each other in perpendicular space have similar local environments, which maps a particular local lattice configuration to a volume in perpendicular space. As projection from higher dimension is a linear mapping, volumes in perpendicular space are proportional to frequencies of corresponding lattice configurations. In this paper, we apply this general idea to label and count localized eigenstates for a specific quasicrystalline model.

We consider the vertex tight binding model on a Penrose lattice (PL)\cite{pen74}. PL can be described as the projection of the intersection of a five dimensional simple cubic lattice with a 2D-plane \cite{bru81}. The real space lattice points have orientational order where nearest neighbors are in one of the five star vector directions. Nearest neighbor bonds define thick and thin rhombuses, which can alternatively be used to define the Penrose lattice through local matching rules. Vertex tight binding lattice considers an s-state at each point of the PL and uniform hopping probability only over the nearest neighbor bonds. This model was investigated first in 1985 where numerical results indicated a density of states peak at zero energy \cite{oda86,cho85}, which was shown to emerge from strictly localized states by Kohmoto and Sutherland \cite{koh86}. In a following publication Arai, Tokihiro, Fujiwara and Kohmoto reported six independent types of such strictly localized states(LS) and their frequencies \cite{ara88}. The method for establishing the independence of these states and counting their frequencies relied on identification of real space structures such as `one three edge rhombus strings' and `bridge sites' combined with inflation-deflation counting of local structures \cite{kum86}.

Here, we show that perpendicular space methods can be used to uniquely label, as well as count the frequencies of the strictly localized states in this model.  This labeling allows us to calculate the overlaps between different localized states and establish the independence of different localized states. While each one of type-1 to type-4 states are independent of each other as a collective set this is not true for type-5 and type-6 states. A combination of one type-5 state with a particular type-2 and a particular type-3 state results in another type-5 state around the same S5 vertex. Thus, the frequencies of type-5 states which are independent of each other is a factor of golden ratio $\tau=(1+\sqrt(5))/2$ higher than frequency of type-5 states independent of the collective set of type-1 to type-5 states. A similar situation plays out for type-6 states. Perpendicular space methods also help identify sites which are forbidden by local connectivity from hosting an eigenstate of zero energy. We show that a finite number of iterations in perpendicular space rules out $f_{forbidden}=1358-839 \tau=46.95\%$ of PL vertices from hosting strictly localized states. This process allows us to prove that a site is either forbidden to host a LS or in the support of  at least one LS.      

The next section contains the basic definitions related to the PL, its perpendicular space and the vertex model. In section \ref{sec:Frequencies}, we identify the perpendicular space areas corresponding to the 6 types of LS and count their frequencies. The overlaps between different localized states are calculated and their independence is established in section \ref{sec:Overlaps}. The following section details the identification and counting of forbidden sites and show that each site is either forbidden or in the perpendicular space are of a LS. We close the paper with a summary of our results and conclusions  in section \ref{sec:Conclusion}.

\section{\label{sec:PerpSpace} Penrose lattice and its Perpendicular Space}

There are various equivalent methods for defining the quasiperiodic PL. Here, we  use the projection definition following de Bruijn \cite{bru81} as our method relies on the perpendicular space of the lattice. Consider the five dimensional real space as spanned by the orthogonal unit vectors $\hat{u}_0,\hat{u}_1,\hat{u}_2,\hat{u}_3,\hat{u}_4$:
\begin{equation}
    \vec{x}= x_0 \hat{u}_0 + x_1 \hat{u}_1 +x_2 \hat{u}_2+ x_3 \hat{u}_3 +x_4 \hat{u}_4.
\end{equation}
This five dimensional space can be partitioned into open unit cubes $k_m-1<x_m<k_m$ where  $k_m$ are integers for $m=0,...,4$.   If we define $\zeta=e^{i \frac{2 \pi}{5}}$ the following five orthogonal vectors also span the five dimensional space
\begin{eqnarray}
\label{eq:C vectors}
\vec{c}_0 &=&\sum_m \hat{u}_i, \\ \nonumber
\vec{c}_1&=&\sum_m Re(\zeta^m) \hat{u}_m ,\\ \nonumber
\vec{c}_2&=&\sum_m Im(\zeta^m) \hat{u}_m ,\\ \nonumber
\vec{c}_3&=&\sum_m Re(\zeta^{2m}) \hat{u}_m, \\ \nonumber
\vec{c}_4&=&\sum_m Im(\zeta^{2m}) \hat{u}_m.
\end{eqnarray}
We  define a two-dimensional plane with the following three equations,
\begin{eqnarray}
\label{eq:Projection constraints}
\vec{x}\cdot \vec{c}_0 = 0 \\ \nonumber
(\vec{x}-\vec{\gamma})\cdot \vec{c}_3 = 0 \\ \nonumber
(\vec{x}-\vec{\gamma})\cdot \vec{c}_4 = 0 
\end{eqnarray}
the intercept vector $\vec{\gamma}=\sum_m \gamma_m \hat{u}_m$ is chosen to be perpendicular to $\vec{c}_0$.

A real space projection of a point in the five dimensional cubic lattice $\vec{R}_5=\sum_m k_m \hat{u}_m$ is in the PL if and only if there is a point in its open unit cube which satisfies Eq.(\ref{eq:Projection constraints}). We further assume that $\vec{\gamma}$ is chosen such that the plane and the cube intersection does not yield a singular lattice \cite{bru81}.  
The points $k_0,...,k_4$ satisfying this condition can be expressed as 
\begin{equation}
    \vec{R}_5=\sum_m k_m \hat{u}_m =\frac{2}{5} \left( x_R \vec{c}_1 + y_R \vec{c}_2 + x_\perp \vec{c}_3 + y_\perp \vec{c}_4 + \frac{N_0}{2} \vec{c}_0\right).
\end{equation}
We  refer to $\vec{r}_R= x_R \hat{i}+ y_R \hat{j}$ as the real space projection and $\vec{R}_\perp= x_\perp \hat{i}_\perp + y_\perp \hat{j}_\perp $ as the perpendicular space image of this point. The unit orthogonal vectors spanning the real space are defined as $\hat{i},\hat{j}$ and similarly perpendicular space is constructed by $\hat{i}_\perp,\hat{j}_\perp$.    $N_0=\vec{R}_5\cdot\vec{c}_0=\sum_m k_m$ is an integer by construction and is called the index of this point. 

\begin{figure}[!htb]
    \centering
    \includegraphics[trim=8mm 8mm 8mm 8mm,clip,width=0.3\textwidth]{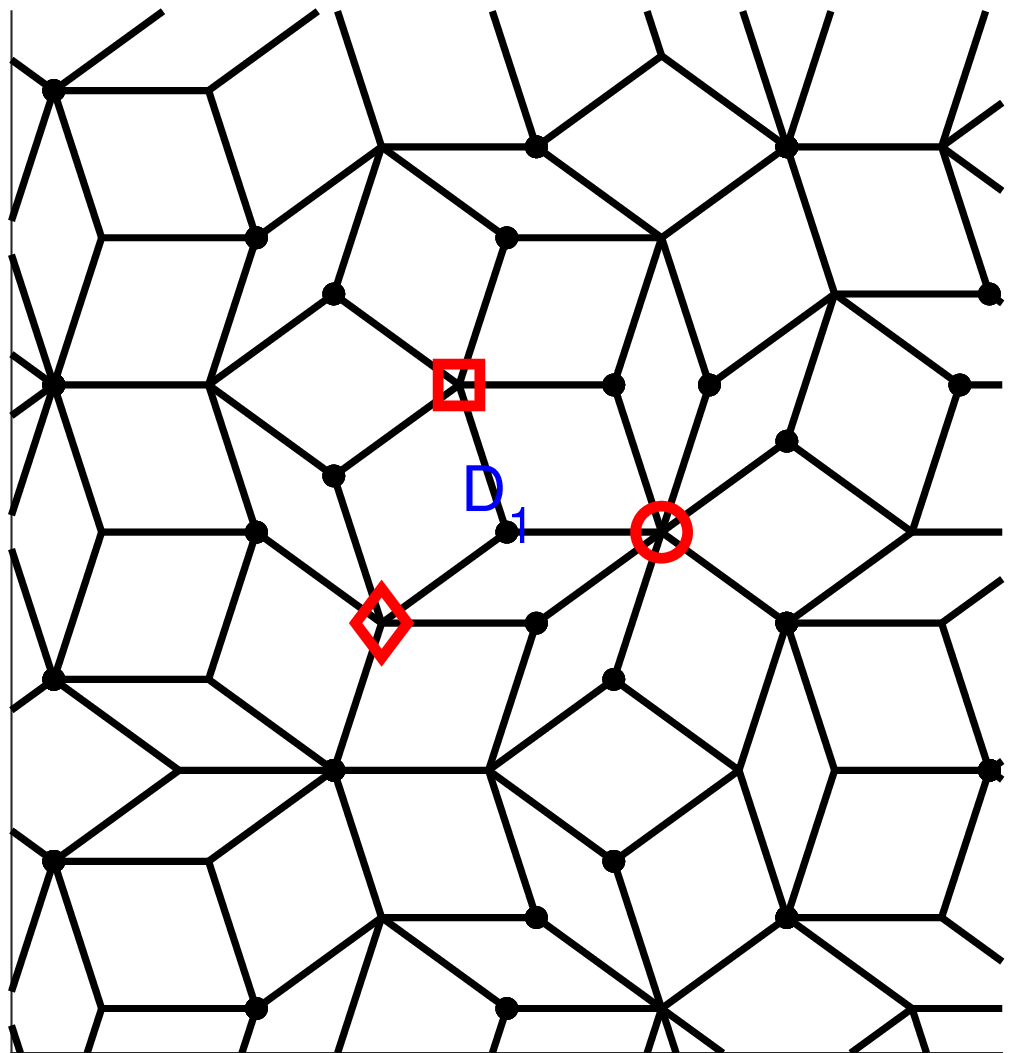}
    \caption{A finite section of the Penrose Lattice in real space. The central point $D_1$ has three neighbors. The perpendicular space projections of $D_1$ and neighbors are shown in \ref{fig:PerpendicularSpaceOfPL}.}
    \label{fig:PenroseLattice}
\end{figure}

A finite section of the PL is shown in figure \ref{fig:PenroseLattice}, where we can equivalently express the positions
    $\vec{r}_R= \sum_m k_m \hat{e}_m$ where the five star vectors are $\hat{e}_m= Re(\zeta^m) \hat{i}+Im(\zeta^m)\hat{j}$. Thus all the nearest neighbors are aligned in the directions $\pm \hat{e}_m$ and there is orientational order throughout the PL. 
    
\begin{figure*}[!htb]
\includegraphics[trim=1mm 1mm 1mm 1mm,clip,width=0.4\textwidth]{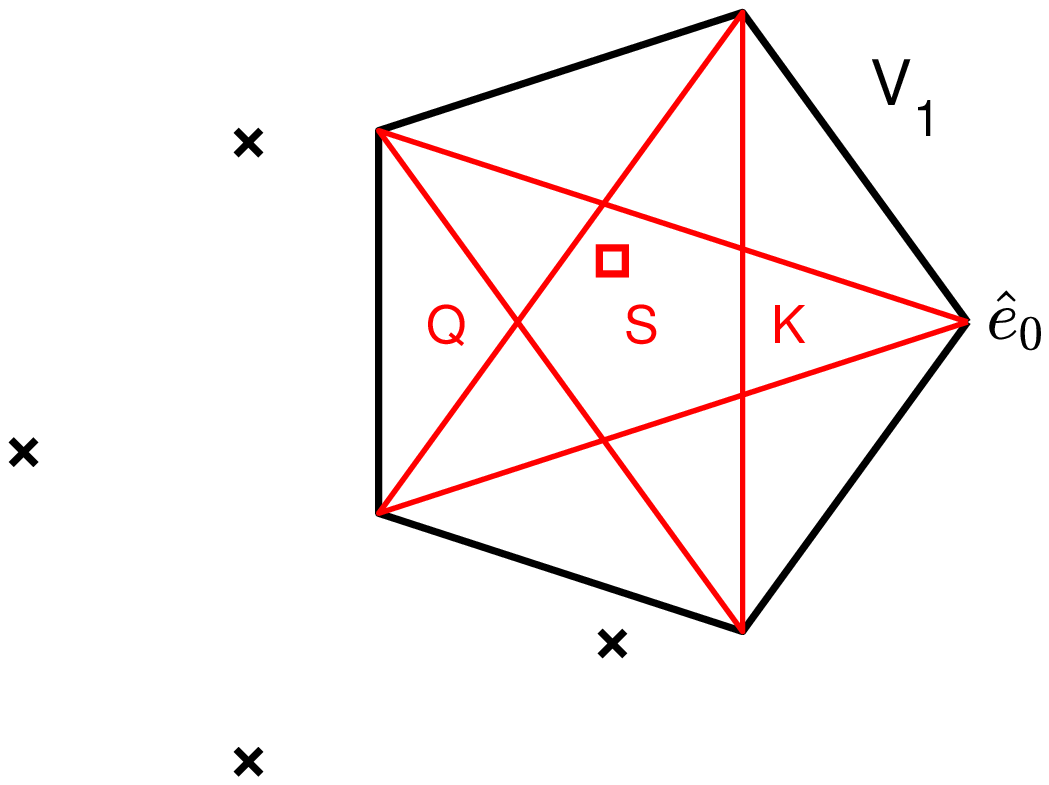} 
\includegraphics[trim=1mm 1mm 1mm 1mm,clip,width=0.4\textwidth]{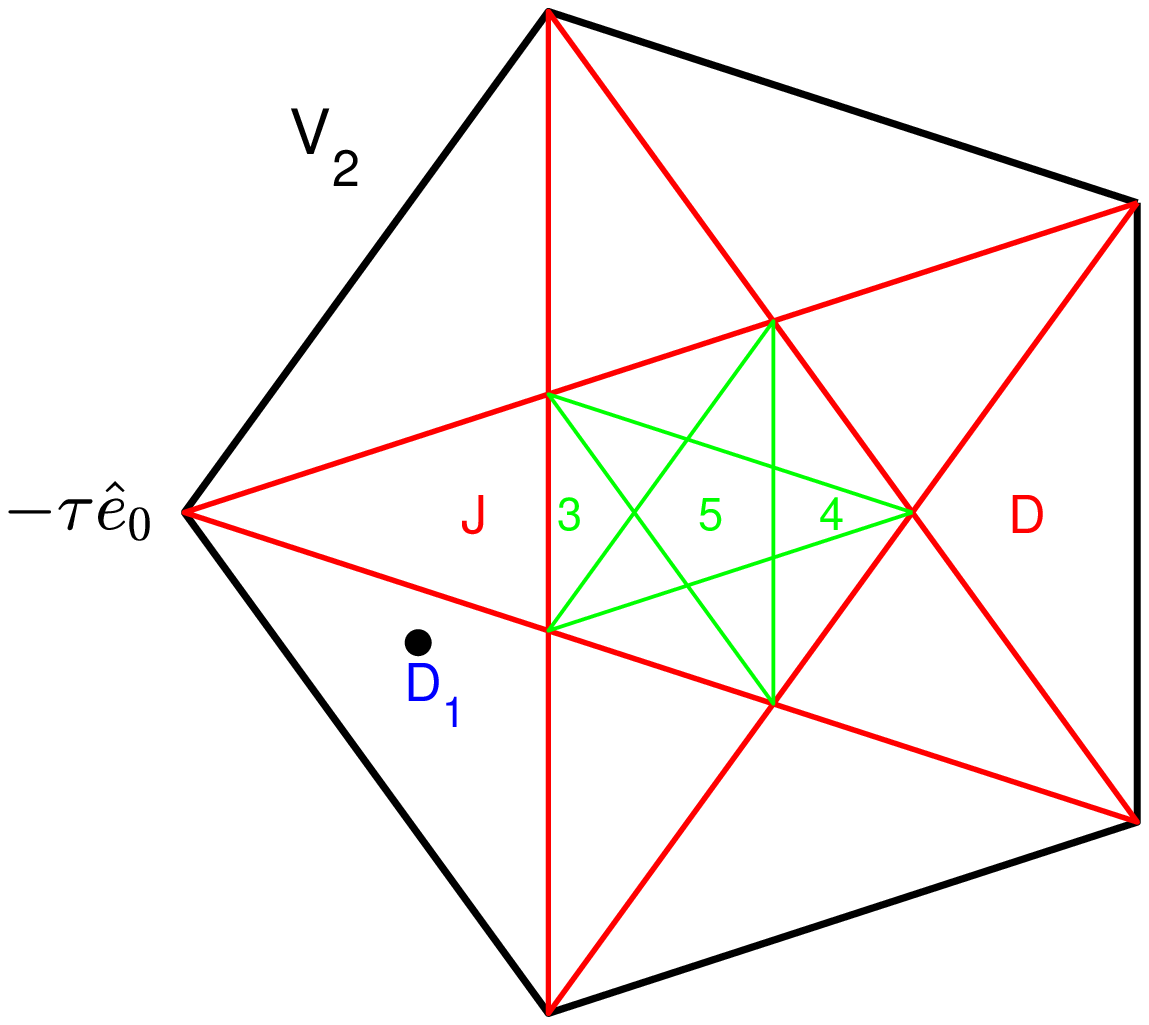}\\
\includegraphics[trim=1mm 1mm 1mm 1mm,clip,width=0.4\textwidth]{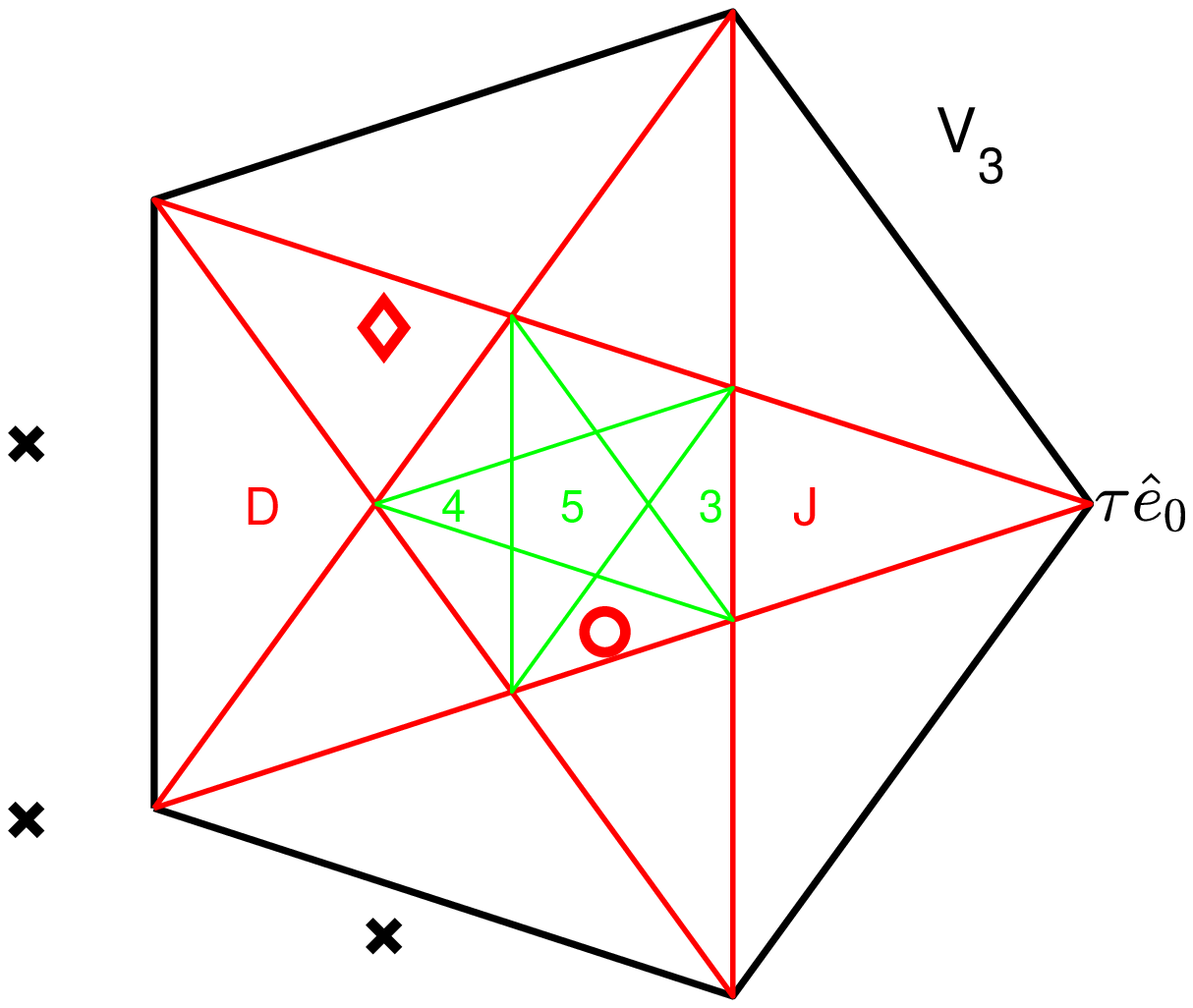} 
\includegraphics[trim=1mm 1mm 1mm 1mm,clip,width=0.4\textwidth]{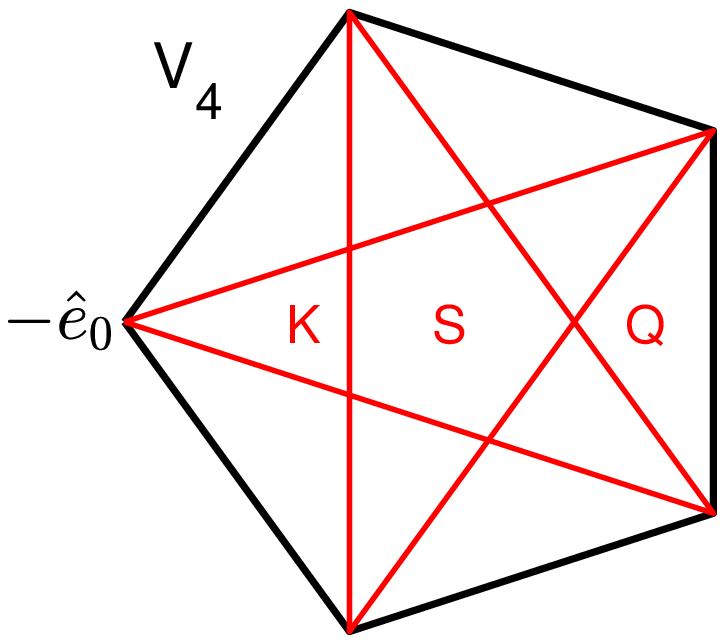} 
\caption{The 4 pentagons $V_1$ to $V_4$ in perpendicular space. Notice that $V_1$ and $V_4$ are scaled up for visual clarity on all subsequent figures. Perpendicular space image of $D_1$ and and its neighbors are shown. The points marked outside the pentagons $V_1$ and $V_3$ differ from $D_1$ by star vectors $\hat{e}_m$ but have no counterparts in real space as they lie outside the pentagons. The regions for $S,Q,K,S5,S4,S3,D,J$ type vertices are marked by S,Q,K,5,4,3,D,J respectively, empty regions have the same vertex type as the marked region they are related by five fold rotational symmetry.}
    \label{fig:PerpendicularSpaceOfPL}
\end{figure*}

The index $N_0$ can only take four values $N_0=1,2,3,4$, and the projection conditions Eq.(\ref{eq:Projection constraints}) define a finite area $V_{N_0}$ which constrains the perpendicular space image of points in the PL. These areas form 4 regular pentagons in perpendicular space and are shown in Fig. \ref{fig:PerpendicularSpaceOfPL}. For indices $N_0=1,4$ the pentagons have circumscribed circles of radius 1, while $N_0=2,3$ have circumscribed circles of radius $\tau=(1+\sqrt(5))/2$. We refer to the radius of the circumscribed circle of a pentagon as the radius of the pentagon throughout the paper. The collection of these 4 pentagons $V_1,V_2,V_3,V_4$ form all of the perpendicular space of the PL. Although the perpendicular space is 3 dimensional, the image of all PL points are restricted to these four 2 dimensional regions\cite{hen86,jar86}. This fact shows that the projection could be carried out from an initial 4 dimensional lattice \cite{baa02}, but we choose not to use such an approach.

It is important to notice that the perpendicular space image of any point contains information about the local structure of the point in real space \cite{kdu85}.  Consider point $D_1$ in Fig. \ref{fig:PenroseLattice} and its perpendicular space image $D_1$ in Fig. \ref{fig:PerpendicularSpaceOfPL}. We see that this point has index two as it is in the $V_2$ pentagon. Any nearest neighbor of $D_1$ can be along either  the $+\hat{e}_m$ or the $-\hat{e}_m$ directions. If a point is reached by a positive $\hat{e}_m$ vector from $D_1$, its perpendicular space image must lie in $V_3$, and if by a negative $\hat{e}_m$ in $V_1$. Furthermore the perpendicular space position of the neighbor is related to which $\hat{e}_m$ vector is used in real space translation. From Eq.(\ref{eq:C vectors}) we find the correspondence
\begin{equation}
\label{eq:eVectorCorrespondence}
\hat{e}_0 \rightarrow \hat{e}_0, \quad 
\hat{e}_1 \rightarrow \hat{e}_2, \quad
\hat{e}_2 \rightarrow \hat{e}_4,  \quad
\hat{e}_3 \rightarrow \hat{e}_1, \quad
\hat{e}_4 \rightarrow \hat{e}_3 
\end{equation}
between real space translations and perpendicular space translations. All five possible perpendicular space points reached from $D_1$ with positive $\hat{e}_m$ are shown in Fig\ref{fig:PerpendicularSpaceOfPL}. Only two of these are inside $V_3$, thus $D_1$ has only two neighbors of index 3. A similar reasoning for negative $\hat{e}_m$ translations show that only 1 out of possible 5 neighbors is in $V_1$, thus $D_1$ has only three neighbors in total. 

The nearest neighbor configurations can be used to classify vertices, and we use the nomenclature of de Bruijn \cite{bru81}. $K$, $Q$ and $S$ vertices have index of $N_0=1,4$ and their perpendicular space regions in $V_1,V_4$ are shown in Fig.\ref{fig:PerpendicularSpaceOfPL}. Similarly $D,J,S3,S4$ and $S5$ vertex regions are marked in $V_2, V_3$. All the points of the PL have a perpendicular space image in one of the four $V_m$ pentagons. The mapping from PL points to perpendicular space is not one to one but injective, there are some points inside the $V_m$ pentagons which do not correspond to any PL point. As an example consider the points on the lines separating regions for distinct vertex types. A point on this line does not correspond to a point of either vertex type, it remains not associated with any PL point as $\tau$ is an irrational number. Still, the points in perpendicular space associated with PL are dense, any such point has infinitely many similar points in any finite neighborhood. The set of points inside the pentagons is a two dimensional analogue of the real line with all rational numbers removed.  

The mapping from 5 dimensional space to  real and perpendicular spaces Eq. (\ref{eq:C vectors}) is a linear transformation. Thus, not only are the PL projected points inside the pentagons dense, their density is uniform inside all four pentagons. The ratio of areas in perpendicular space give the ratio of corresponding vertex sites in real space. As an example consider the frequency of $S$ vertices in the PL.  We see that the area corresponding to $S$ in $V_1$ is a pentagon of radius $\tau^{-2}$. As $V_1$ and $V_2$ have radii of $1$ and $\tau$ the ratio of the area of $S$ regions to total area of pentagons is
\begin{equation}
    f_{S}=\frac{\tau^{-4}}{1+\tau^2}=\frac{18-11 \tau}{5} \simeq 0.04032,
\end{equation}
as found in the literature \cite{jar86}.
We use the same reasoning to count and classify the LS in the following.

The vertex tight binding model is defined on the PL with uniform hopping strength on all nearest neighbor bonds
\begin{equation}
\label{eq: Hamiltonian}
    {\cal H}=-\sum_{<ij>} |\vec{R}_i\rangle \langle \vec{R}_j |,
\end{equation}
where the sum is only on neighbors which are connected with $\hat{e}$ vectors and $|\vec{R}\rangle$ is a state localized at PL real space point $\vec{R}$ which form an orthonormal set $\langle \vec{R}|\vec{R'}\rangle = \delta_{\vec{R},\vec{R'}}$. The hopping strength sets the energy scale and by taking it as 1 we report dimensionless quantities throughout the paper. The eigenstates $|\Psi\rangle=\sum \Psi(\vec{R}) |\vec{R}\rangle$ satisfy
\begin{equation}
\label{eq:Schrodinger}
    {\cal H} |\Psi\rangle = E |\Psi \rangle.
\end{equation}
The spectrum of this problem has only been calculated numerically. Initial numerical work on finite systems showed that the density of states near zero energy had a significant peak \cite{cho85,oda86}. The reason for the zero energy peak was first clearly identified by Kohmoto and Sutherland,  the vertex model has an infinitely degenerate manifold of strictly localized states\cite{koh86}. These eigenstates have non-zero wavefunction only on a finite number of sites of the PL, and are confined by local connectivity. Destructive interference between different sites with non-zero wavefunction prevents probability leakage to any neighboring sites.  We consider only these states with zero energy throughout the rest of the paper, and refer to them as localized states (LS) for convenience, there should not be confusion with eigenstates which are exponentially localized.

The PL is bipartite, the vertices with index 1,3 have only nearest neighbors with indices 2 or 4. Thus, the vertex model Eq.(\ref{eq: Hamiltonian}) has a spectrum which is symmetric under sign change of energy. The LS have energy zero and bipartite symmetry enables us to  split the degenerate manifold of LS into states which are defined on the (1-3) sublattice and states defined on the (2-4) sublattice. As there are equal number of these states,  we  consider the (1-3) sublattice in all our calculations. All LS we display on $V_1,V_3$ have a corresponding state in $V_4,V_2$ which can be obtained by inversion (both in real and perpendicular space). 

Arai, Tokihiro, Fujiwara and Kohmoto reported six independent types of LS and their frequencies\cite{ara88}. In the following section, we use perpendicular space projections to label and count these states.  

\section{\label{sec:Frequencies} Frequencies and labeling  of strictly localized states}

\begin{figure}[!htb]
    \centering
    \includegraphics[trim=1mm 1mm 1mm 1mm,clip,width=0.4\textwidth]{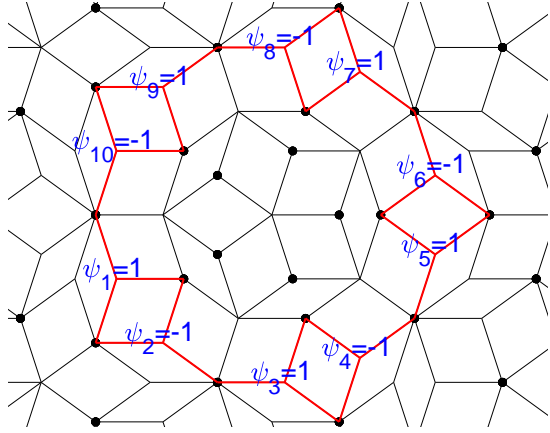}
    \caption{Type-1 LS support contains 10 points forming a ring around an S vertex. All the points in the support have index 3, points in the (2-4) sublattice are marked with black dots.  The wavefunctions $\psi$ alternate as $\pm 1$, all other points in the (1-3) sublattice have zero wavefunction. }
    \label{fig:Type1RealSpace}
\end{figure}

\begin{figure}[!htb]
    \centering
    \includegraphics[trim=1mm 1mm 1mm 1mm,clip,width=0.4\textwidth]{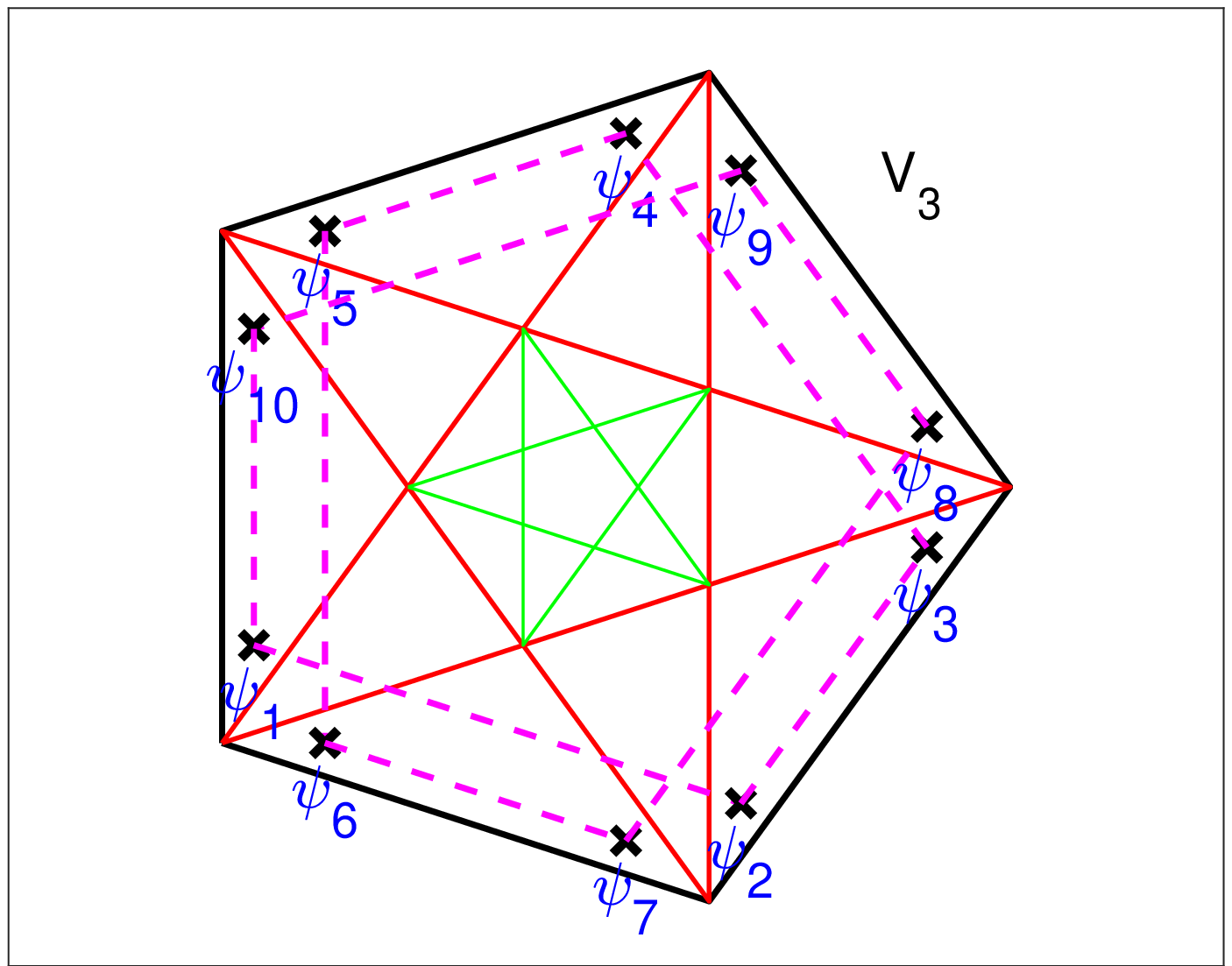}
    \caption{The perpendicular space images of the points shown in figure \ref{fig:Type1RealSpace}. Nearest neighbors are connected by vectors $\hat{e}_i-\hat{e}_j$. If all 10 points lie in $V_3$ as shown in the figure type-1 LS exists on this set. }
    \label{fig:Type1PerpSpace}
\end{figure}

\begin{figure}[!htb]
    \centering
    \includegraphics[trim=5mm 5mm 5mm 5mm,clip,width=0.4\textwidth]{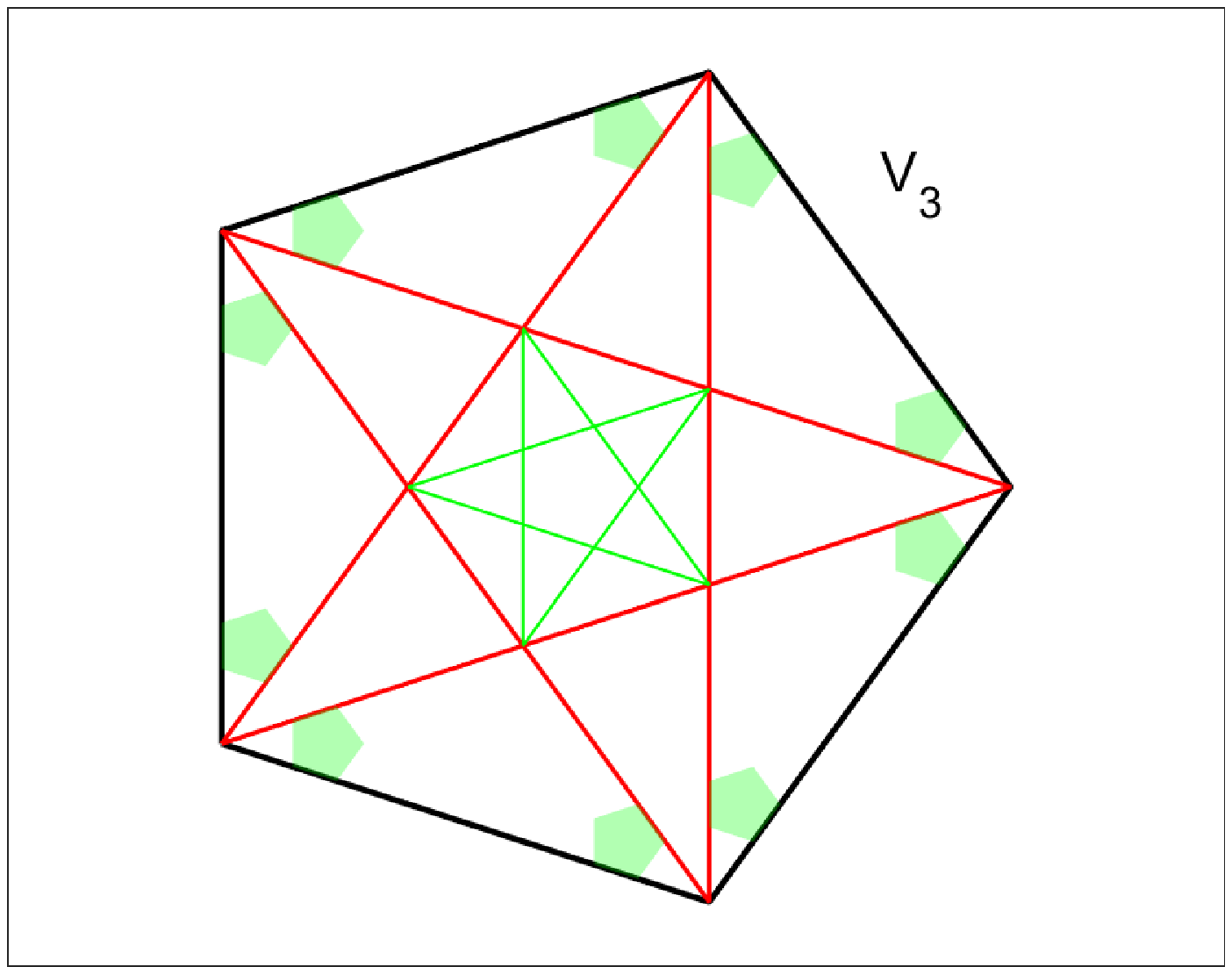}
    \caption{The allowed area for Type-1 vertices. Choosing the position of one vertex in one of the pentagons fixes the position of the remaining vertices as in Fig. \ref{fig:Type1PerpSpace}. Each pentagon has radius $\tau^{-4}$. All vertices in the support are D vertices. }
    \label{fig:Type1PerpSpaceArea}
\end{figure}

A strictly localized state (LS) is an eigenstate of the tight--binding Hamiltonian Eq.(\ref{eq: Hamiltonian}) which has non-zero density only on a finite number of lattice sites. When the support, {\it i.e.} the set of sites with non-zero wavefunction, is finite, there is a radius R beyond which the wavefunction is exactly zero. The condition for obtaining such states is more stringent than the usual definition of localized states which have exponentially decaying tails going off to infinity. 

The support of the LS must satisfy certain conditions, any one of the nearest neighbors of a point in the support must be a nearest neighbor of at least one other point in the support. Furthermore, the wavefunction defined on the support must be such that it must interfere destructively on all the nearest neighbor points. Thus, local configuration and connectivity of sites play a decisive role on the existence of LS.  Strict localization does not take place in periodic lattices, unless the LS is confined to a single unit cell. However similar strict localization can be obtained by imposing an external magnetic field on the lattice and forming flux cages \cite{vid98}.

The vertex model on the PL is bipartite, the sites with index $N_0=1,3$ (the (1-3) sublattice) are connected only to sites with index $N_0=2,4$ (the (2-4) sublattice). Thus, the existence of an eigenstate with energy $E>0$ implies that there is another eigenstate with energy $-E$. These two eigenstates can be obtained from each other by multiplying the wavefunction on one of the sublattices by -1. All LS in the vertex model have 0 energy, forming a massively degenerate manifold. As the first step, we divide this manifold into two using the bipartite symmetry of PL. If a LS $\Psi_1$ has non-zero wavefunction in both sublattices it must be degenerate with another LS $\Psi_2$ which is obtained by multiplying $\Psi_1$ with $(-1)^{N_0}$ at each lattice site. The difference $\Psi_1-\Psi_2$ is a LS defined in the (1-3) sublattice, while the sum $\Psi_1+\Psi_2$ resides on the (2-4) sublattice. Thus, all LS can be chosen to lie only in one sublattice.

We present results for states defined on the (1-3) sublattice, thus all the sites in the support of LS have perpendicular space images in $V_1$ and $V_3$, while the nearest neighbors are on $V_2$ and $V_4$. We follow the nomenclature of Ref.\cite{ara88}, who have found six real space configurations for LS. We use perpendicular space images of the points in the support of each one of these LS to count and label them.
 
A type-1 LS is formed by 10 points which form a ring around an S vertex as shown in Fig.\ref{fig:Type1RealSpace}. Each one of the points in the support are the second nearest neighbors of two more points in the support and the wavefunction alternates sign going from one point to its second nearest neighbor. Now consider the perpendicular space images of all the points in the support as shown in  Fig.\ref{fig:Type1PerpSpace}. 

In real space a point in the support is connected to its second nearest neighbor by a vector $\hat{e}_i - \hat{e}_j$, the minus sign showing that all the points in the support have the same index. Thus, the vectors connecting the perpendicular space images also have the same form  $\hat{e}_i - \hat{e}_j$, obtained by Eq.(\ref{eq:eVectorCorrespondence}). As shown in Fig.\ref{fig:Type1PerpSpace} the connecting vectors define a decagon with 5 short and 5 long sides in perpendicular space. One can imagine moving one of the vertices of this shape will also move all the other vertices if all the connecting vectors are held rigid. As long as all the vertices of the decagon are still inside $V_3$ a type-1 LS can be defined on these points. The mapping between real space and perpendicular space is dense and uniform as discussed in the previous section, thus the area covered by moving one of the vertices while keeping all the vertices inside $V_3$ is proportional to the frequency of type-1 states. 

The allowed region for the movement of any one of the vertices is shown by the ten identical green pentagons in Fig.\ref{fig:Type1PerpSpaceArea}. These pentagons have radius $\tau^{-4}$, thus the frequency of type-1 states can be calculated by dividing the area of one of these pentagons with the sum of the areas of $V_1$ and $V_3$
\begin{equation}
    f_{T1}=\frac{\tau^{-8}}{1+\tau^2}=\frac{123-76\tau}{5}  \simeq 5.883 \times 10^{-3},
\end{equation}
which is the result obtained by Arai et.al \cite{ara88}.

The perpendicular space position of any one of the vertices inside the pentagons can be used to label and distinguish different type-1 states.  Equivalently, we can use the perpendicular space position of the central S vertex which lies in $V_1$ to label the states.  Thus  a type one state is uniquely labeled by a vector $|T1,\vec{R}_\perp\rangle$.  Here the vector $\vec{R}_\perp$ must lie inside a regular pentagon of radius $\tau^{-4}$ centered at the origin. This region has smaller area than the area of all S vertices, so only $\tau^{-4}\simeq 14.59\% $ of S vertices have a type-1 state around them. 
The perpendicular space positions of points in the support can be reached by adding $\hat{e}_0+\hat{e}_2+\hat{e}_3-\hat{e}_1$ or $\hat{e}_0+\hat{e}_2+\hat{e}_3-\hat{e}_4$ and their five fold rotations to $\vec{R}_\perp$. 
We should also stress that $\vec{R}_\perp$ is not defined on the set of real numbers $\mathbb{R}^2$, but any vectors that have rational inner products with star vectors are removed from this set. Nonetheless, this vector uniquely defines the local neighborhood in real space and the density of similar local configurations can be counted as above.
\begin{figure}[!htb]
    \centering
    \includegraphics[trim=1mm 1mm 1mm 1mm,clip,width=0.4\textwidth]{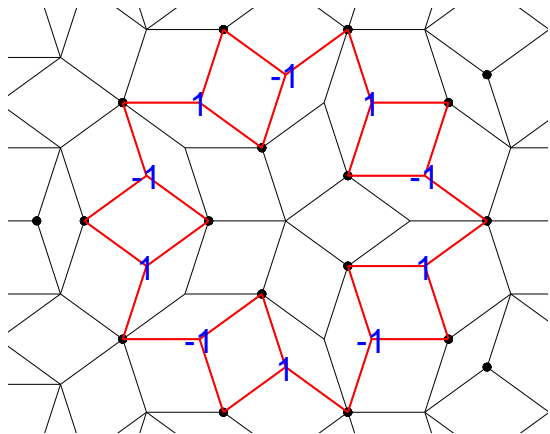}
    \caption{Type-2 LS has a support of 10 points around an S5 vertex. All points in the support have index 3 and are D vertices. The non-zero wavefunctions are written over the lattice sites.}
    \label{fig:Type2RealSpace}
\end{figure}
\begin{figure}[!htb]
    \centering
    \includegraphics[trim=5mm 5mm 5mm 5mm,clip,width=0.4\textwidth]{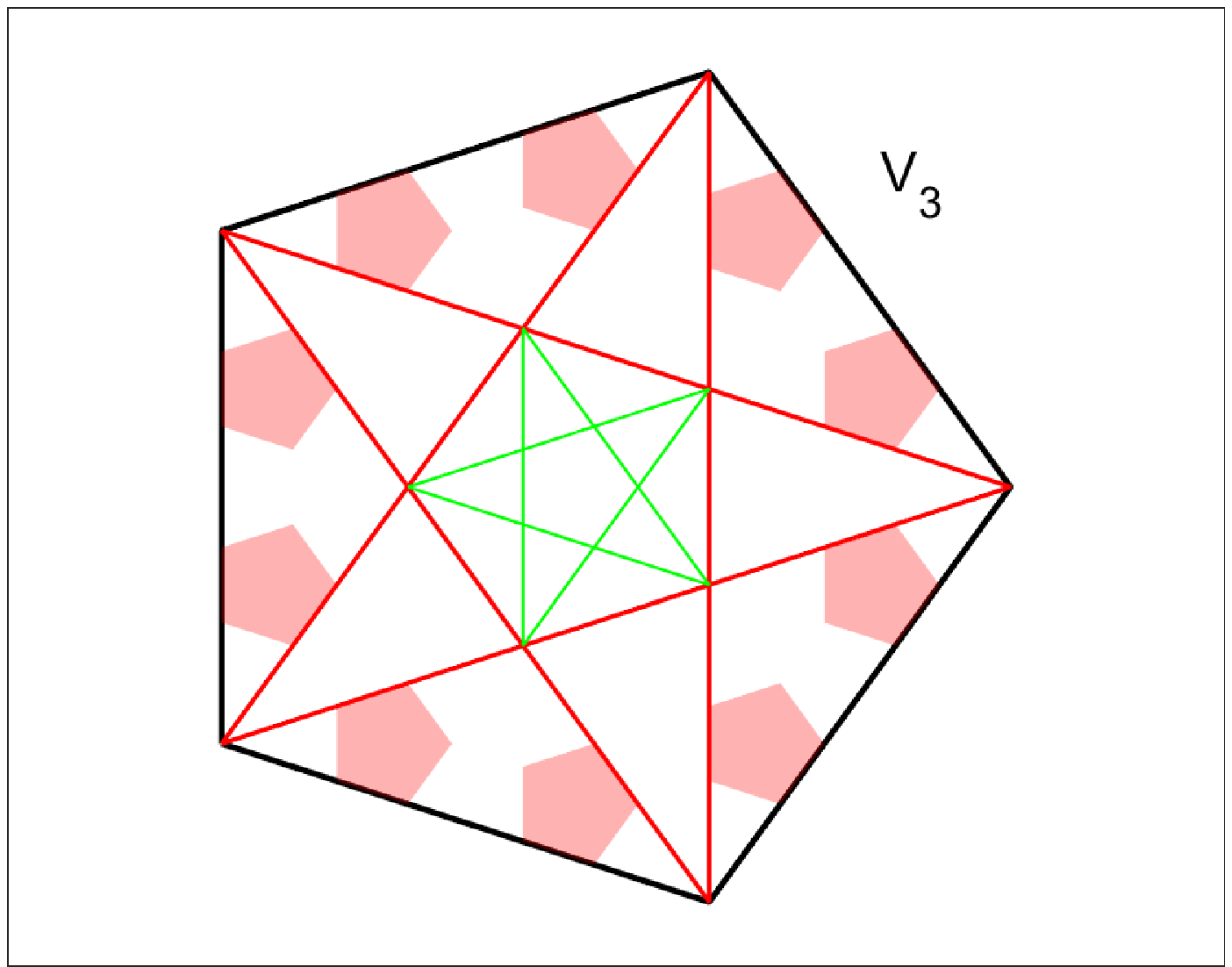}
    \caption{Allowed areas for the ten points in the support of type-2 LS. Each pentagon is inside a D region and has radius $\tau^{-3}$.}
    \label{fig:Type2PerpSpaceArea}
\end{figure}
\begin{figure}[!htb]
   \centering
    \includegraphics[trim=1mm 1mm 1mm 1mm,clip,width=0.3\textwidth]{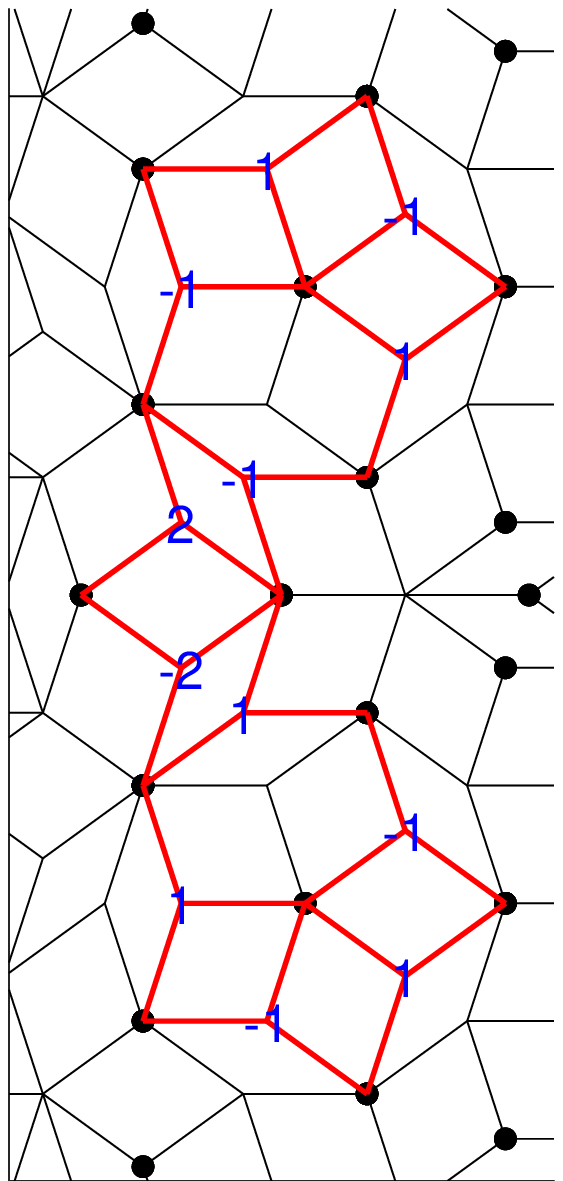}
    \caption{Type-3 LS with $\hat{e}_0$ orientation. There are 12 points in the support, 2 of which have index 1.}
    \label{fig:Type3RealSpace}
\end{figure}
\begin{figure}[!htb]
    \centering
    \includegraphics[trim=5mm 5mm 5mm 5mm,clip,width=0.4\textwidth]{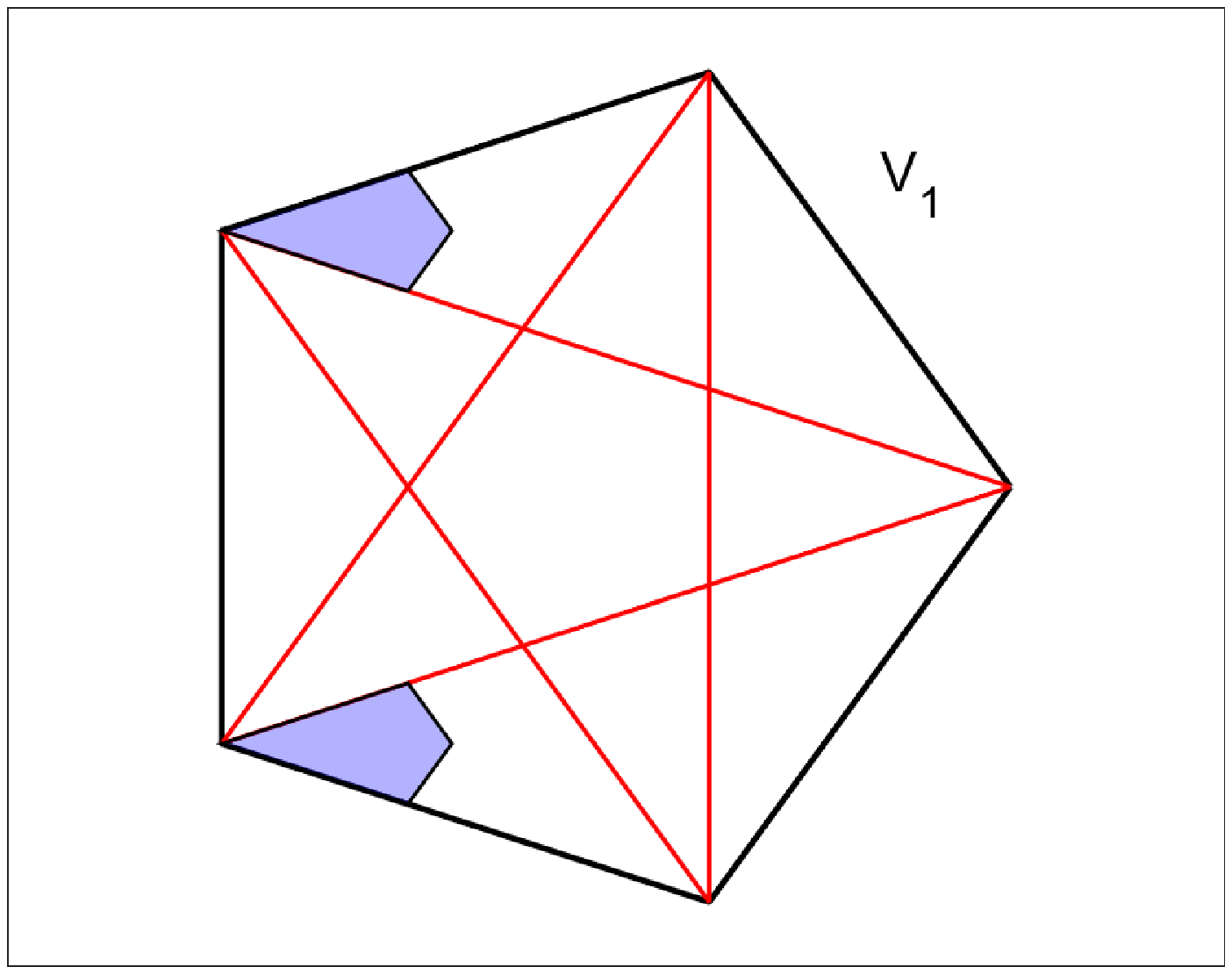}
    \includegraphics[trim=5mm 5mm 5mm 5mm,clip,width=0.4\textwidth]{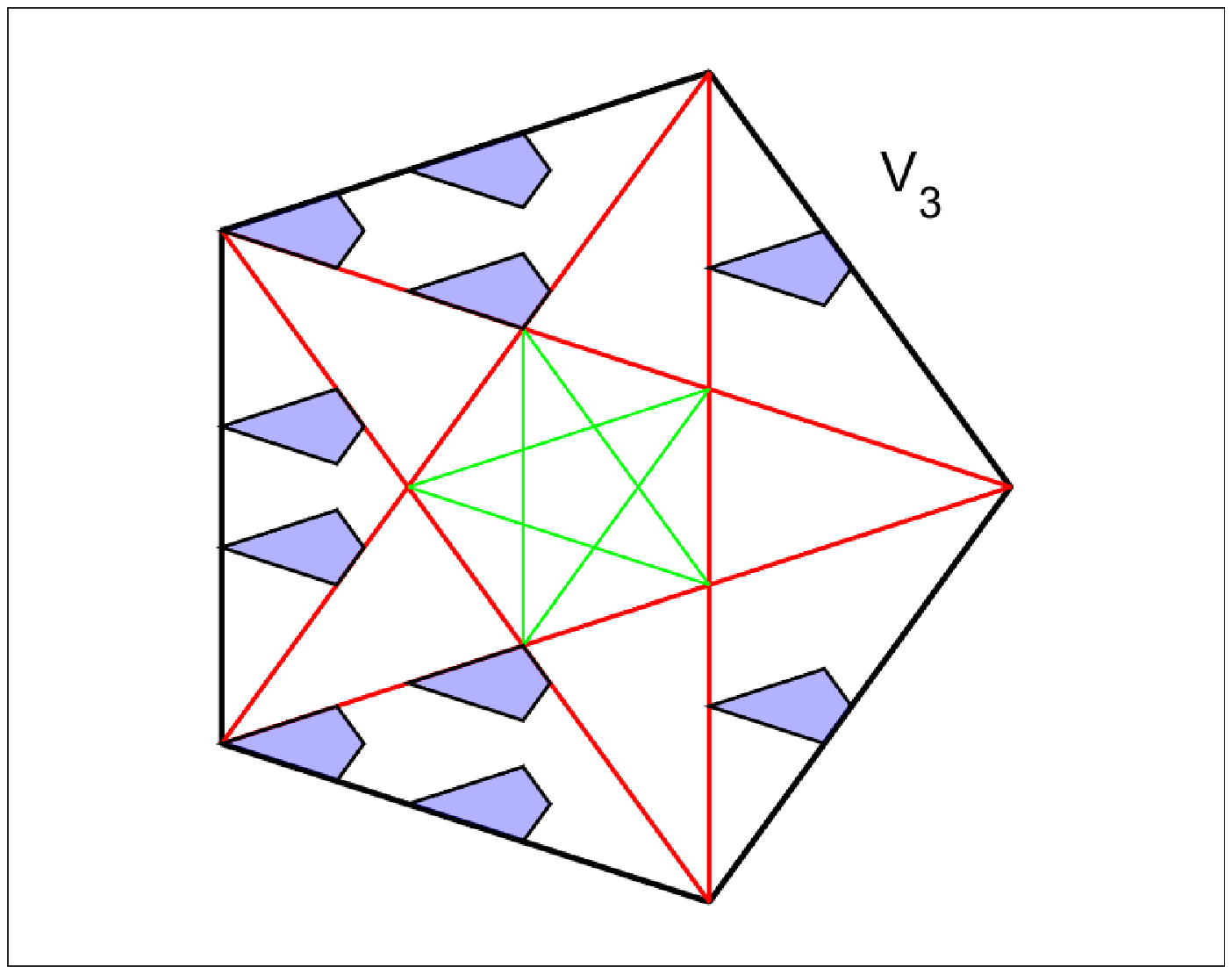}
     \includegraphics[trim=7mm 7mm 7mm 7mm,clip,width=0.6\textwidth]{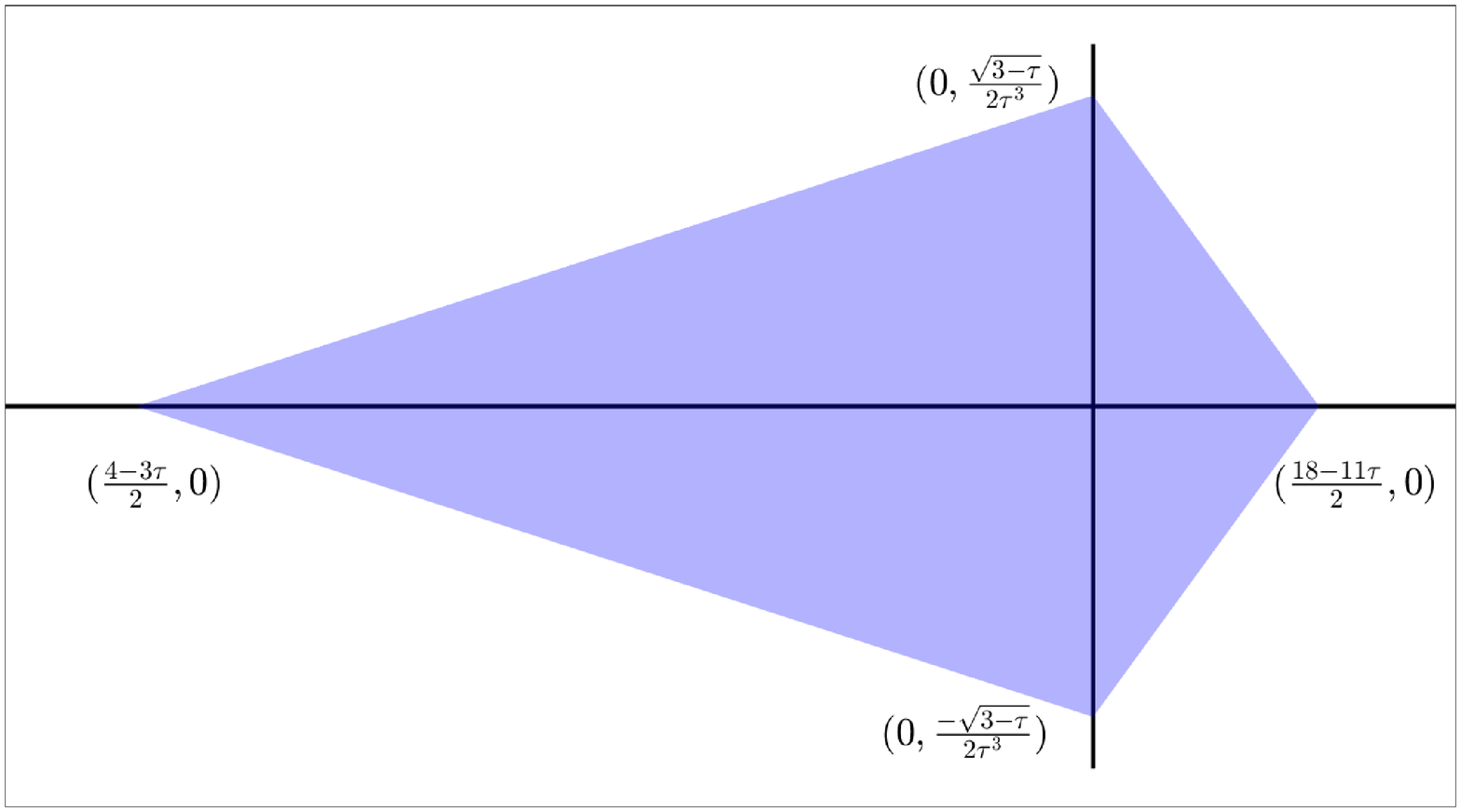}
    \caption{Allowed areas for the vertices of type-3 with $\hat{e}_0$ orientation. $V_1$ is scaled by $\tau$ for visual clarity, all 10 kites have the same size. The perpendicular space vector $R_\perp$ used to label type-3 vertices is chosen to lie in the kite shown in panel (c).}
    \label{fig:Type3PerpSpaceArea}
\end{figure}
\begin{figure}[!htb]
    \centering
    \includegraphics[trim=5mm 5mm 5mm 5mm,clip,width=0.4\textwidth]{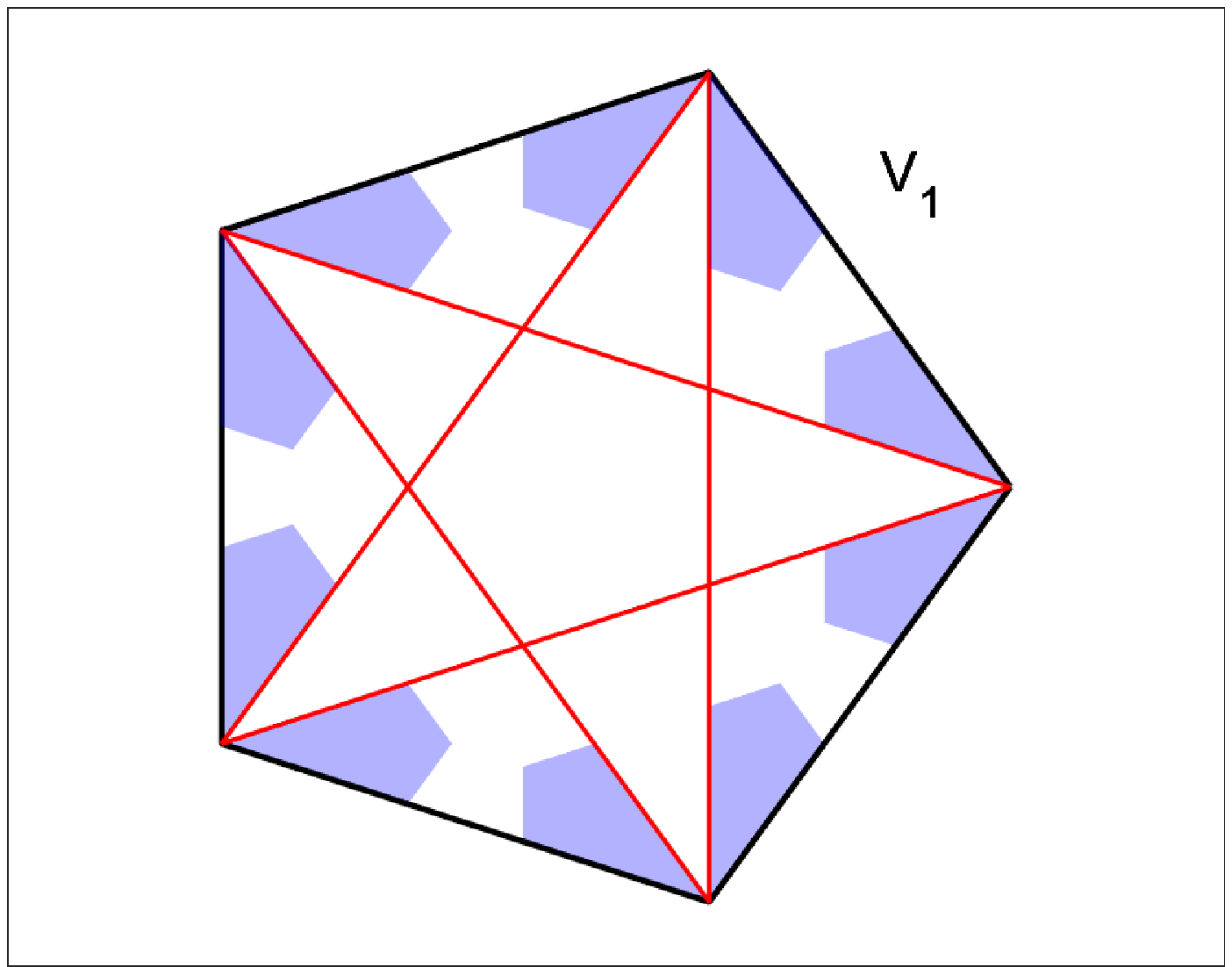}
    \includegraphics[trim=5mm 5mm 5mm 5mm,clip,width=0.4\textwidth]{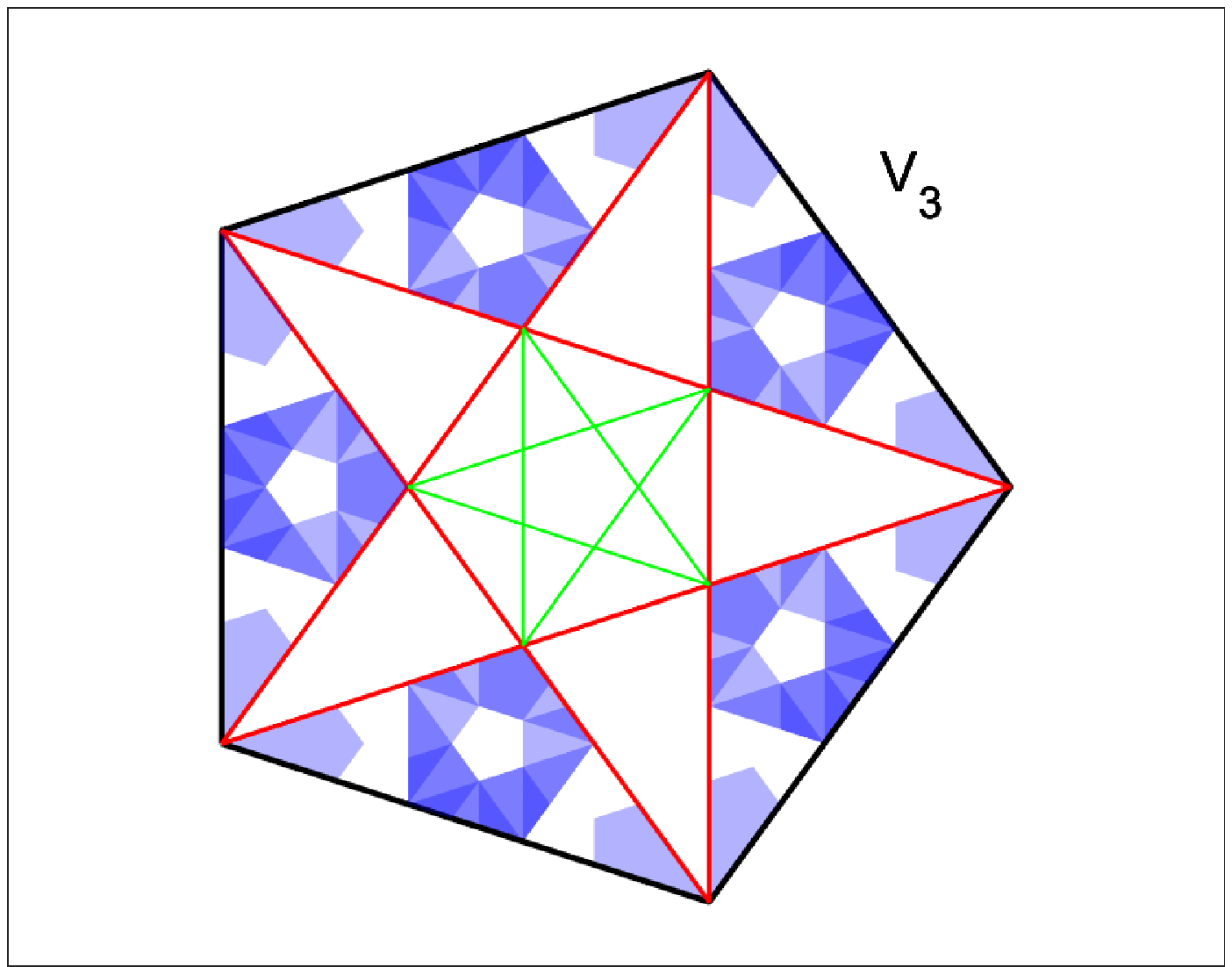}
    \caption{Allowed areas for all orientations of type-3 states. While 8 out of the 12 areas have overlaps between type-3 states of different orientation 4 sites remain unique to each orientation.}
    \label{fig:Type3PerpSpaceAreaRotated}
\end{figure}

Type-2 LS are similar to type-1 as their support contains 10 points arranged in a second nearest neighbor ring with sign alternating wavefunction, see Fig.\ref{fig:Type2RealSpace}. However, the ring is centered at an S5 vertex instead of an S vertex. Plotting the perpendicular space positions of the sites in the support, we observe that these points create a regular decagon. Moving this decagon with the constraint that all the vertices stay inside $V_3$ defines the allowed areas of type-2. As shown in Fig.\ref{fig:Type2PerpSpaceArea}, the allowed areas are pentagons of radius $\tau^{-3}$. The ratio of the area of one of these pentagons to the total area of $V_1$ and $V_3$ gives the type-2 frequency
\begin{equation}
    f_{T2}=\frac{\tau^{-6}}{1+\tau^2}=\frac{47-29\tau}{5} \simeq 1.540 \times 10^{-2},
\end{equation}
in agreement with Ref.\cite{ara88}. We label the states once again with a perpendicular space vector $| T2,\vec{R}_\perp\rangle$, and use the perpendicular space position of the central S5 vertex as $\vec{R}_\perp$. Thus, $\vec{R}_\perp$ must lie in a regular pentagon of radius $\tau^{-3}$, which exactly is the perpendicular space area of S5 vertices, showing that all S5 vertices have a type-2 LS encircling them.  The perpendicular space positions of the sites in the support of the LS can be reached from $\vec{R}_\perp$ by adding $-\hat{e}_0+\hat{e}_4$ , $-\hat{e}_0+\hat{e}_1$ and their five-fold rotations.

Type-3 LS are different from type-1 and type-2 in two important regards. First they break the five-fold rotational symmetry, thus five different orientations of type-3 states exist. Second, the support of type-3 contains 2 Q vertices which have index 1, so the perpendicular space image of the type-3 support has a component in $V_1$ as well as $V_3$.  One orientation of type-3 is displayed in Fig.\ref{fig:Type3RealSpace}. Once again one can calculate the perpendicular space image of the 12 points in the support with respect to each other. Moving these points with the constraint that all of them remain in  $V_1$ and $V_3$ define the allowed areas for this particular orientation of type-3 as shown in Fig.\ref{fig:Type3PerpSpaceArea}. Each one of the 12 regions is a kite with three $\frac{3 \pi}{5}$ and one $\frac{\pi}{5}$ angles.   The long diagonal of the kite has length $7-4\tau$ and the short diagonal $\sqrt{3-\tau}\tau^{-3}$, which gives the fraction of the one particular orientation of type-3 as
\begin{equation}
\begin{split}
    f_{T3,0} =& \frac{(7-4\tau)2\tau^{-3}}{5\tau(1+\tau^2)}\\ \nonumber =& \frac{68-42 \tau}{5} \\ \nonumber
    \simeq& 8.514 \times 10^{-3}.
\end{split}
\end{equation}
\begin{figure}[!htb]
    \centering
    \includegraphics[trim=2mm 1mm 1mm 1mm,clip,width=0.3\textwidth]{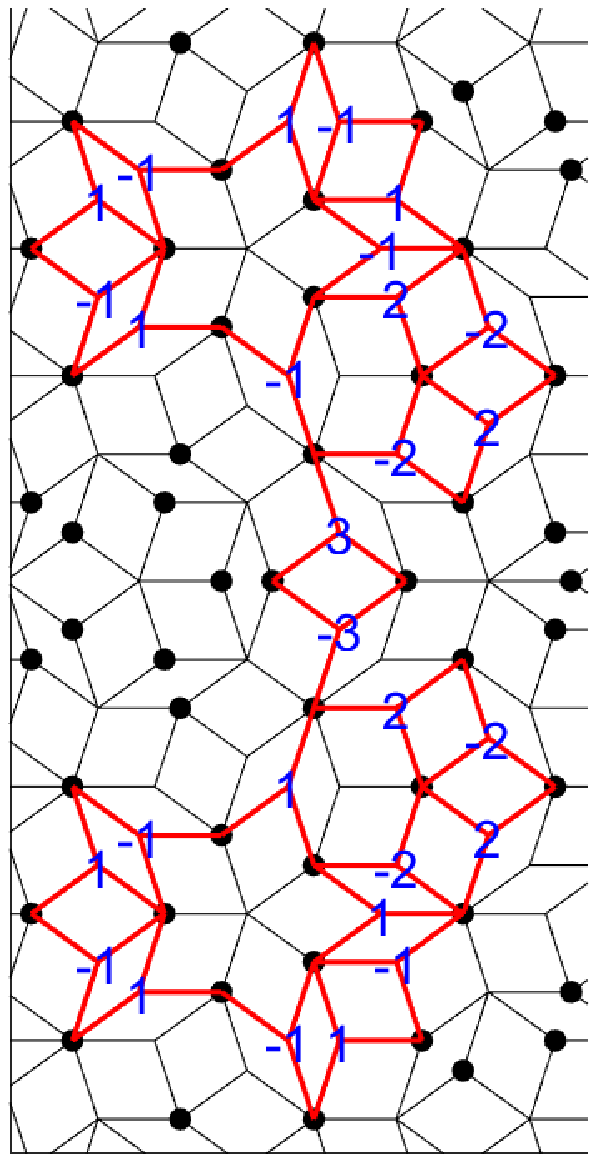}
    \caption{Type-4 localized state with $\hat{e}_0$ orientation.There are 18 sites with index 3, and 10 sites with index 1 in the support.}
    \label{fig:Type4RealSpace}
\end{figure}
\begin{figure}[!htb]
    \centering
    \includegraphics[trim=5mm 5mm 5mm 5mm,clip,width=0.4\textwidth]{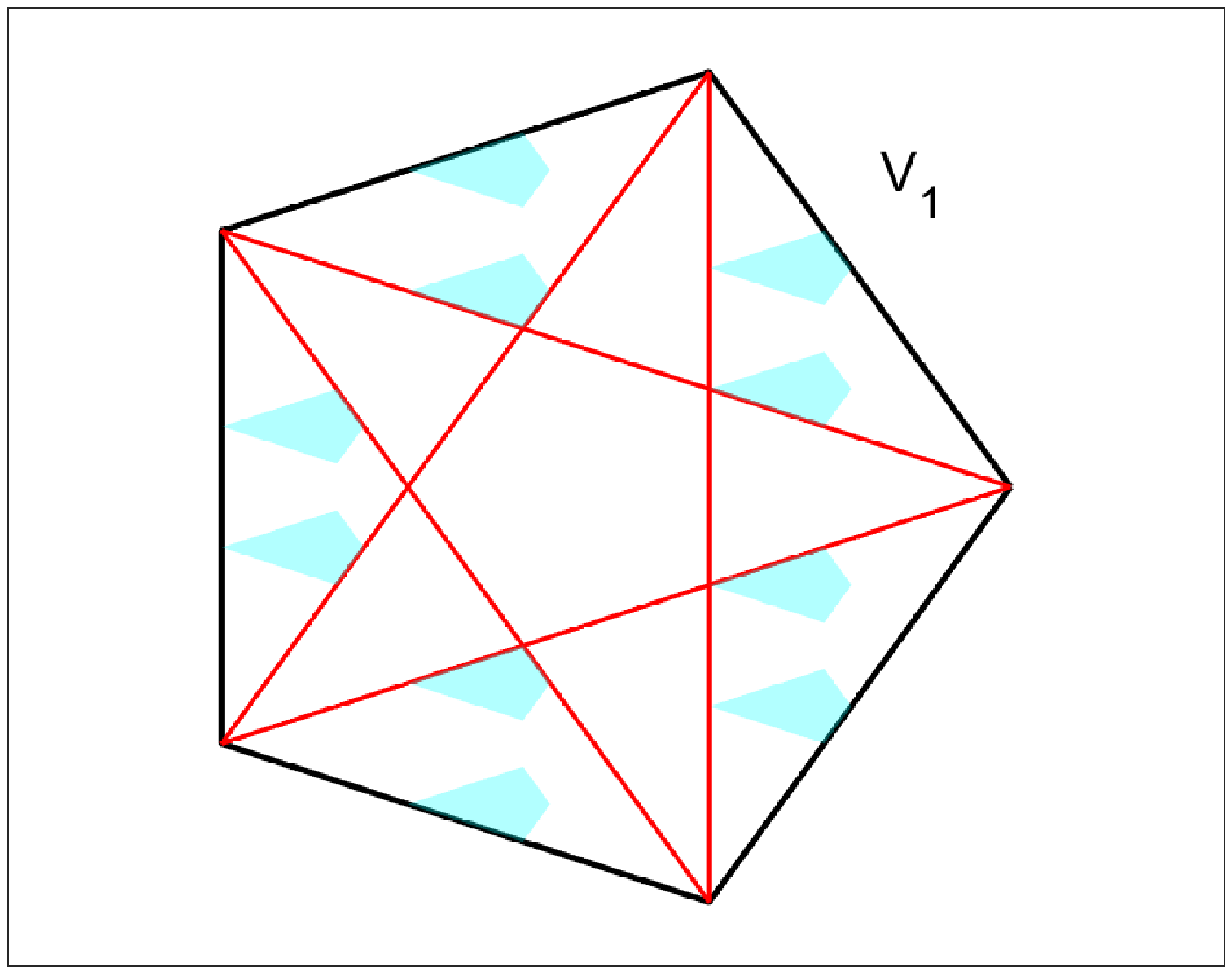}
    \includegraphics[trim=5mm 5mm 5mm 5mm,clip,width=0.4\textwidth]{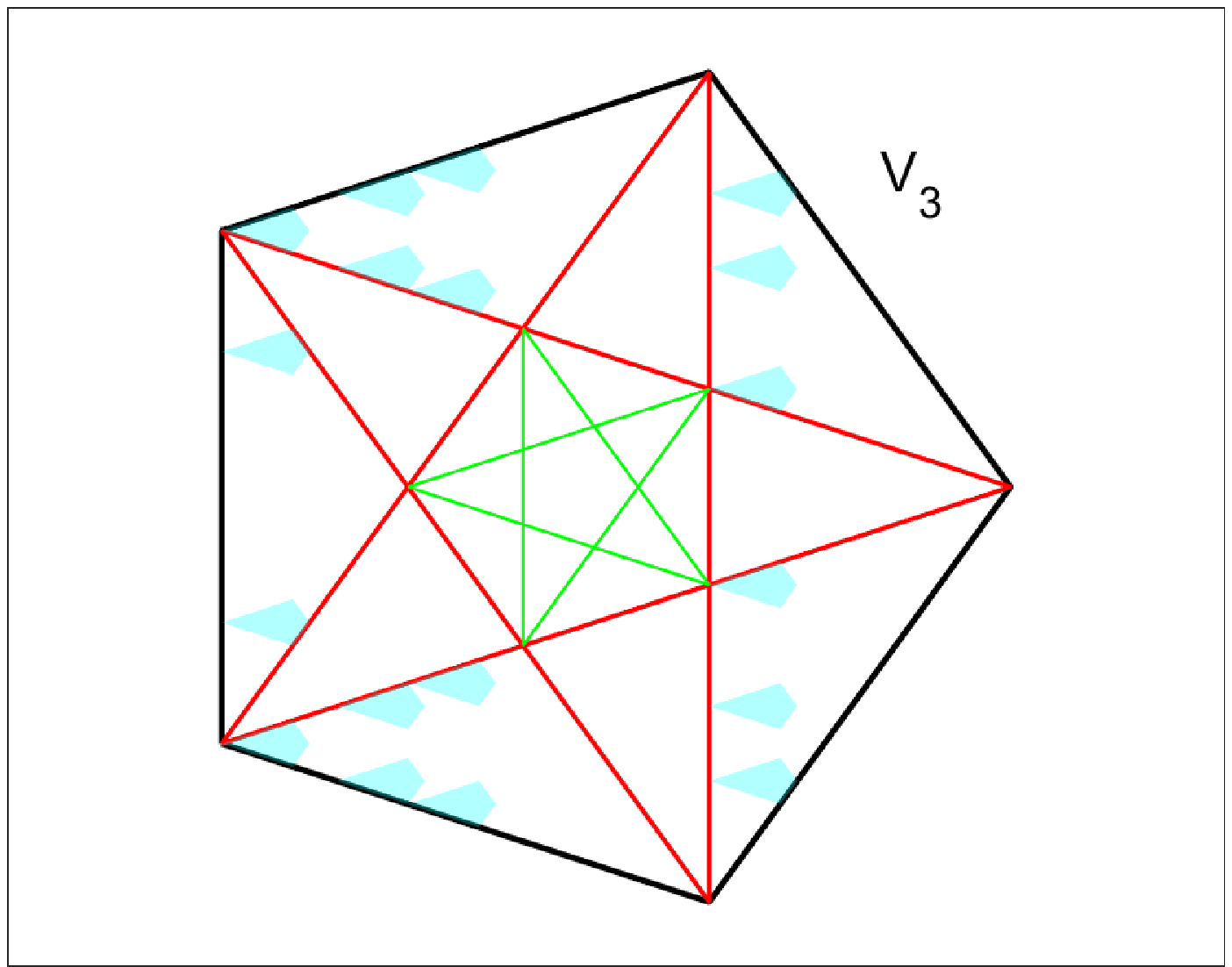}
    \caption{Allowed areas for the vertices of type-4 LS with $\hat{e}_0$ orientation. Each kite is smaller by a factor of $\tau$ compared to the type-3 areas in Fig.\ref{fig:Type3PerpSpaceArea}. The labeling inside the kite is done similarly to type-3 states as in Fig.\ref{fig:Type3PerpSpaceArea}}
    \label{fig:Type4PerpSpaceArea}
\end{figure}
\begin{figure}[!htb]
    \centering
  \includegraphics[trim=5mm 5mm 5mm 5mm,clip,width=0.4\textwidth]{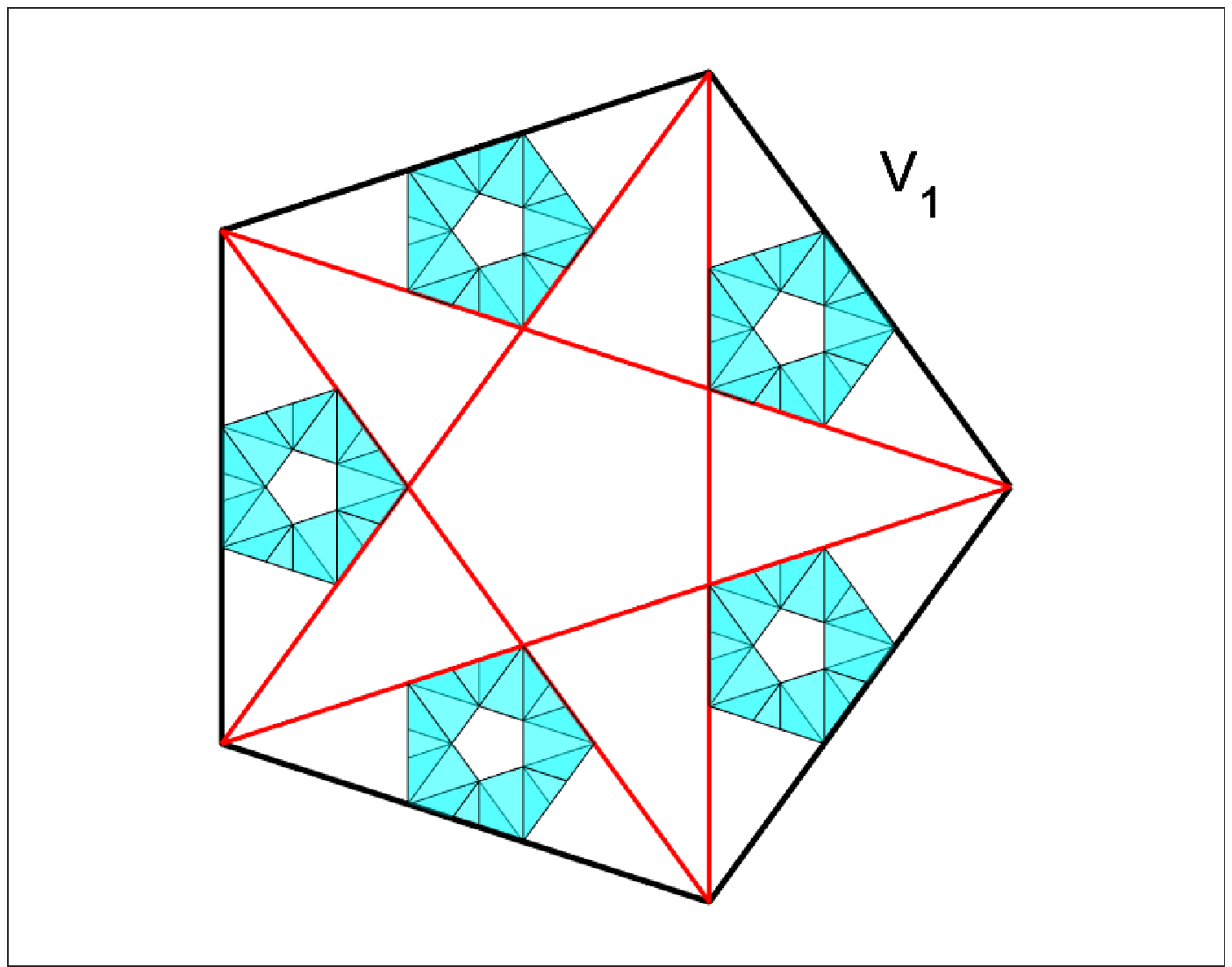}
    \includegraphics[trim=5mm 5mm 5mm 5mm,clip,width=0.4\textwidth]{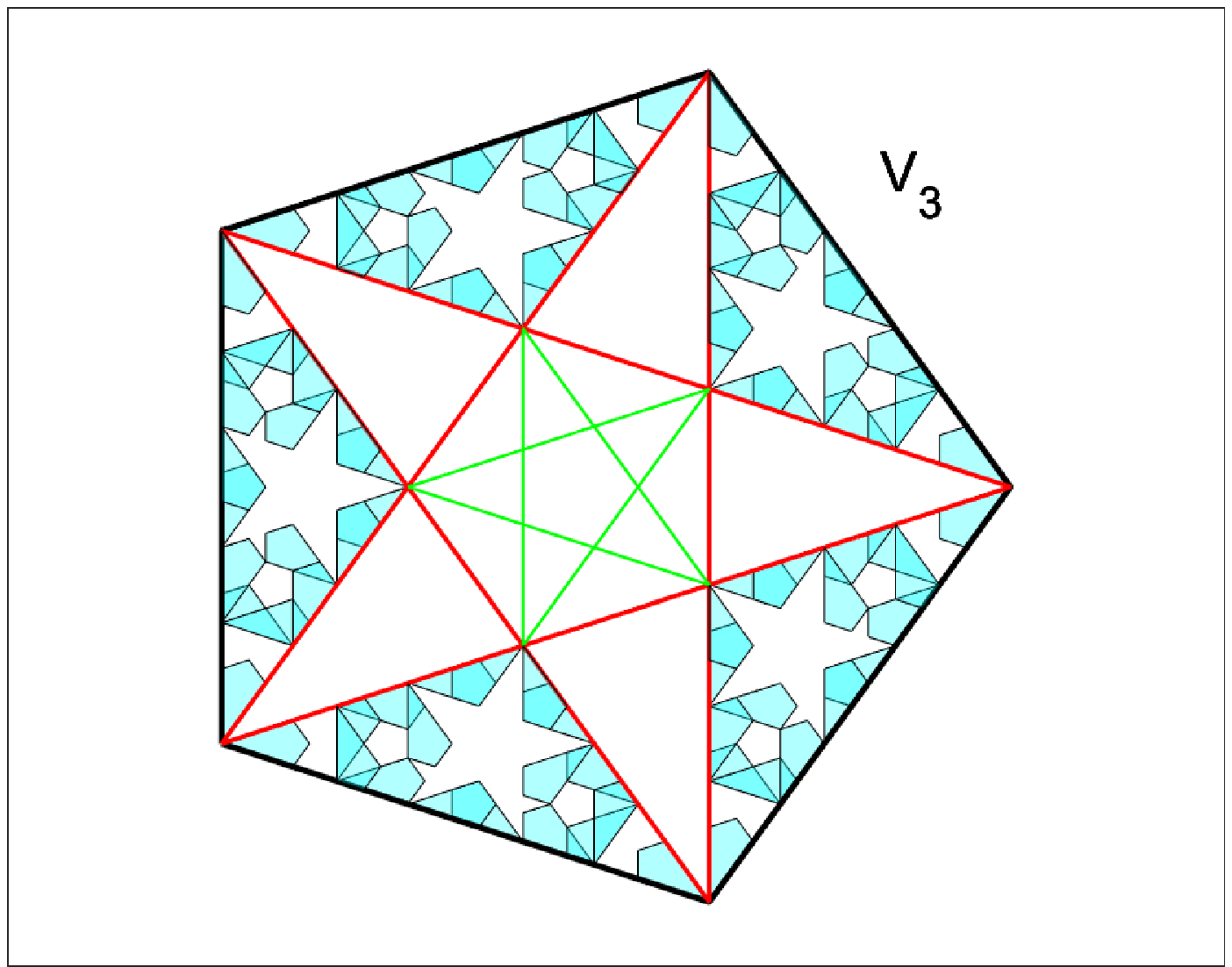}
    
    \caption{Allowed areas for Type-4 LS of all orientations superimposed. There are only two unique sites for each orientation which are in $V_3$.}
    \label{fig:Type4PerpSpaceAreaRotated}
\end{figure}
All other orientations of type-3 can be obtained from $\hat{e}_0$ orientation by $2\pi/5$ rotations, and their allowed regions follow the same pattern, only with the correspondence in Eq.(\ref{eq:eVectorCorrespondence}). In Fig.\ref{fig:Type3PerpSpaceAreaRotated} we show allowed areas for all the orientations. It can be seen that regions in $V_3$ for two different orientations can overlap. The overlap in perpendicular space shows that two different orientations of type-3 in real space can have the same point in their support. The real space overlaps of type-3 can be used to define LS which have linked second-neighbor support forming longer LS, as first proposed by Kohmoto and Sutherland \cite{koh86}, but such ring states  will not be independent from the LS defined here. The independence and overlaps of different LS will be explored in the next section. Even if different type-3 states are not automatically orthogonal they remain independent and the total type-3 fraction is
\begin{equation}
    f_{T3}=\sum_{m=0}^{4}f_{T3,m} = 68-42 \tau \simeq 4.257 \times 10^{-2},
\end{equation}
 in agreement with Ref\cite{ara88}. Type-3 states can be labeled by their orientation and the relative position of the point within the allowed perpendicular space region $|T3,\hat{e}_m,\vec{R}_\perp\rangle$ where $\hat{e}_m$ denotes the orientation and $\vec{R}_\perp$ is limited inside the quadrilateral defined by $(4-3\tau)/2 \hat{e}_m$,  $(18-11\tau)/2 \hat{e}_m $ and $\frac{\pm 1}{2\tau^3} \sqrt{3-\tau} \hat{z}\times\hat{e}_m$, see Fig.\ref{fig:Type3PerpSpaceArea}.

Type-4 has 28 sites in its support as seen in Fig.\ref{fig:Type4RealSpace}, and it breaks the five-fold rotational symmetry. The allowed areas for one orientation are displayed in Fig.\ref{fig:Type4PerpSpaceArea}, and one can notice that the allowed area is a scaled version of the type-3 kite by a factor of $\tau^{-1}$. Which allows us to write
\begin{equation}
    f_{T4,0}=\tau^{-2} f_{T3,0}=  \frac{178-110\tau}{5} \simeq 3.252 \times 10^{-3}.
\end{equation}
Considering all five orientations, we see that there is significant overlap between different orientations of type-4 particularly in $V_1$ as seen in Fig.\ref{fig:Type4PerpSpaceAreaRotated}. We establish the independence of type-4 states from previous types and each other in the next section. The total frequency is 
\begin{equation}
    f_{T4}=\sum_{m=0}^{4}f_{T4,m} = 178-110 \tau \simeq 1.626 \times 10^{-2},
\end{equation}
in agreement with Ref\cite{ara88}. We label type-4 states in a similar manner to type-3, $|T4,\hat{e}_m,\vec{R}_\perp\rangle$, however the kite constraining the perpendicular space region is smaller by a factor of $\tau$, given by the points $ -(7-4\tau)/2 \hat{e}_m $,  $(18\tau-29)/2 \hat{e}_m $ and $\frac{\pm 1}{2\tau^{4}} \sqrt{3-\tau} \hat{z}\times\hat{e}_m$.

\begin{figure}[!htb]
    \centering
    \includegraphics[trim=1mm 1mm 1mm 1mm,clip,width=0.4\textwidth]{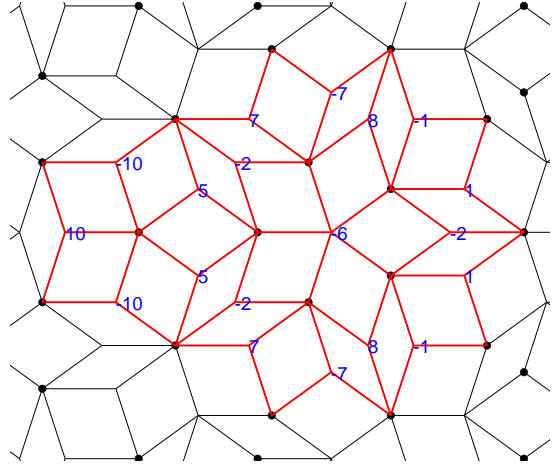}
    \caption{Type-5 LS with $\hat{e}_0$ orientation in real space. Five of the sites in the support have index 1, while the remaining 14 have index 3. }
    \label{fig:Type5RealSpace}
\end{figure}

\begin{figure}[!htb]
    \centering
    \includegraphics[trim=5mm 5mm 5mm 5mm,clip,width=0.4\textwidth]{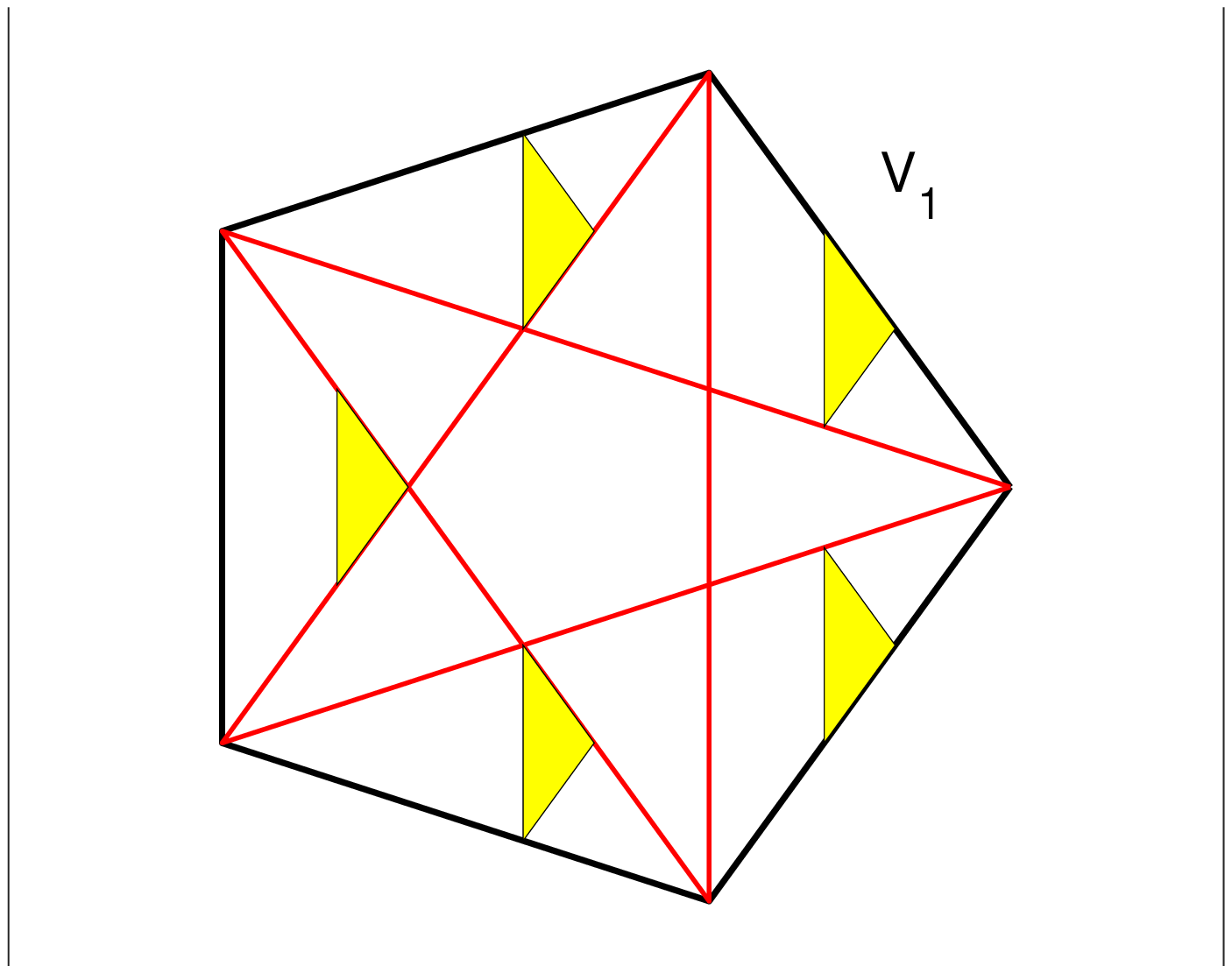}
    \includegraphics[trim=5mm 5mm 5mm 5mm,clip,width=0.4\textwidth]{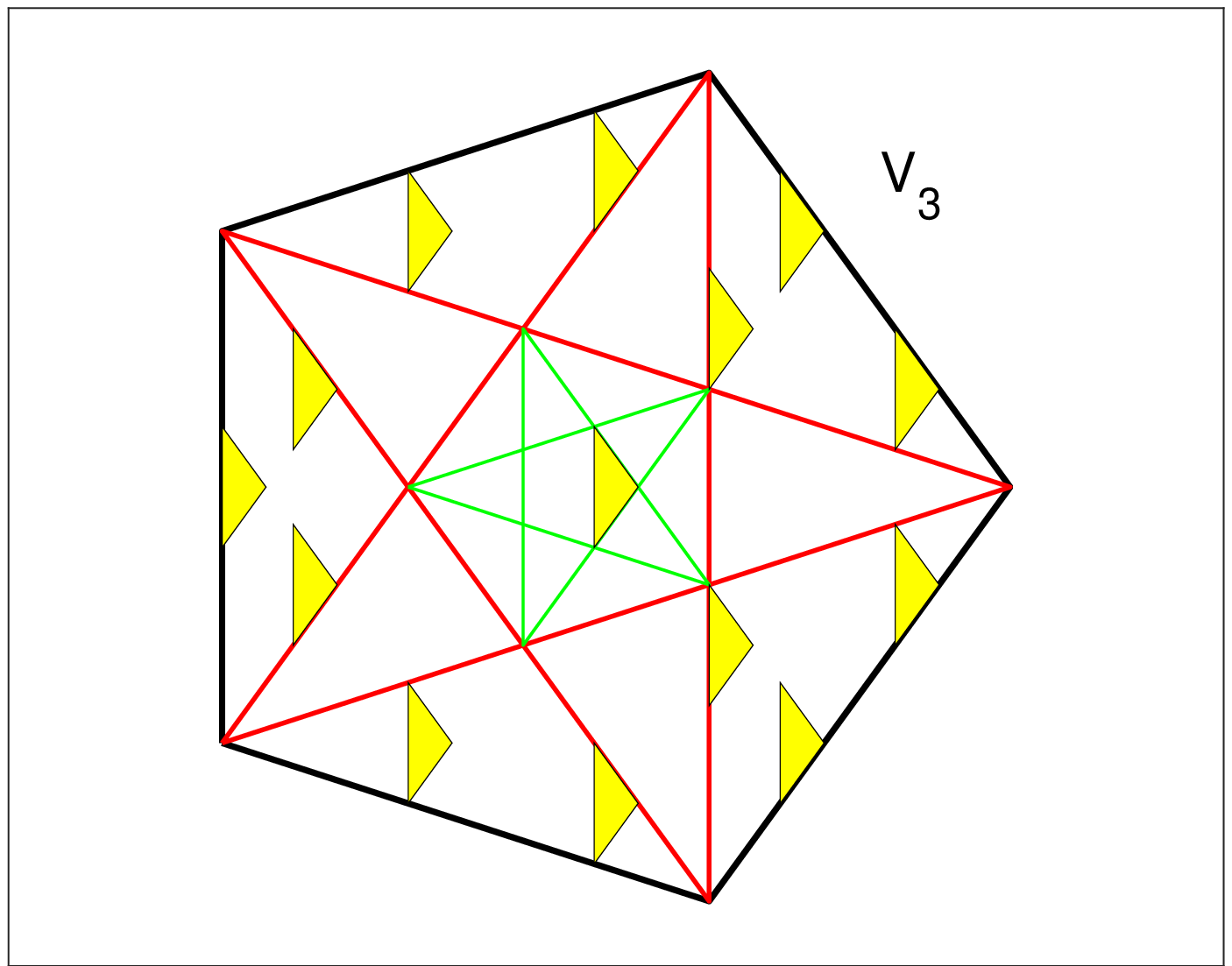}
    \caption{Allowed areas for vertices of type-5 LS with $\hat{e}_0$ orientation are triangles. The central area in $V_3$ is used to label type-5 states.}
    \label{fig:Type5PerpSpaceArea}
\end{figure}

\begin{figure}[!htb]
    \centering
     \includegraphics[trim=5mm 5mm 5mm 5mm,clip,width=0.4\textwidth]{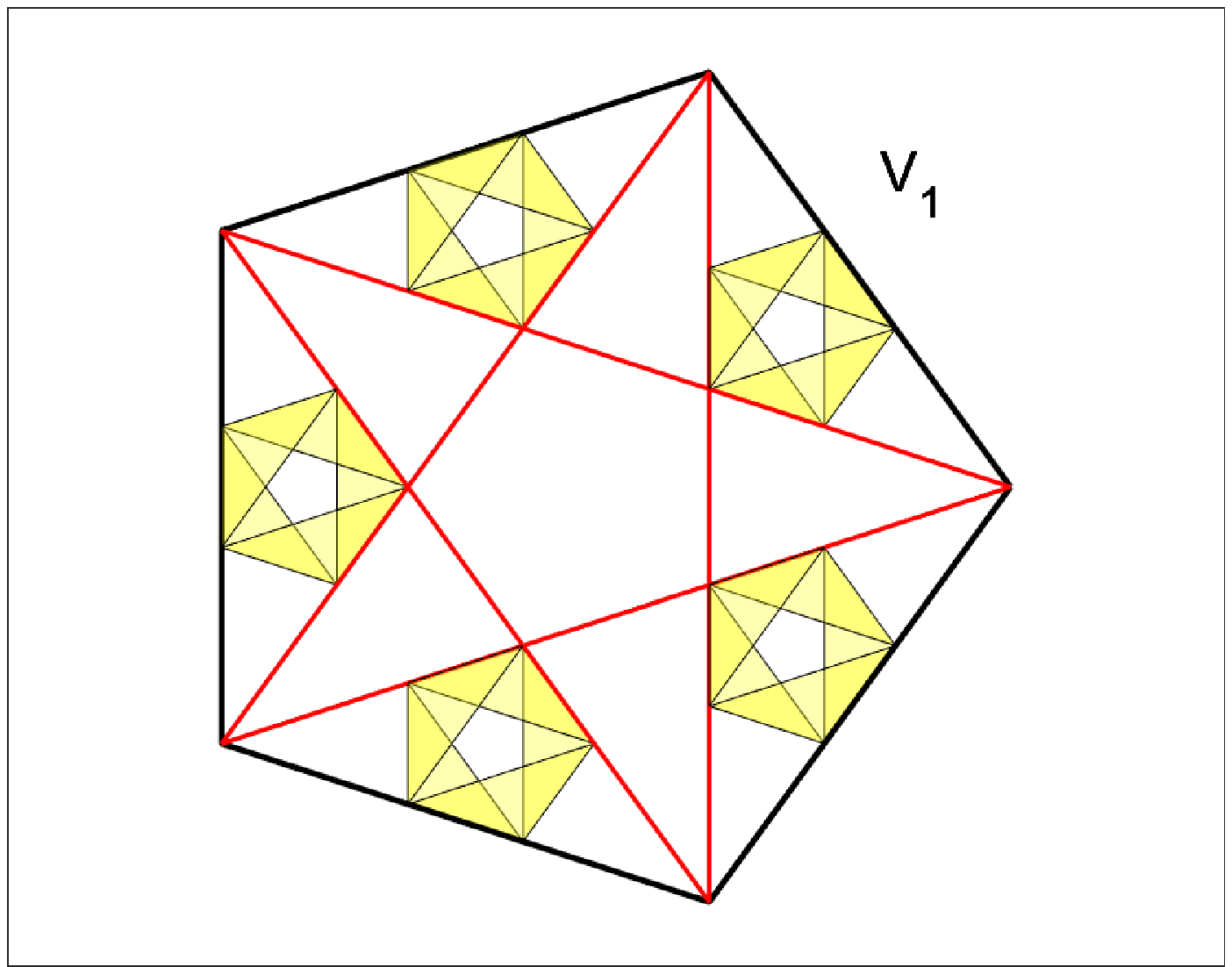}
    \includegraphics[trim=5mm 5mm 5mm 5mm,clip,width=0.4\textwidth]{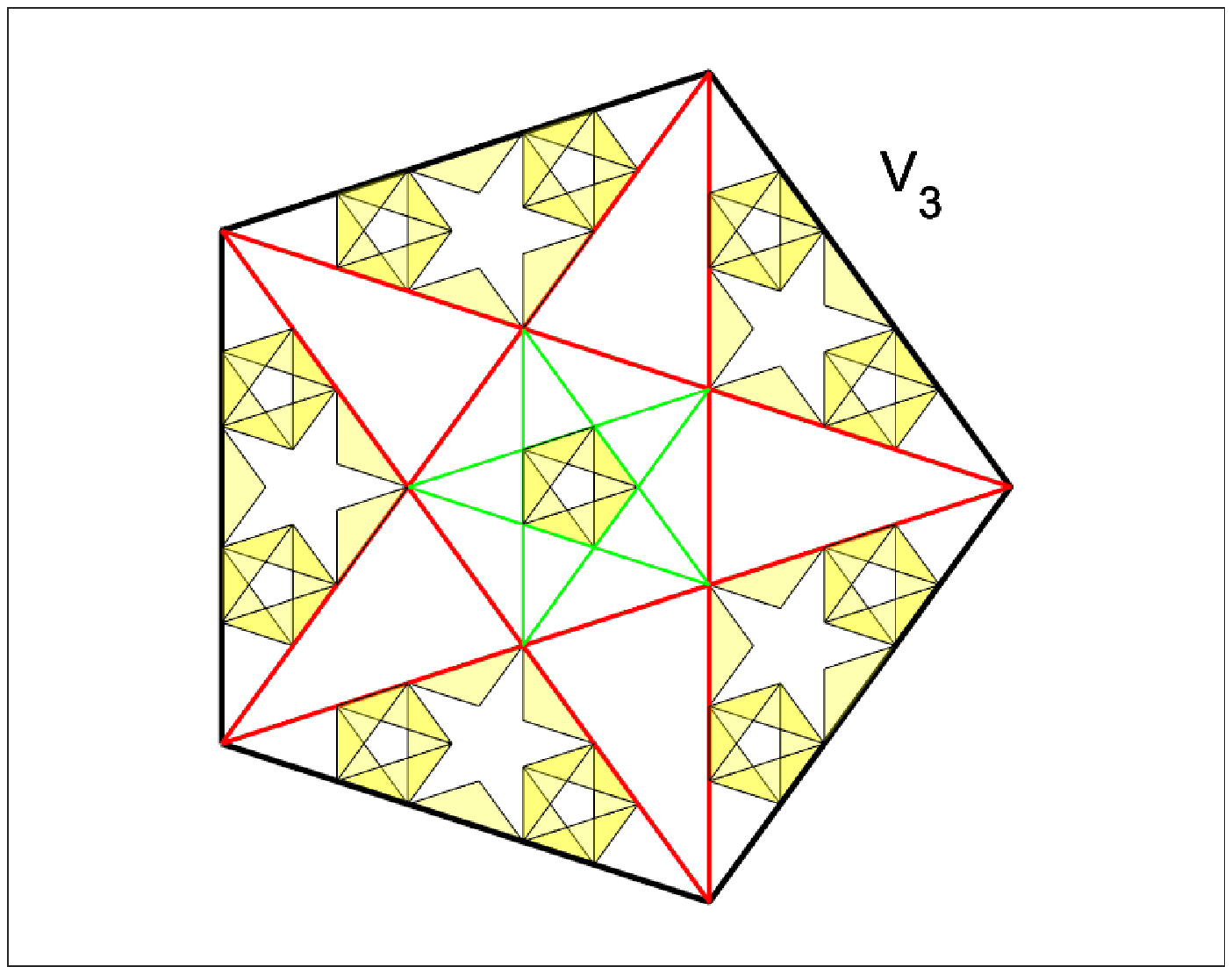}
    \caption{Allowed areas for all orientations of type-5. Notice that two allowed areas for different orientations can overlap in both $V_1$ and $V_3$. }
    \label{fig:Type5PerpSpaceAreaRotated}
\end{figure}

Type-5 LS have 19 sites in their support and they break the fivefold symmetry as shown in Fig\ref{fig:Type5RealSpace}. These are the only LS which have a density on a S5 vertex, and the triangular area in the S5 region shown in Fig.\ref{fig:Type5PerpSpaceArea} is used to label them. Plotting all five possible orientations in Fig.\ref{fig:Type5PerpSpaceAreaRotated}, we see that not all S5 vertices support a type-5 LS. An S5 site can support one  type-5 LS or two different orientations of type-5.  The allowed region for the state $|T5,\hat{e}_0,\vec{R}_\perp\rangle$ with $\hat{e}_0$ orientation is a triangle with vertices at $\hat{e}_0/\tau^3$,$\hat{e}_1/\tau^3$ and $\hat{e}_4/\tau^3$, which gives an area
\begin{equation}
    f_{T5,0}=\frac{1}{5\tau^8}= \frac{34-21\tau}{5} \simeq 4.257\times 10^{-3}.
\end{equation}
When we consider all orientations of the type-5 states the frequency becomes
\begin{equation}
    f_{T5}=5 f_{T5,0}= \frac{1}{\tau^8}= 34-21 \tau \simeq 2.128 \times 10^{-2}.
\end{equation}

This expression is a factor of $\tau$ larger than the value reported in Ref\cite{ara88}.  This frequency gives us the number of type-5 states that are independent of each other, but in the next section we show that when other types of LS are taken into account, two type-5 states sharing the same S5 vertex are not independent from each other. The number of type-5 states that are independent from each other and types 1 to 4 is in agreement with the previously reported value.


\begin{figure}[!htb]
    \centering
     \includegraphics[trim=5mm 1mm 1mm 1mm,clip,width=0.4\textwidth]{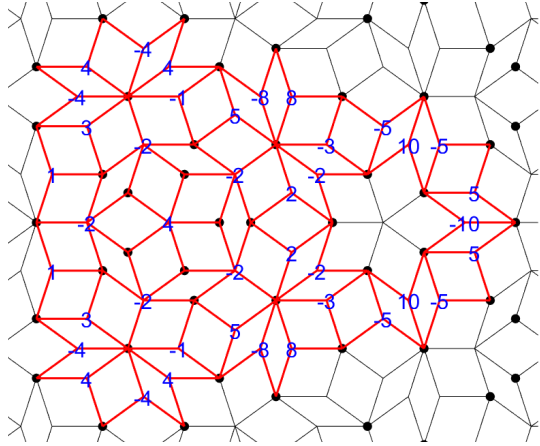}
    \caption{Type-6 LS has 12 points with index 1 and 29 points with index 3. LS with orientation $\hat{e}_0$ is displayed.}
    \label{fig:Type6RealSpace}
\end{figure}

\begin{figure}[!htb]
    \centering
     \includegraphics[trim=5mm 5mm 5mm 5mm,clip,width=0.4\textwidth]{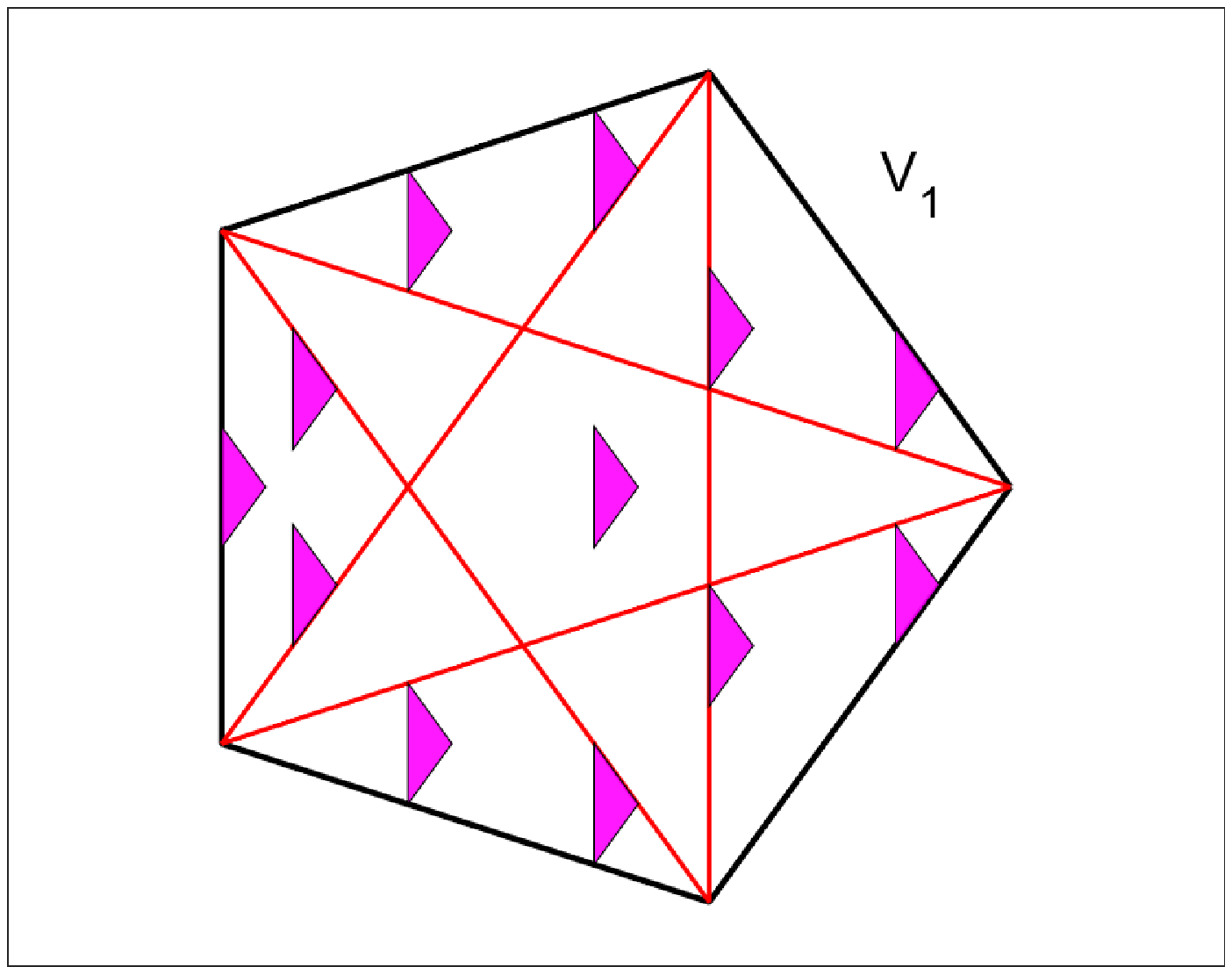}
    \includegraphics[trim=5mm 5mm 5mm 5mm,clip,width=0.4\textwidth]{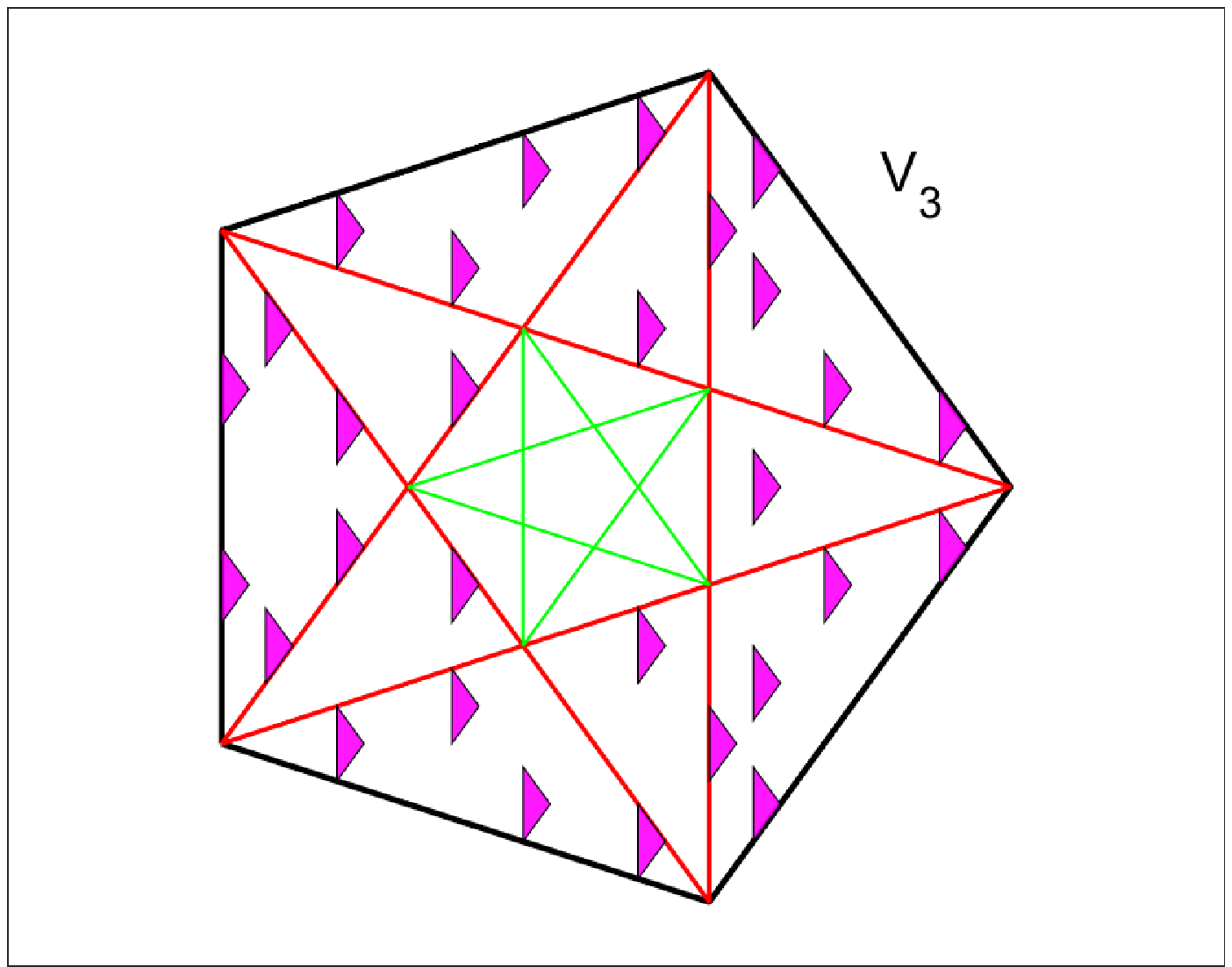}
    \caption{Allowed areas for type-6 vertices are triangles which a scaled down by a factor of  $\tau$ compared to the allowed areas of type-5.}
    \label{fig:Type6PerpSpaceArea}
\end{figure}

\begin{figure}[!htb]
    \centering
    \includegraphics[trim=5mm 5mm 5mm 5mm,clip,width=0.4\textwidth]{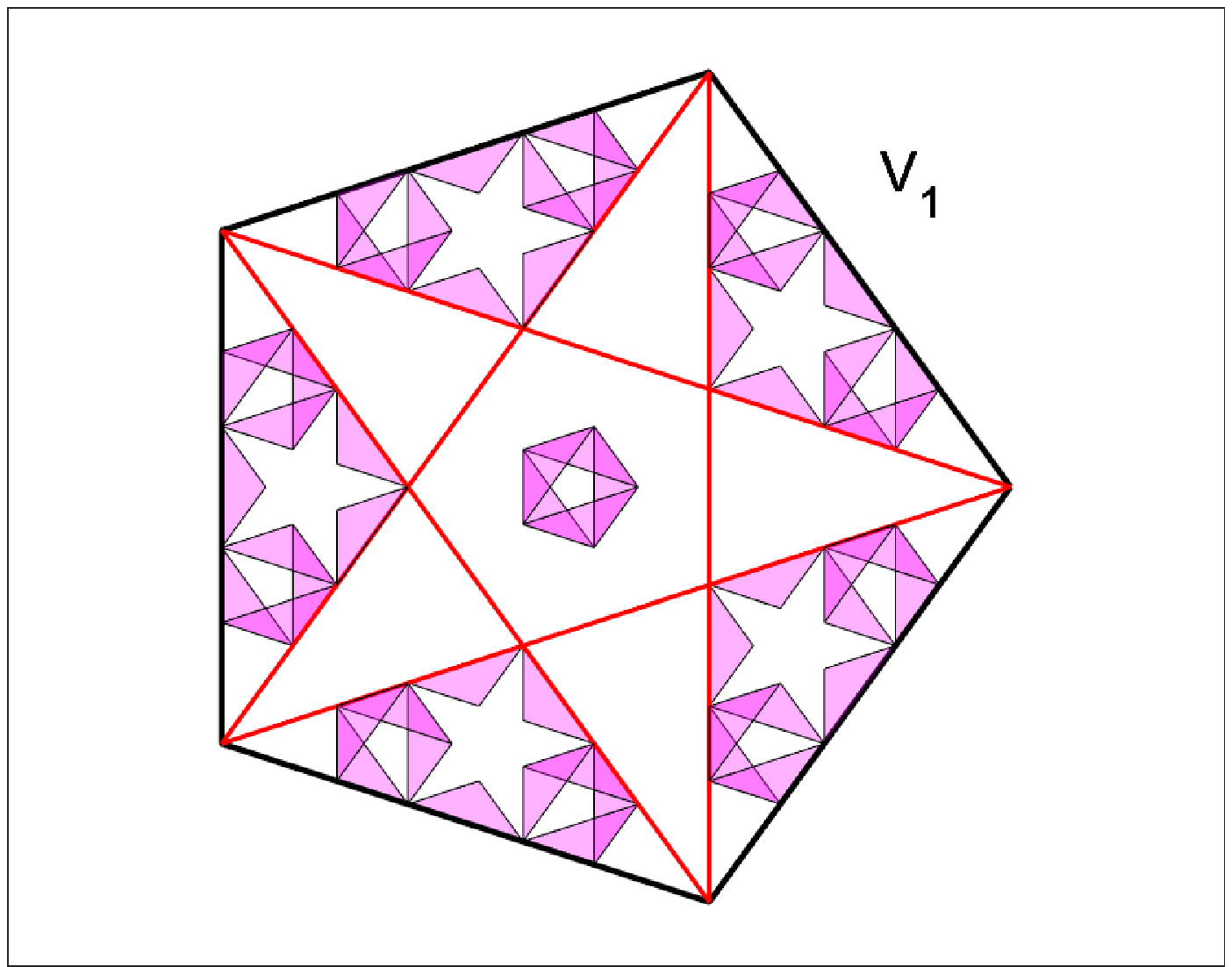}
    \includegraphics[trim=5mm 5mm 5mm 5mm,clip,width=0.4\textwidth]{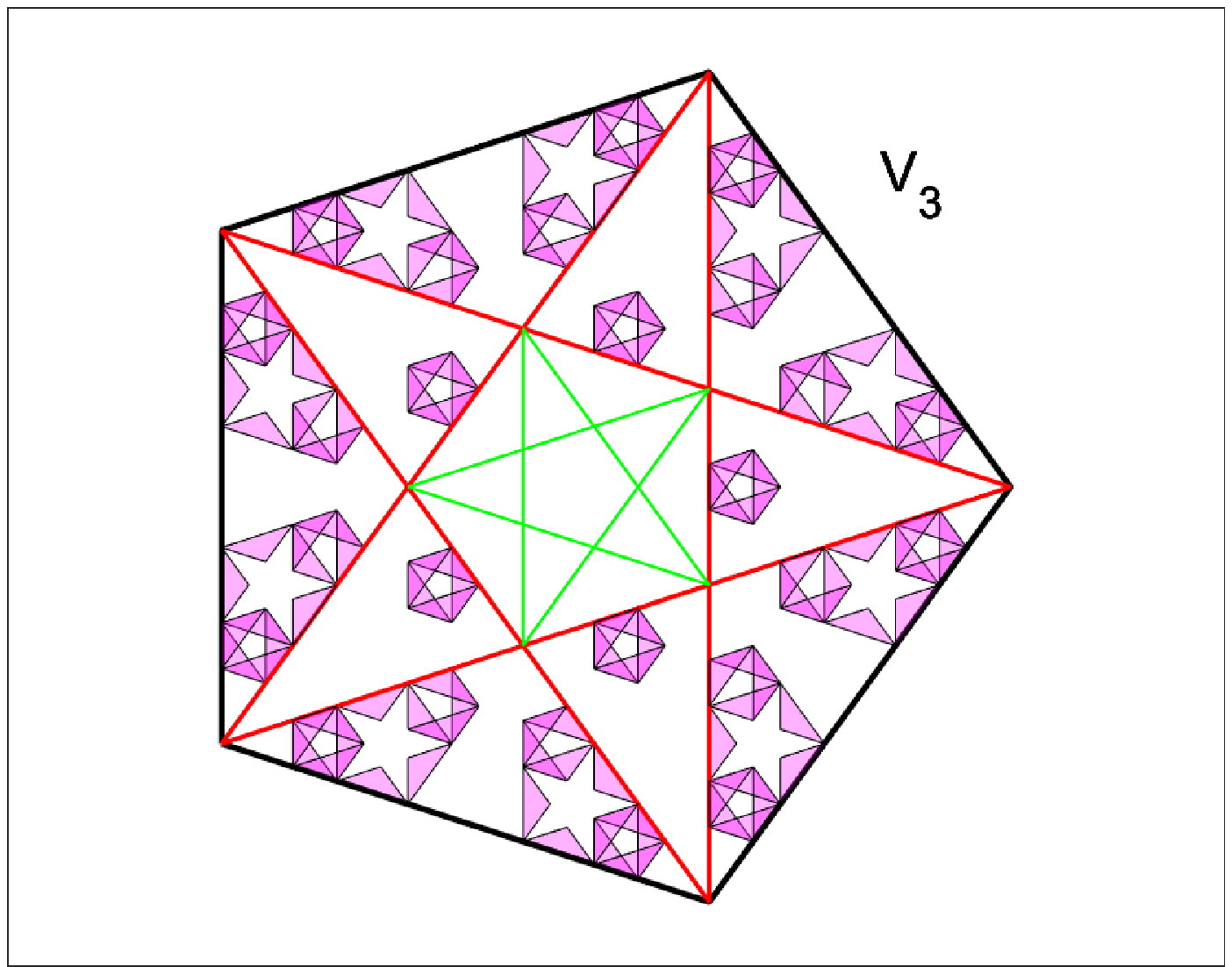}
    \caption{Type-6 allowed areas for  all orientations. Notice that both the central S in $V_1$, and the surrounding five J sites in $V_3$ can be shared by two type-6 states with different orientations.}
    \label{fig:Type6PerpSpaceAreaRotated}
\end{figure}

Finally, the real space configuration for type-6 is displayed in Fig.\ref{fig:Type6RealSpace}, and corresponding perpendicular space pictures of one orientation and all five orientations are given in Figs. \ref{fig:Type6PerpSpaceArea} and \ref{fig:Type6PerpSpaceAreaRotated} respectively. Type-6 LS is the only LS with support containing an S site. Only some of the S sites which have type-1 LS around them can support type-6 LS. Similar to type-5 some S sites support only one orientation of type-6 while some of them support two different orientations. The allowed areas are triangles which can be obtained by scaling type-5 allowed area with $\tau$, so type-6 LS $|T6,\hat{e}_0,\vec{R}_\perp\rangle$ with $\hat{e}_0$ orientation  must have $\vec{R}_\perp$ lie in a triangle with vertices at $\hat{e}_0/\tau^4$,$\hat{e}_1/\tau^4$ and $\hat{e}_4/\tau^4$. The frequencies for a single orientation and all orientations are given by
\begin{equation}
    f_{T6,0}=\frac{f_{T5,0}}{\tau^2}= \frac{1}{5\tau^{10}}= \frac{89-55\tau}{5} \simeq 1.626 \times10^{-3},
\end{equation}
\begin{equation}
    f_{T6}=5 f_{T6,0}= \frac{1}{\tau^{10}}= 89-55 \tau \simeq 8.131\times 10^{-3}.
\end{equation}
This total frequency is again larger by a factor of $\tau$ from the value reported in Ref\cite{ara88}. This discrepancy is resolved in the next section by showing that the two type-6 LS sharing the same S vertex are not independent when type-1 to 4 LS are taken into account.


\section{\label{sec:Overlaps} Overlaps and independence of localized states}

The LS form a massively degenerate manifold, thus the sum of any two LS we considered in the previous section is also a LS.
We labeled LS eigenstates of the PL vertex model but did not consider if these eigenstates are orthogonal, or independent from each other.  In this section, we use perpendicular space methods to establish independence of LS types and calculate overlaps between LS of different types.

\begin{figure}[!htb]
    \centering
 \includegraphics[trim=5mm 5mm 5mm 5mm,clip,width=0.39\textwidth]{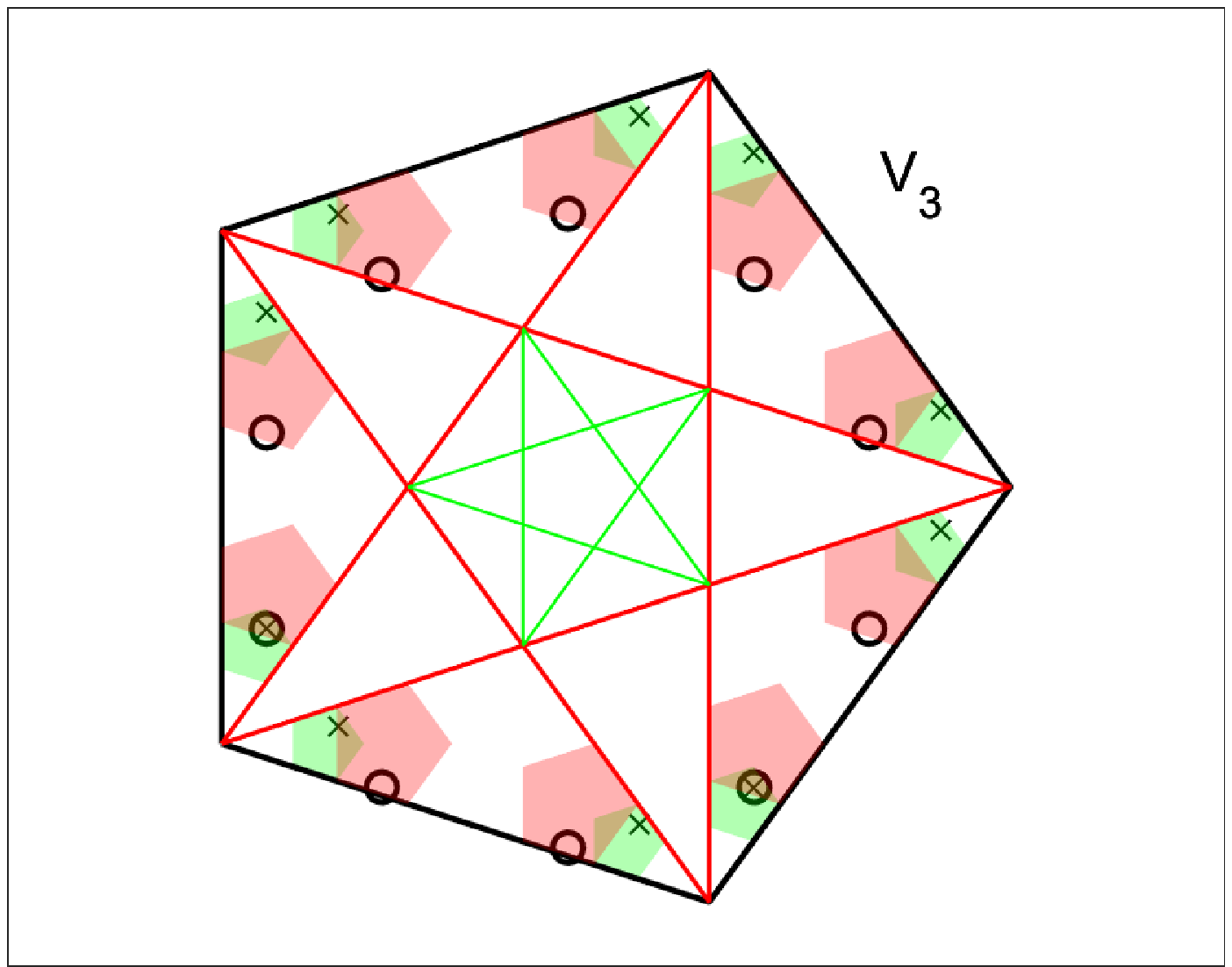}
    \caption{Perpendicular space overlap of type-1 and type-2 allowed regions. Selecting a point in the overlap region for both type-1 and type-2 determines the remaining 9 points for each LS. Two points out of ten are shared by type-1 and type-2.}
    \label{fig:T1T2OverlapPerpSpace}
\end{figure}

\begin{figure*}[!htb]
    \centering
  \includegraphics[trim=5mm 1mm 1mm 1mm,clip,width=0.6\textwidth]{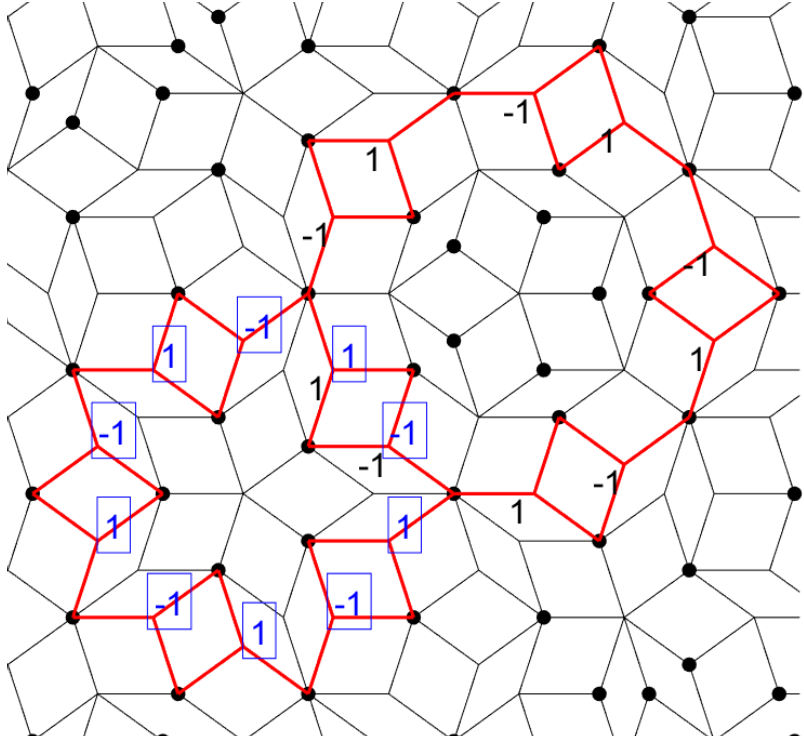}
    \caption{Real space images of the  points shown in the previous figure. Wavefunction values for type-1 are displated in black (unframed), for type-2 in blue (framed). Overlap of type-1 and type-2 states shows that these states are not orthogonal.}
    \label{fig:T1T2OverlapRealSpace}
\end{figure*}

The perpendicular space allowed areas for type-1 states in Fig.\ref{fig:Type1PerpSpaceArea} show that any two type-1 states are orthogonal as they correspond to  two different points in each pentagonal region. The same logic applied to Fig.\ref{fig:Type2PerpSpaceArea} yields the orthogonality of any two type-2 states. However, if we plot the allowed areas for both types simultaneously as in Fig.\ref{fig:T1T2OverlapPerpSpace} we observe that there is a significant region of overlap between the two allowed areas. Picking one point in the overlap region and translating it through type-1 and type-2 second neighbor vectors we see that a type-1 and a type-2 LS can share two sites in their support. This situation can be visualized in real space as in Fig.\ref{fig:T1T2OverlapRealSpace}. The overlap between the two states can be calculated as
\begin{equation}
\begin{split}
   \frac{\langle T1,\vec{R}_{1}|T2,\vec{R}_{2}\rangle}{\sqrt{\langle T1,\vec{R}_{1}|T1,\vec{R}_{1}\rangle\langle T2,\vec{R}_{2}|T2,\vec{R}_{2}\rangle}} \\ \nonumber
   =0.2 \sum_{m=0}^4 \delta \left(\vec{R}_{1}-\vec{R}_{2}-\frac{\hat{e}_m}{\tau^3} \right). 
\end{split}
\end{equation}

Investigating the overlap region area in Fig.\ref{fig:T1T2OverlapPerpSpace} we see that $\frac{1}{\tau^4}=14.59\%$ of type-1 states do not have an overlap with type-2, $\frac{5}{\tau^3(1+\tau^2)}=32.62\%$ of type-1 states overlap with only one type-2 state, while the remaining $\frac{5}{\tau^2(1+\tau^2)}=52.79\%$ have overlaps with 2 type-2 states. As at most 4 points of the support in any type-1 state is covered by type-2 states all type-1 and type-2 states remain linearly independent from each other. Similarly, $\frac{6\tau-8}{(1+\tau^2)}=47.21\%$ of type-2 states do not have any overlap with type-1, while $\frac{5}{\tau^2(1+\tau^2)}=52.79\%$ have an overlap with only one type-1 state.  

Type-3 states are in general not orthogonal to type-1 or type-2. Furthermore, two type-3 states with different orientations can have an overlap as seen in Fig.\ref{fig:Type3PerpSpaceAreaRotated}.   However it is easy to establish the linear independence of any type-3 states from not only types 1 and 2 but also from other orientations of type-3. Investigating the the $V_1$ pentagon in Fig.\ref{fig:Type3PerpSpaceArea} and Fig.\ref{fig:Type3PerpSpaceAreaRotated} we see the top-left allowed Q region is unique to the  $\hat{e}_0$ orientation. Similarly all orientations have unique Q sites which can not be covered by any sum of type-1, type-2 and other orientations of type-3 states. Thus, the collection of all type-1, type-2 and type-3 LS is a linearly independent but not orthogonal set. In Ref\cite{koh86} sum of LS from this set were defined to be ring states. While it may be possible to construct another basis for the space spanned by types 1 to 3, we do not see any advantage to such an approach.

\begin{figure*}[!htb]
    \centering
 \includegraphics[trim=7mm 5mm 7mm 5mm,clip,width=0.40\textwidth]{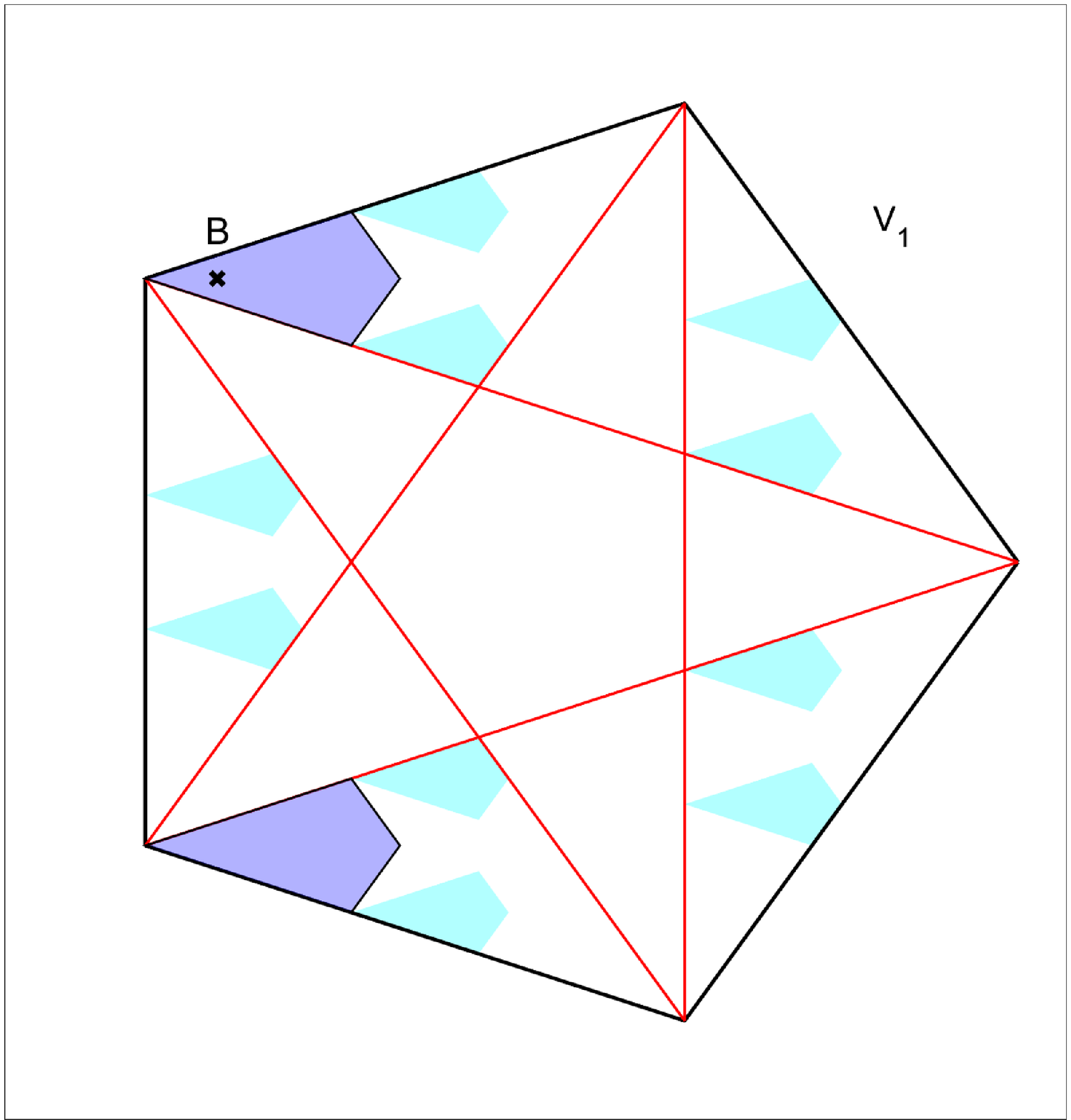}
 \includegraphics[trim=7mm 5mm 7mm 5mm,clip,width=0.40\textwidth]{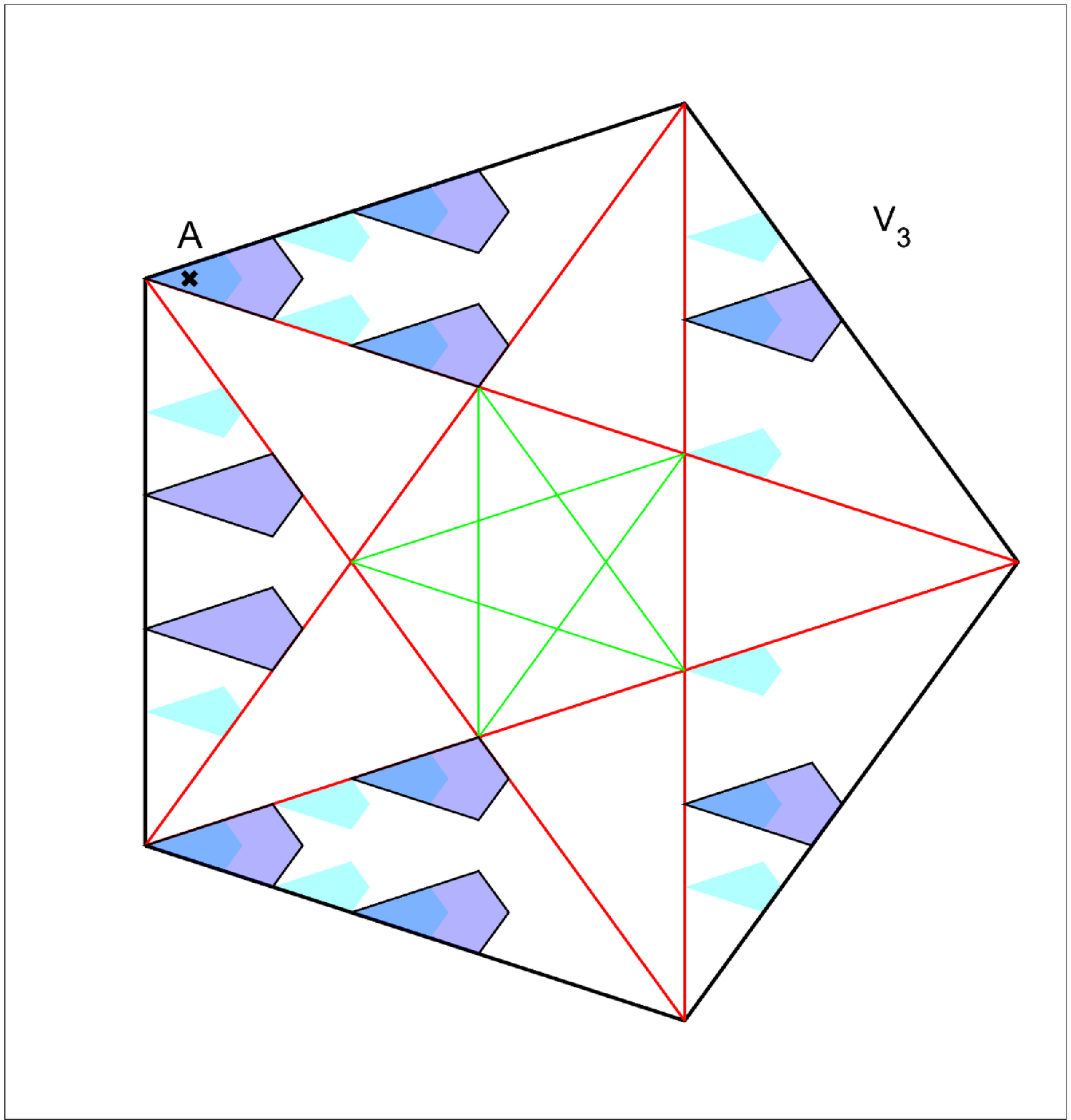} \\
 \includegraphics[trim=7mm 5mm 7mm 5mm,clip,width=0.40\textwidth]{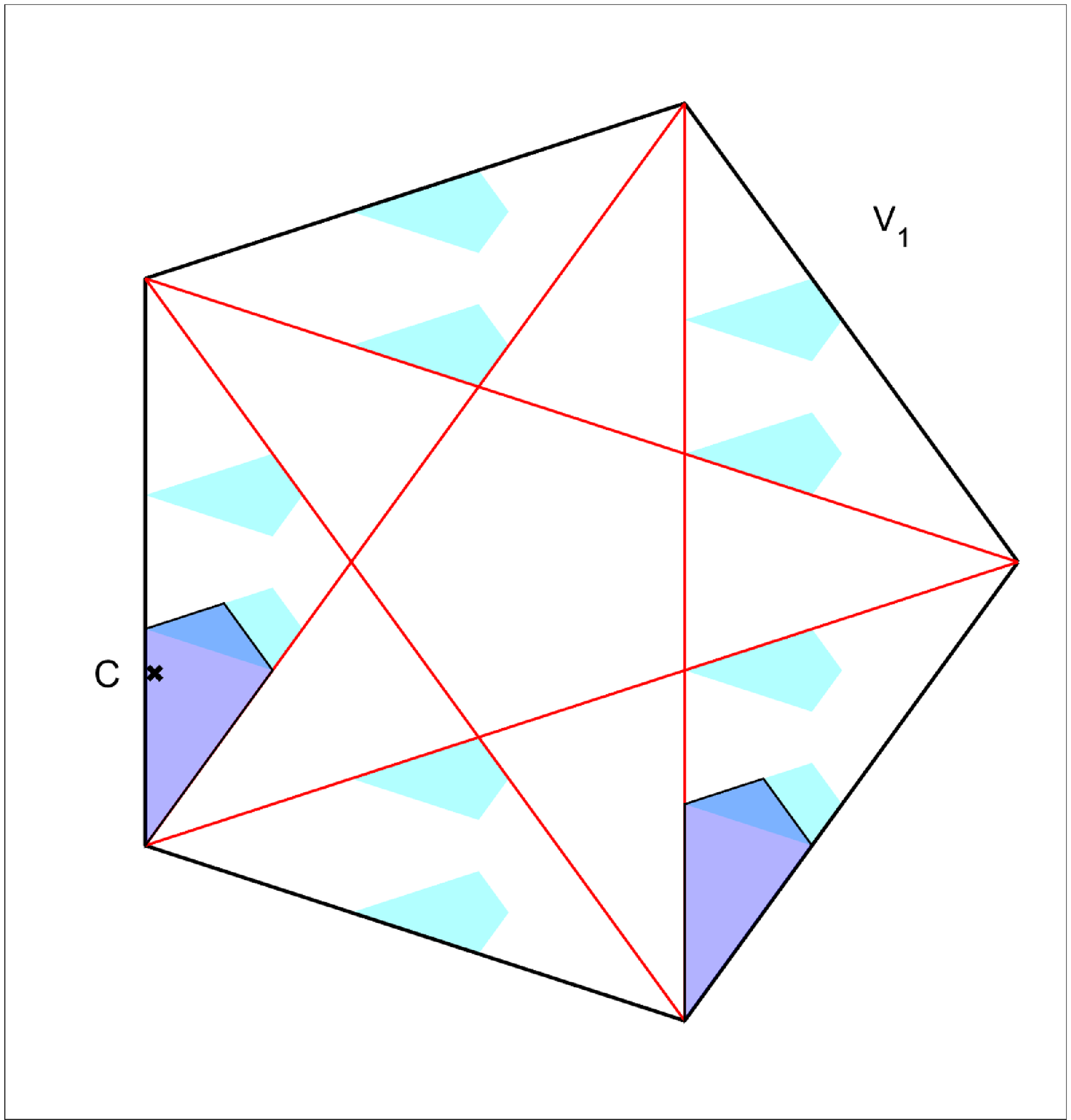}
 \includegraphics[trim=7mm 5mm 7mm 5mm,clip,width=0.40\textwidth]{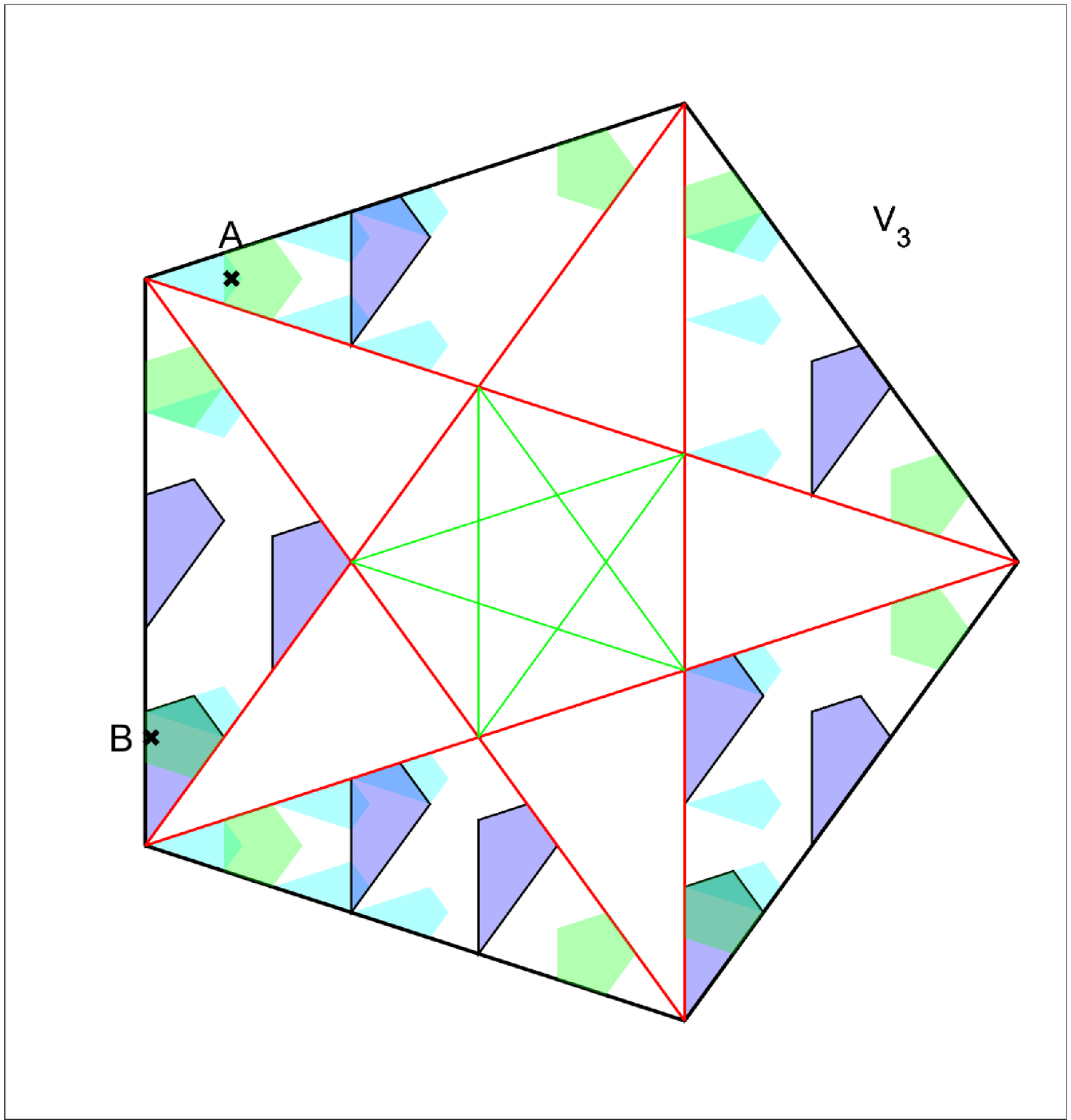}
 
    \caption{Perpendicular space images establishing the independence of type-4 LS. Blue, cyan and green regions correspond to type-4, type-3 and type-1 LS respectively. The existence of point A in the upper panel requires point B which is unique for a single orientation of type-3 LS. In the lower panel point A is covered by a type-1 state, which creates point B. Point B requires the existence of point C, which is again unique. Thus, type-4 states of any orientation are linearly independent of all other LS.}
    \label{fig:IndependenceT4}
\end{figure*}

Establishing the independence of type-4 states requires more work as one cannot find a unique site which is covered only by a type-4 state of a given orientation. In other words all the areas in Fig.\ref{fig:Type4PerpSpaceArea}, are covered by either types 1-3 or other orientations of type-4. To establish independence, we focus on the type-4 allowed region in the top-left corner of $V_3$ as shown in Fig.\ref{fig:IndependenceT4}. A point $A$ in this area can only be covered by a type-3 or type-1 state. If point $A$ is covered by a type-3 state, this implies the existence of a point $B$ in $V_1$ which is unique to the type-3 state. As the wave function in point $B$ can not be nullified by adding other states and point $B$ is not in the support of type-4, point $A$ should not be covered by type-3.  If point $A$ is covered by a type-1 LS as shown in the lower panels of Fig.\ref{fig:IndependenceT4}, this implies the existence of a point $B$ in a different region of $V_3$. The amplitude at $B$ can only be cancelled by adding a type-3 state. However such a type-3 state will generate an amplitude in a unique region of $V_1$ at point $C$. Thus, it is impossible to write a type-4 state as a linear combination of types 1 to 3 and other type-4 states. 

The independence of any type-5 state from previous types is easily established by noticing that type-5 states have non-zero amplitude on an S5 site. Thus, the frequency of type-5 states which are independent from all other types is $f_{T5}=\frac{1}{\tau^8}$. This number is a factor of $\tau$ higher than the frequency reported in Ref\cite{ara88}, a difference which can be discerned with large finite lattice numerical calculations.  

\begin{figure*}[!htb]
    \centering
   \includegraphics[trim=5mm 1mm 1mm 1mm,clip,width=0.8\textwidth]{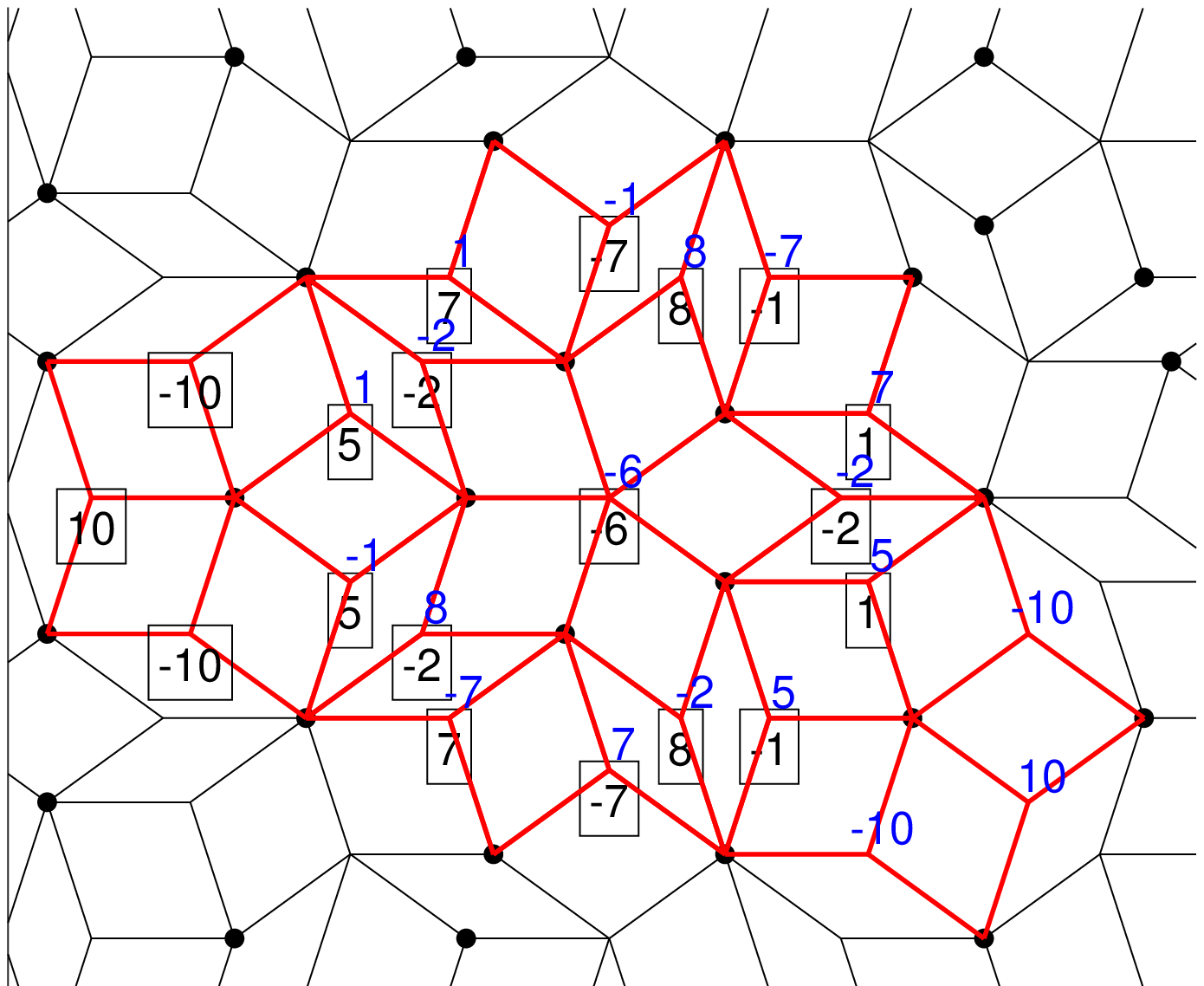}
    \caption{Type-5 states of two different orientations can share the same S5 vertex and overlap in real space. The wavefunction values for $\hat{e}_0$ orientation are shown in black (framed), and $\hat{e}_4$ orientation are shown in blue (unframed).}
    \label{fig:T5T5OverlapRealSpace}
\end{figure*}

\begin{figure}[!htb]
    \centering
   \includegraphics[trim=5mm 5mm 5mm 5mm,clip,width=0.4\textwidth]{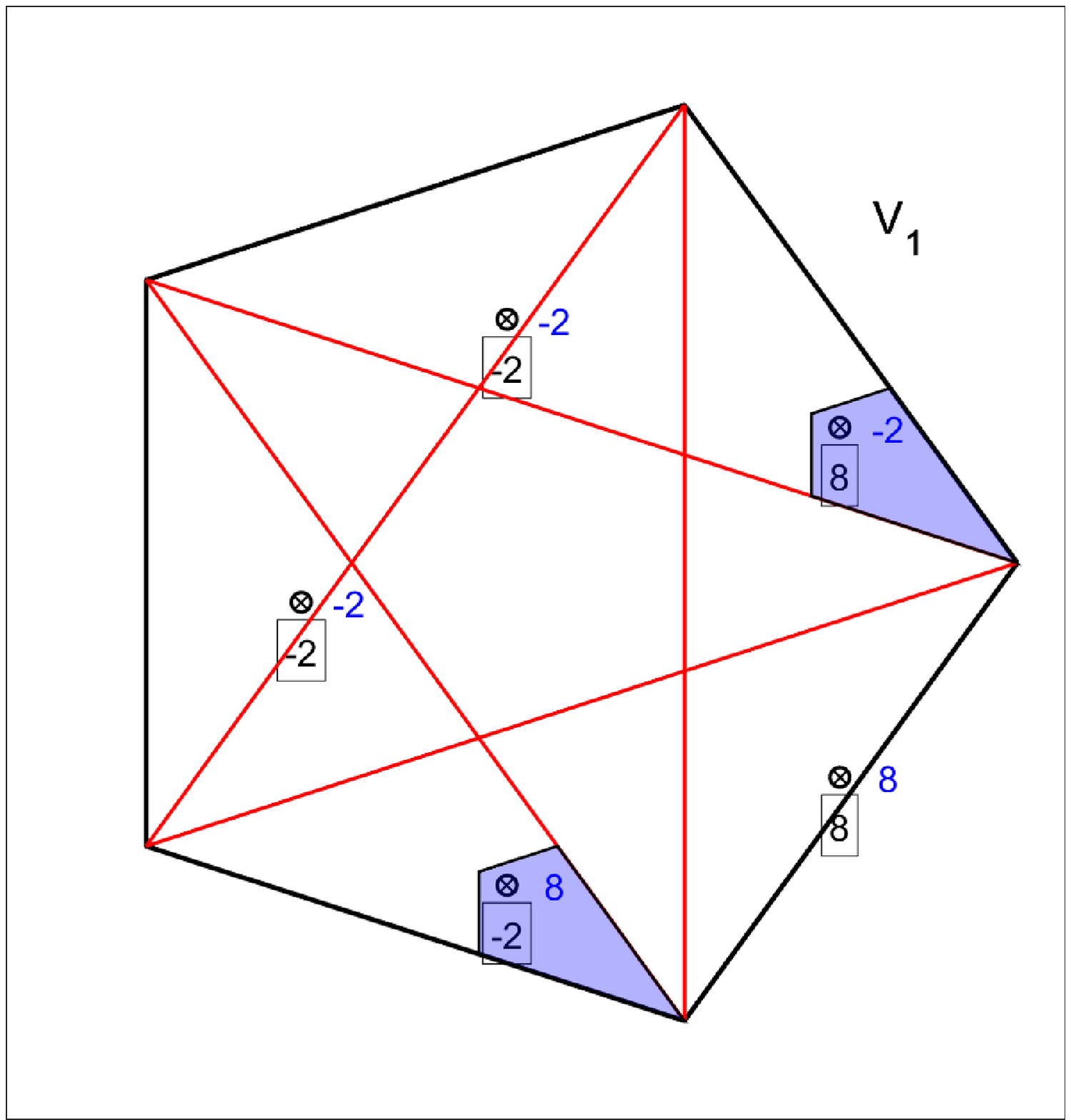}
    \includegraphics[trim=5mm 5mm 5mm 5mm,clip,width=0.4\textwidth]{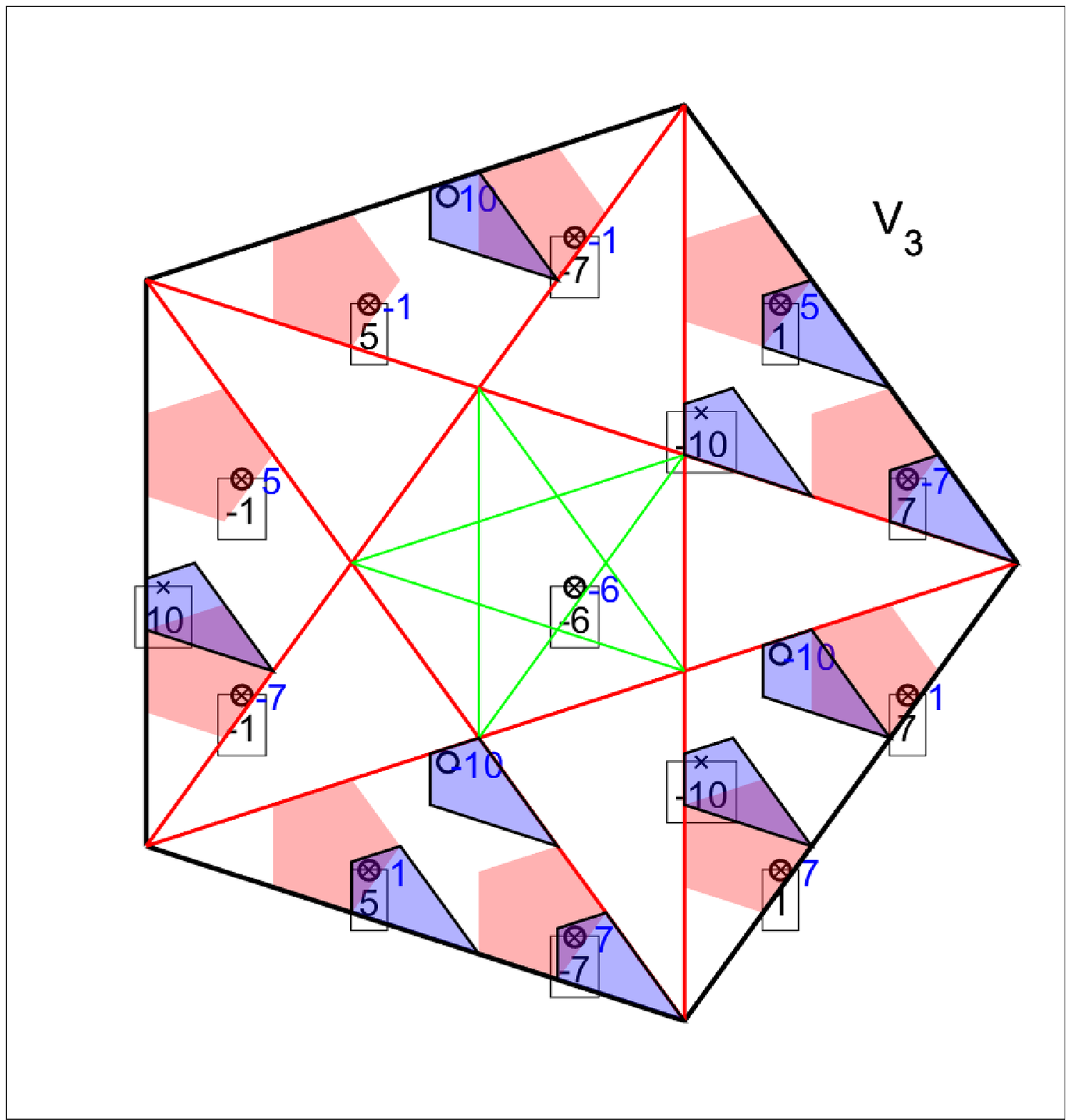}
    \caption{Perpendicular space image of the two wavefunctions shown in the previous figure. All the points for which the difference between the wavefunctions are non-zero can be covered by the allowed areas of type-3 and type-2 states as shown. Type-2 regions are (magenta) pentagons while type-3 regions are blue kites. }
    \label{fig:T5T5OverlapPerpSpace}
\end{figure}

\begin{figure*}[!htb]
    \centering
   \includegraphics[trim=2mm 1mm 1mm 1mm,clip,width=0.8\textwidth]{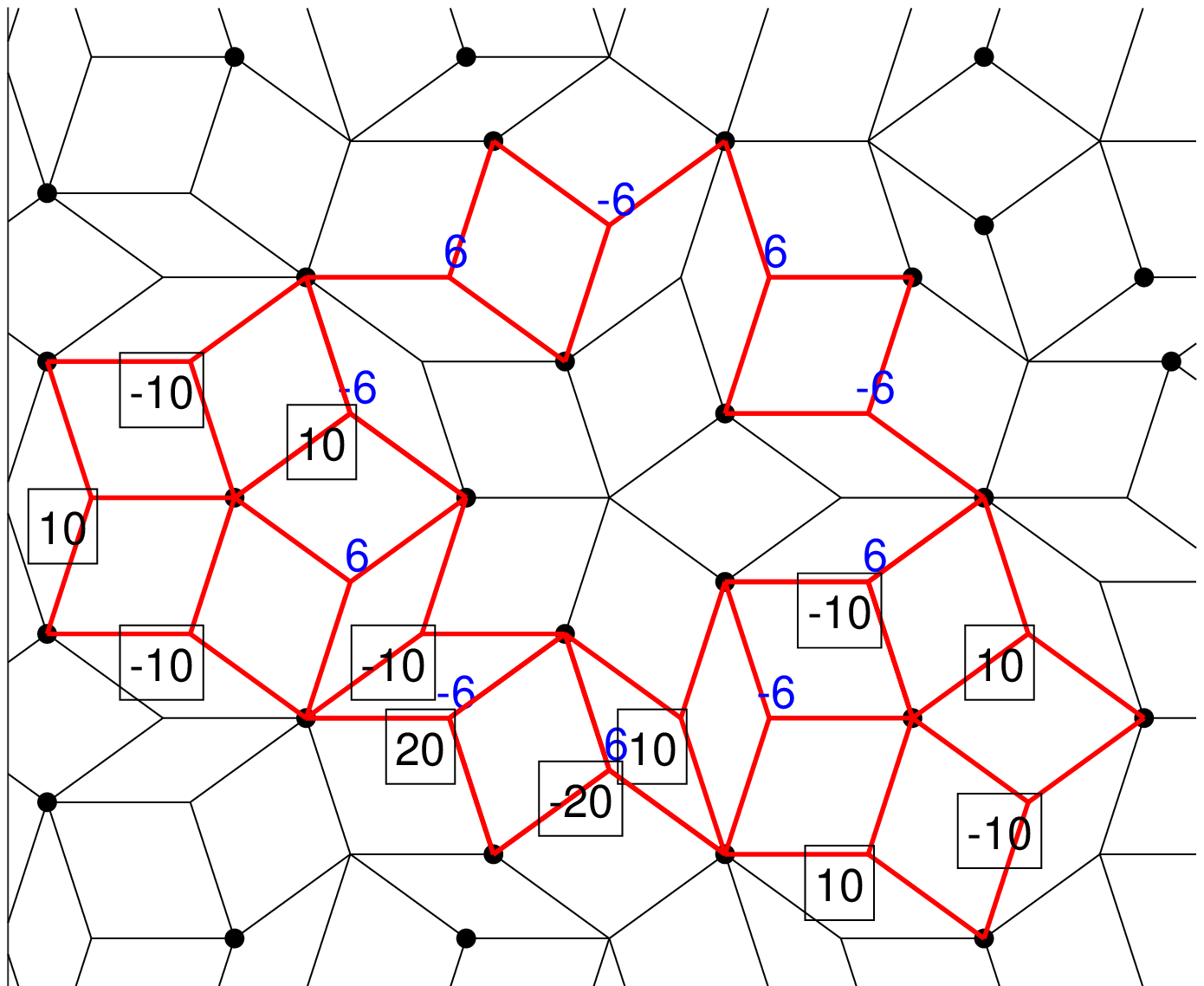}
    \caption{A type-3 (framed) and a type-2 (unframed) state. The difference of the two wavefunctions shown in the Fig.\ref{fig:T5T5OverlapRealSpace} is equal to the sum of the wavefunctions shown in this figure. Different orientations of type-5 are not linearly independent when other types are taken into account.}
    \label{fig:T5Expansion}
\end{figure*}

It is possible for two type-5 states of different orientation to have the same S5 point in their support. This can be seen by the overlapping allowed areas in Fig. \ref{fig:Type5PerpSpaceAreaRotated}, and picking a point in the allowed region we can visualize the two wavefunctions in real space: Fig. \ref{fig:T5T5OverlapRealSpace}. These two wavefunctions are obviously independent from each other, and each one of them is independent of all LS type 1 to 4. However, the question is if one of the two type-5 states can be written as a sum of the other type-5 state with other LS types. To answer this question, we plot the perpendicular space images of the points in the support of both states in Fig.\ref{fig:T5T5OverlapPerpSpace}. The two states are not orthogonal
\begin{widetext}
\begin{equation}
\frac{\langle T5,\hat{e}_0,\vec{R}_\perp|T5,\hat{e}_4,\vec{R}_\perp\rangle}{\sqrt{\langle T5,\hat{e}_4,\vec{R}_\perp|T5,\hat{e}_4,\vec{R}_\perp\rangle\langle T5,\hat{e}_0,\vec{R}_\perp|T5,\hat{e}_0,\vec{R}_\perp\rangle}} = \frac{6}{726}\simeq 0.0008264 .
\end{equation}
We see that the difference of two wavefunctions is non-zero on 18 sites. All these 18 sites can be covered by a type-2 state and and a type-3 state of a single orientation. Resulting expansion gives us 
\begin{equation}
    |T5,\hat{e}_4,\vec{R}_\perp\rangle=|T5,\hat{e}_0,\vec{R}_\perp\rangle - 6 |T2,\vec{R}_\perp \rangle + 10 \left|T3,\hat{e}_2,\vec{R}_\perp-\frac{\hat{e}_0+\hat{e}_4}{2\tau^3}\right\rangle
\end{equation}
The expansion can be checked by the real space wavefunctions given in Fig. \ref{fig:T5Expansion}. 
\end{widetext}

We see that the two type-5 states sharing the same S5 site are linearly dependent when other types are taken into account. Any S5 site is in the support of 0, 1 or 2 type-5 LS as can be seen by investigating the center of $V_3$ in Fig.\ref{fig:Type5PerpSpaceAreaRotated}. If we want to count the number of linearly independent type-5 wavefunctions we need to consider the same area without any regard to how many times that area is covered. The resulting area is the difference of a pentagon of radius $\tau^{-3}$ and the inside pentagon of radius $\tau^{-5}$,
\begin{equation}
    f^{indep}_{T5}=\frac{\tau^{-6}-\tau^{-10}}{1+\tau^2}=\tau^{-9}=\frac{f_{T5}}{\tau}
\end{equation}
One can form a linearly independent set of LS by adding one type-5 orientation per S5 site to the set of all type-1 to type-4 states. The frequency of such type-5 states is given by $f^{indep}_{T5}$ which is in agreement with Ref.\cite{ara88}. The distinction between linearly independent and not-independent type-5 states were not made in the literature, but it is remarkable that the frequency of the independent type-5 states coincides with the frequency of S5 sites which can host a type-5 state.

\begin{figure*}[!htb]
    \centering
     \includegraphics[trim=5mm 1mm 1mm 1mm,clip,width=0.8\textwidth]{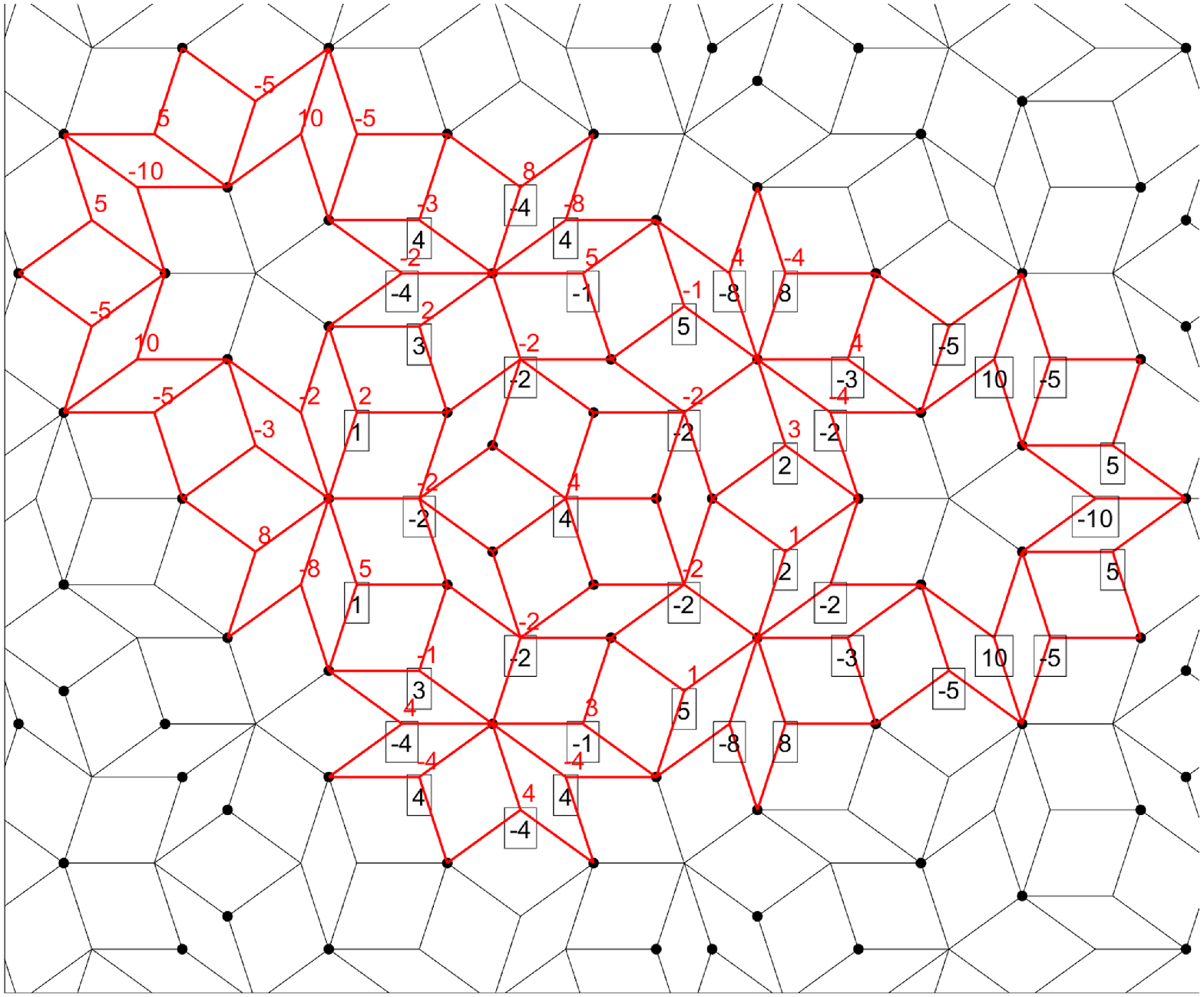}
    \caption{Type-6 LS of two different orientations overlap in real space. The wavefunctions of $\hat{e}_0$ red (unframed)  and $\hat{e}_4$  black (framed) are shown.}
    \label{fig:T6T6OverlapRealSpace}
\end{figure*}

\begin{figure*}[!htb]
    \centering
    \includegraphics[trim=1mm 1mm 1mm 1mm,width=0.50\textwidth]{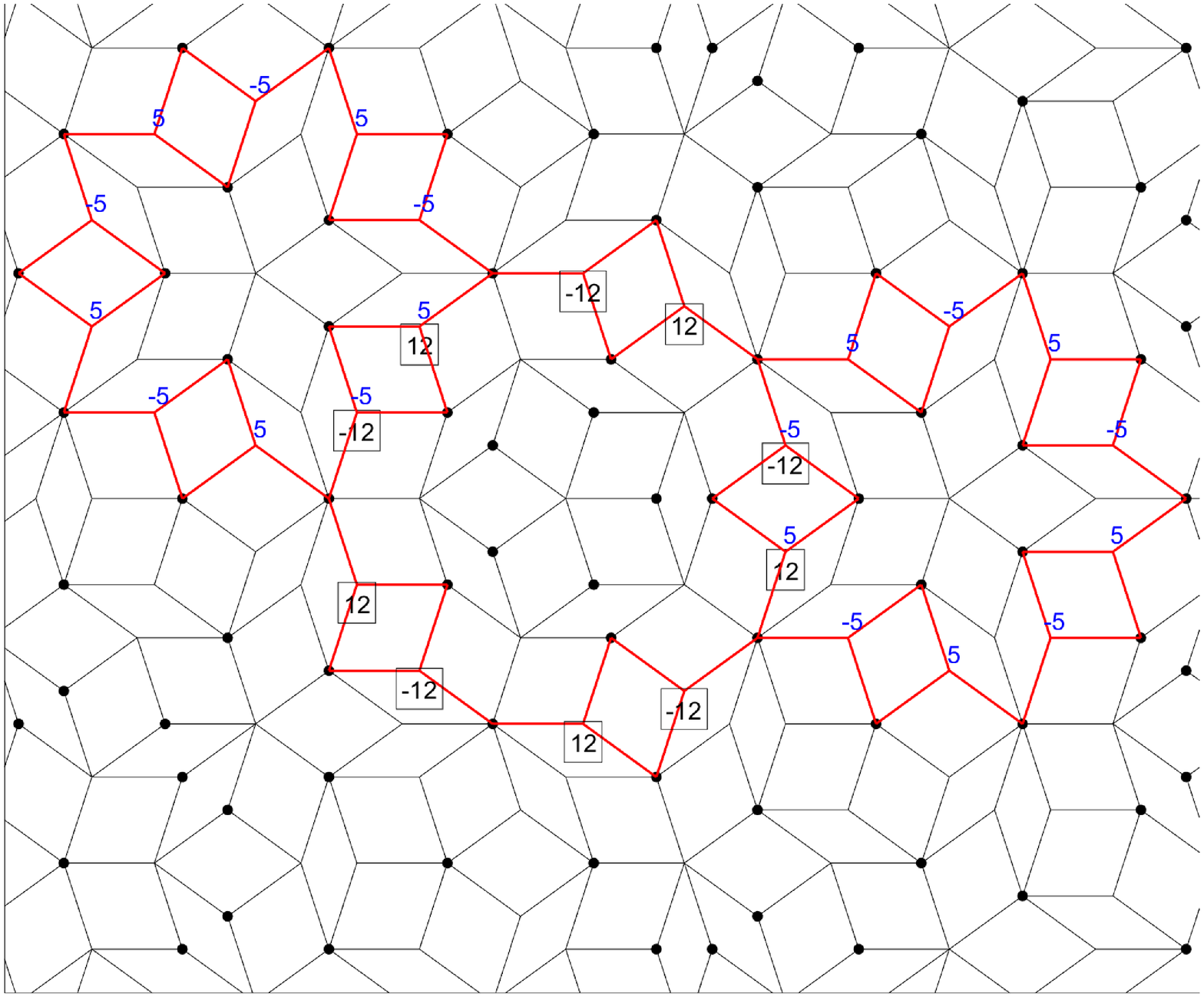} \vspace{1mm}\\
    \includegraphics[trim=1mm 1mm 1mm 1mm,width=0.50\textwidth]{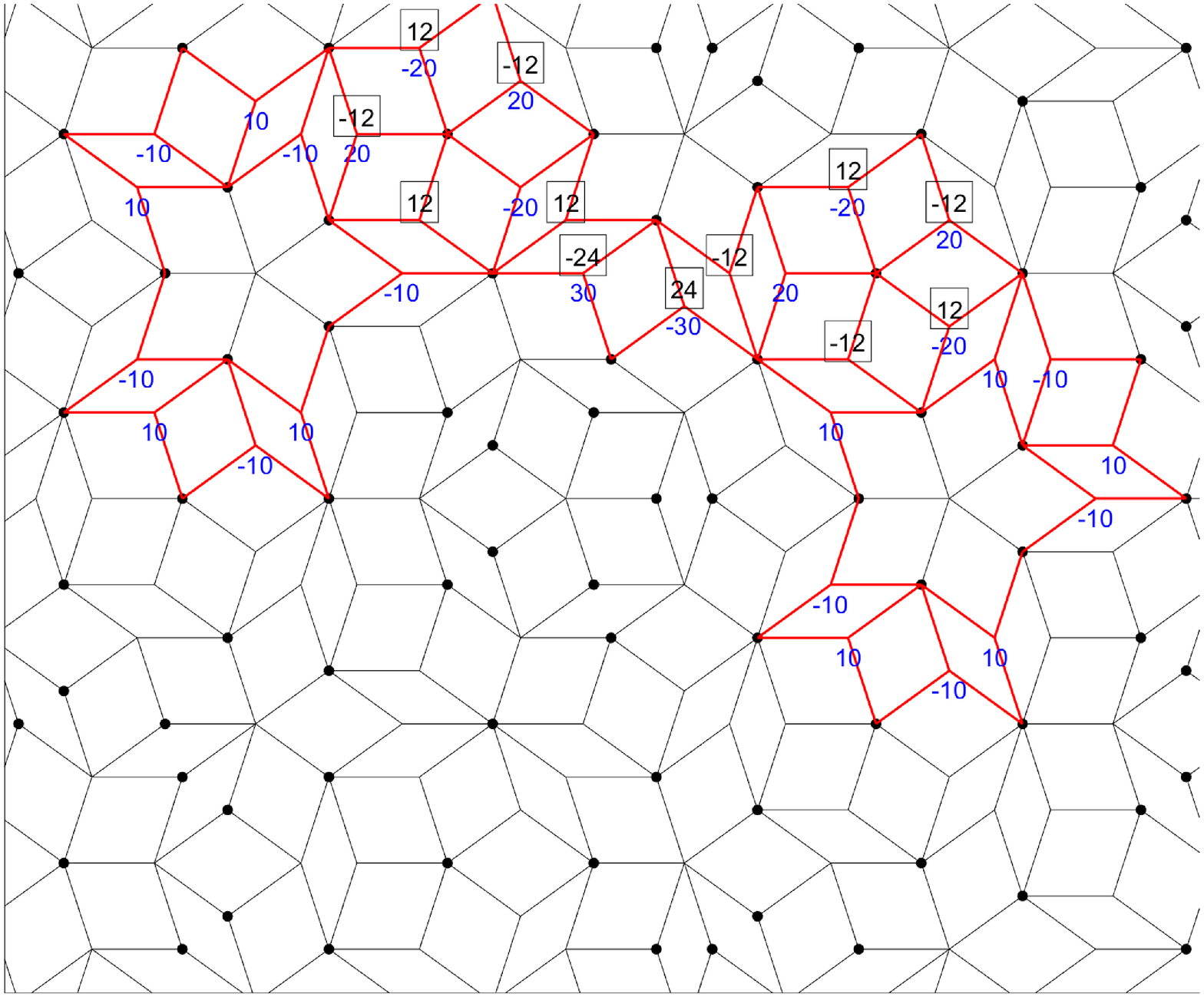}  \vspace{1 mm} \\
    \includegraphics[trim=1mm 1mm 1mm 1mm,width=0.50\textwidth]{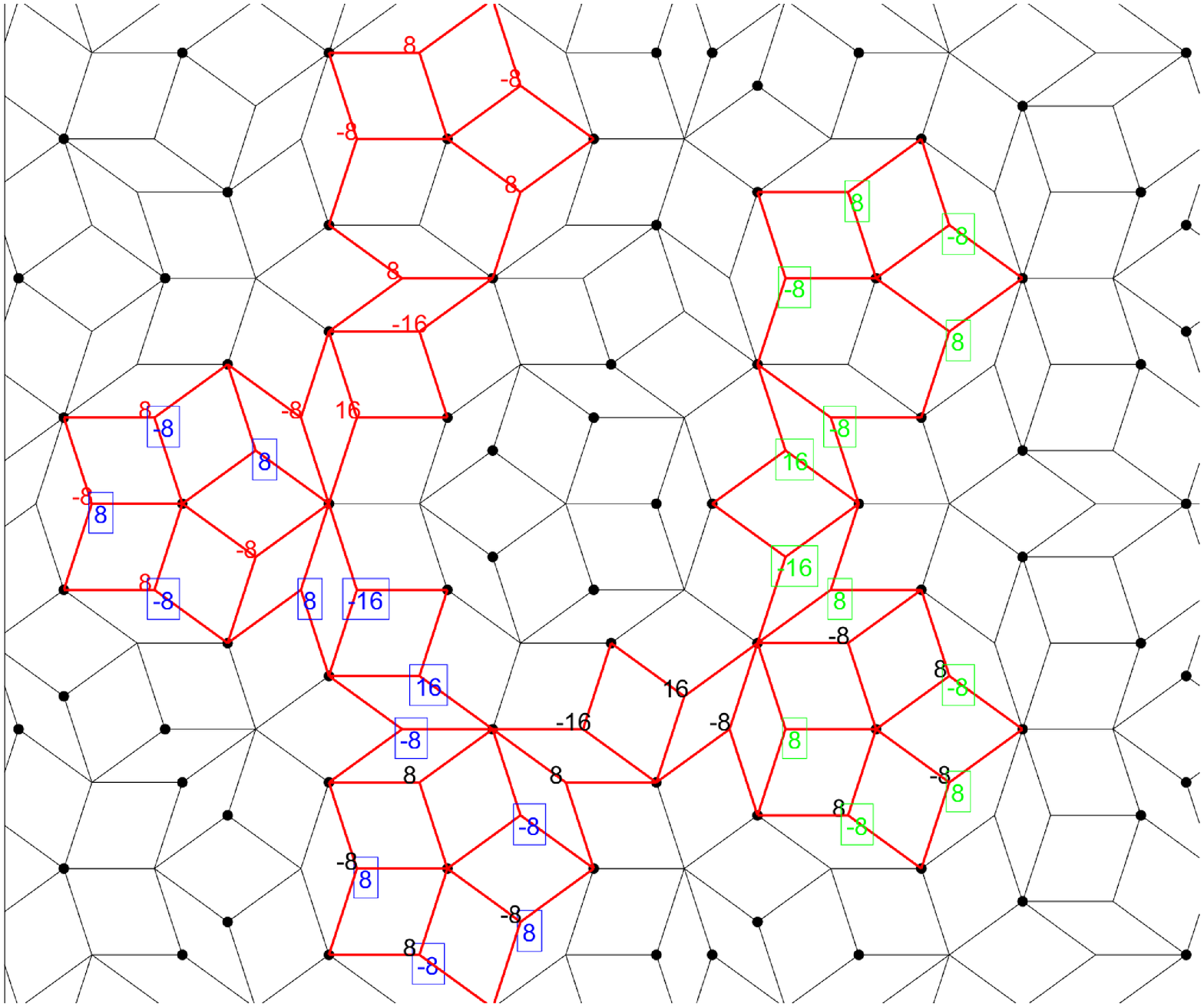}
    \caption{Real space wavefunctions of the terms in Eq.(\ref{eq:Type6Expansion}). Each panel displays one line in the right hand side of expansion Eq.(\ref{eq:Type6Expansion}). The sum of all wavefunctions shown in the three panels equal the difference of the two wavefunctions in the previous figure. Two different orientations of type-6 are not linearly independent.}
    \label{fig:T6Expansion}
\end{figure*}

A similar scenario plays out for type-6 states. The only LS which have amplitudes on S and J vertices are type-6 states, thus each type-6 state is linearly independent of LS types 1 to 5. However an S site can host two type-6 states as can be seen in the center of $V_1$ in Fig.\ref{fig:Type6PerpSpaceAreaRotated}. The real space picture is displayed in Fig.\ref{fig:T6T6OverlapRealSpace}, and we see the overlap is
\begin{widetext}
\begin{equation}
\frac{\langle T6,\hat{e}_0,\vec{R}_\perp|T6,\hat{e}_4,\vec{R}_\perp\rangle}{\sqrt{\langle T6,\hat{e}_4,\vec{R}_\perp|T6,\hat{e}_4,\vec{R}_\perp\rangle\langle T6,\hat{e}_0,\vec{R}_\perp|T6,\hat{e}_0,\vec{R}_\perp\rangle}} = -\frac{154}{976} \simeq -0.1578 .
\end{equation}

While the expansion in terms other LS types is more involved, it can be done systematically by perpendicular space images to yield
\begin{equation}
\label{eq:Type6Expansion}
\begin{split}
    |T6,\hat{e}_0,\vec{R}_\perp\rangle-|T6,\hat{e}_4,\vec{R}_\perp\rangle =& 12 |T1,\vec{R}_\perp\rangle - 5 |T2,\vec{R}_\perp-\frac{\hat{e}_0}{\tau^3}\rangle - 5 |T2,\vec{R}_\perp-\frac{\hat{e}_4}{\tau^3}\rangle \\
    &+ 10 |T4,\hat{e}_2,\vec{R}_\perp-\frac{\hat{e}_0+\hat{e}_4}{2 \tau^4}\rangle -12 |T3,\hat{e}_2,\vec{R}_\perp-\frac{\hat{e}_2}{2 \tau^5}\rangle \\
    &+8 \sum_{m\neq 2} |T3,\hat{e}_m,\vec{R}_\perp-\frac{\hat{e}_m}{2 \tau^5}\rangle 
\end{split}
\end{equation}\end{widetext}
The real space contributions of the terms in this expansion are given in Fig.\ref{fig:T6Expansion}. The difference of the two wavefunctions in Fig.\ref{fig:T6T6OverlapRealSpace} is equal to the sum of all the wavefunctions in Fig.\ref{fig:T6Expansion} for each site.

The linear dependence of the two type-6 LS sharing the same S site reduces the number of independent basis vectors contributed to the LS manifold
\begin{equation}
     f^{indep}_{T6}=\frac{\tau^{-4}-\tau^{-8}}{\tau^4(1+\tau^2)}=\tau^{-11}=\frac{f_{T6}}{\tau}
\end{equation}
which is exactly the value reported in Ref\cite{ara88}, without making a distinction between linearly independent and dependent states.

Considering all the above LS states we see that certain parts of the perpendicular space are left completely empty. In the next section, we consider sites which are forbidden by local connectivity to host any LS and their perpendicular space images.

\section{\label{sec:Forbidden} Forbidden sites}

A LS is defined on a set of sites all of which are in the (1-3) sublattice, and all the sites in this set have at least two second nearest neighbors which are also in this set. Each site in this set, the support of the LS, has a non-zero value of the wavefunction. Any nearest neighbor of the sites in the support lies in the (2-4) sublattice, and defines an equation for the values of the wavefunction on its nearest neighbors.  In general the number of nearest neighbors of any collection of sites is larger than the number of sites in the collection. Thus, generally the set of equations coming from nearest neighbors define an overdetermined set, and a LS state with such a support becomes impossible. 
Local connectivity of the PL can cause even further restrictions, there are sites which cannot have a non-zero value of the wavefunction for any LS. Following Ref.\cite{ara88} we call such sites forbidden sites, and determine the perpendicular space areas associated with them as follows. 

\begin{figure}[!htb]
    \centering
 \includegraphics[trim=1mm 1mm 1mm 1mm,clip,width=0.4\textwidth]{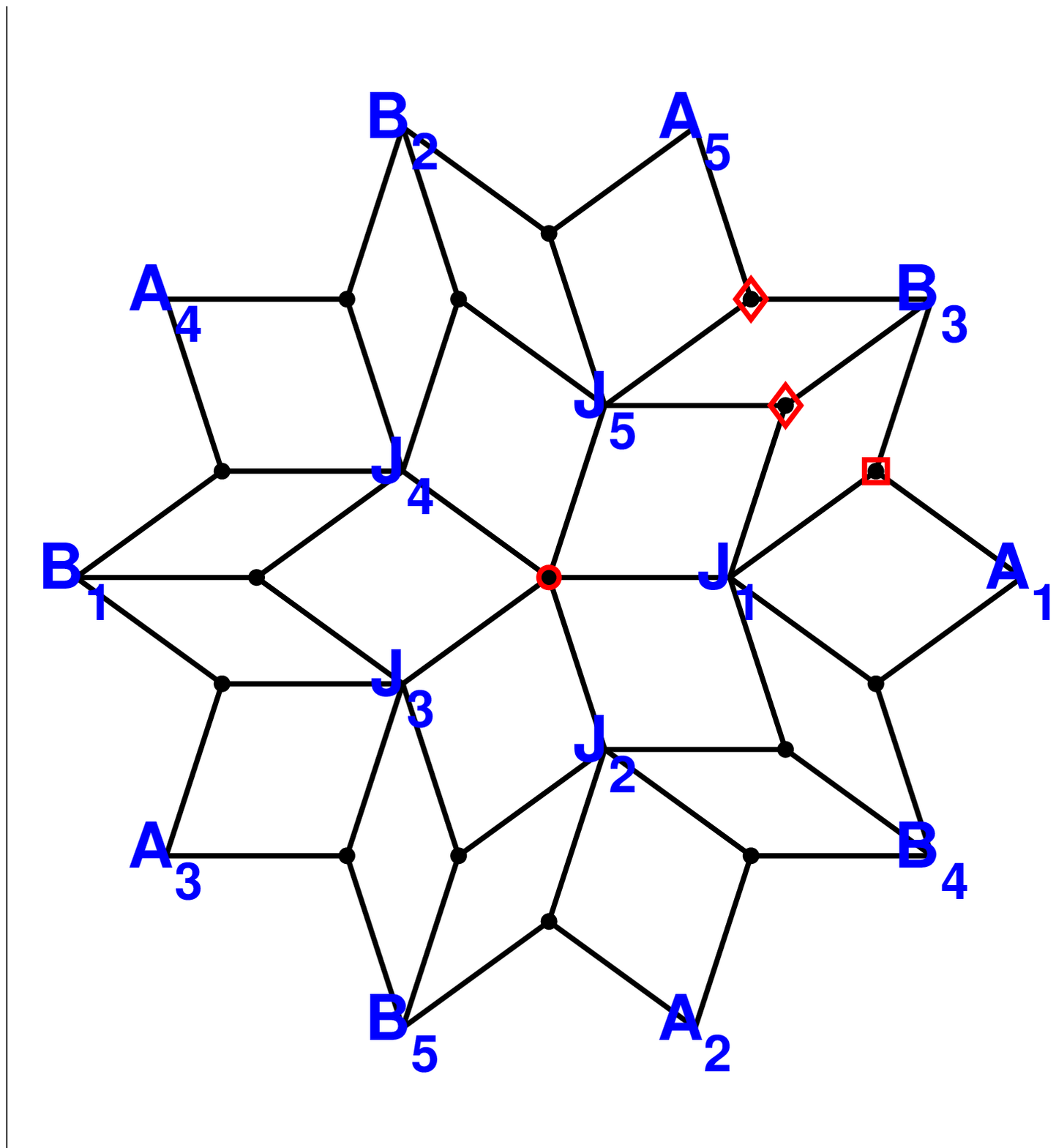}
    \caption{Local neighborhood up to third nearest neighbors of an S5 vertex with index 2. All S5 vertices in $V_2$ have the same neighborhood, the positions of third nearest neighbors are fixed, but their vertex type can change. Two sites labeled by red diamonds provide equations which require $A_5=J_1$. The site labeled by the red square requires $B_3$ site to be forbidden if $A_1$ and $J_1$ are forbidden.  }
    \label{fig:RealSpaceForbiddenS5V2}
\end{figure}

First, we consider an S5 vertex with index-2. All S5 vertices in $V_2$ have the same local neighborhood up to their third nearest neighbor as shown in Fig.\ref{fig:RealSpaceForbiddenS5V2}. To be more precise, the vertex types of first and second nearest neighbors of an S5 in $V_2$ are fixed,  while only the positions of the third nearest neighbors are fixed. The sites labeled by A and B in Fig.\ref{fig:RealSpaceForbiddenS5V2}  have more links than the two displayed, and the number and orientation of these links will depend on the position of central S5 in perpendicular space.

As we are considering LS defined on the (1-3) sublattice, the central S5 and its second nearest neighbors host no wavefunction. Instead the Schrodinger equation  Eq.\ref{eq:Schrodinger} at each one of these sites give an equation for the wavefunctions defined in the (1-3) sublattice sites. Let's call the value of the wavefunction on the five J sites surrounding S5 $J_1,..,J_5$, and similarly label the boundary sites with $A_1,...,A_5$ and $B_1,...,B_5$.  As we are looking for solutions with zero energy, the two equations provided by the sites shown with red diamonds in Fig.\ref{fig:RealSpaceForbiddenS5V2} are
\begin{eqnarray}
A_5+J_5+B_3 &=0, \\ \nonumber
J_1+J_5+B_3 &=0.
\end{eqnarray}
These equations are satisfied only if $A_5=J_1$. Similarly one can conclude using two other sites that $A_5=J_4$. Repeating this procedure we arrive at the conclusion $J_1=. ..=J_5=A_1=...=A_5$. Now the equation provided by the central S5 is $J_1+J_2+J_3+J_4+J_5=0$, leading  to 
\begin{equation}
    J_n=0,\quad A_n=0 , \quad n=1,..,5,
\end{equation}
hence all sites labeled with $J$ and $A$ in Fig.\ref{fig:RealSpaceForbiddenS5V2} are forbidden sites. 
\begin{figure}[!htb]
    \centering
    \includegraphics[trim=5mm 5mm 5mm 5mm,clip,width=0.4\textwidth]{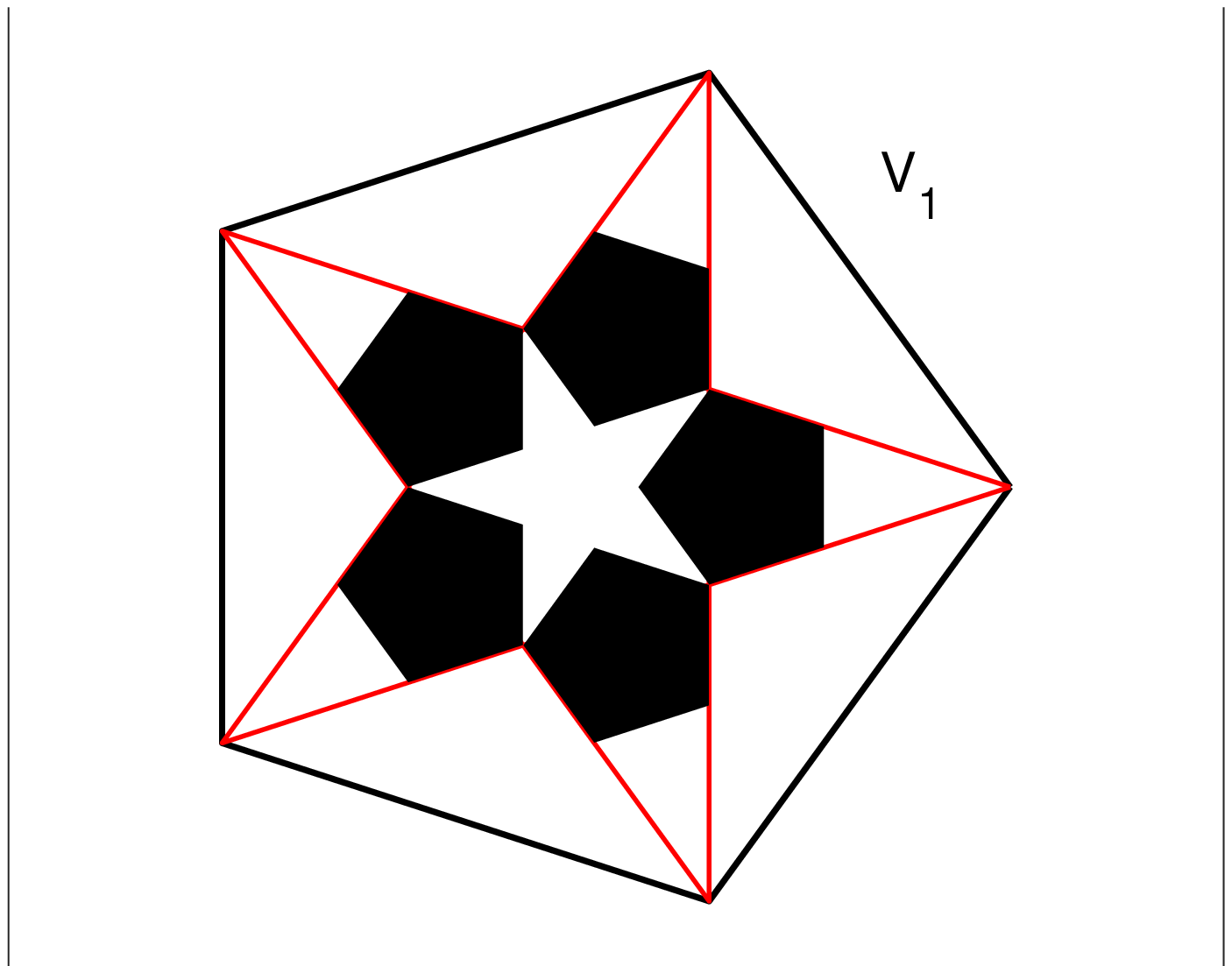}
    \includegraphics[trim=5mm 5mm 5mm 5mm,clip,width=0.4\textwidth]{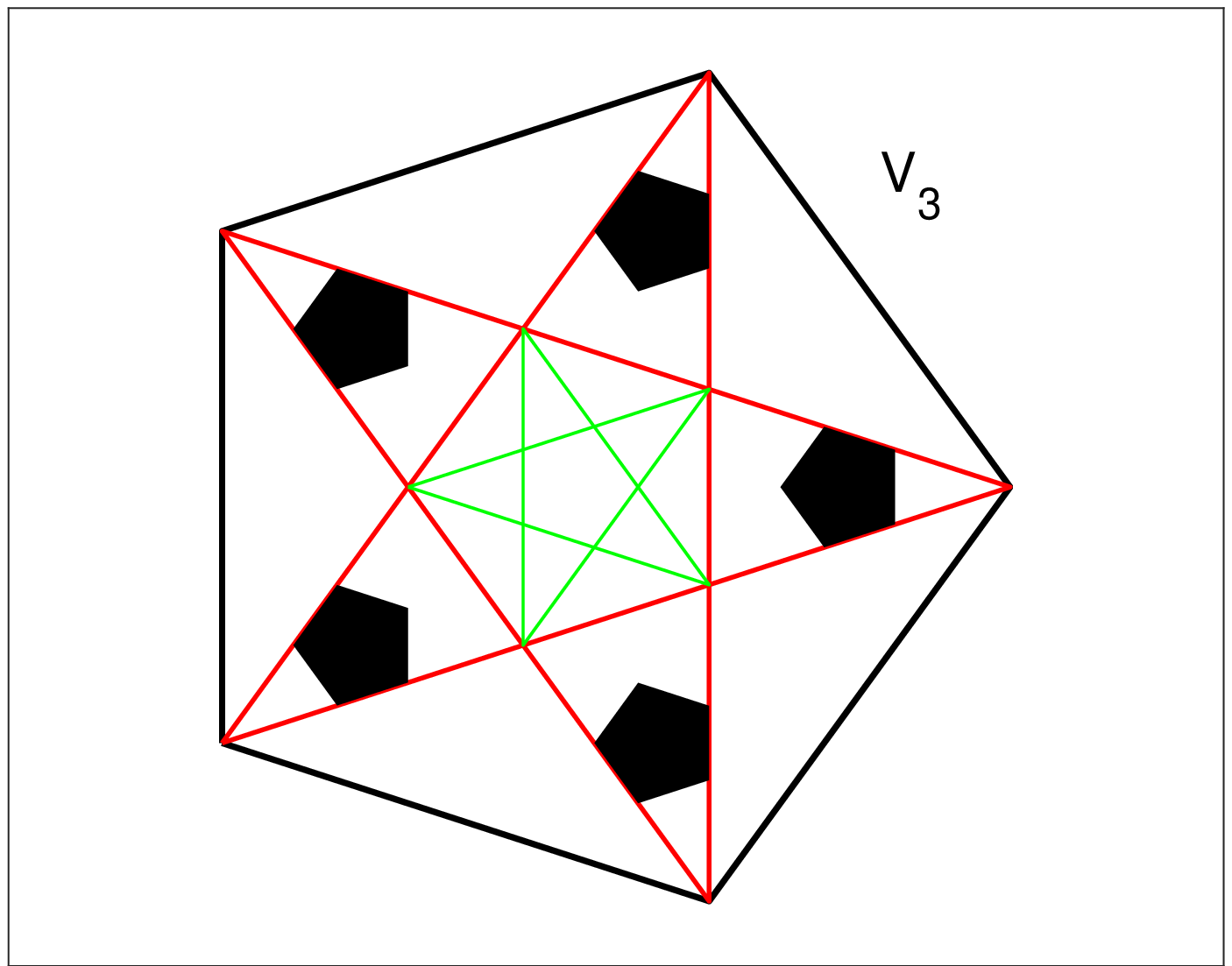}
    \caption{The forbidden regions in $V_1$ and $V_3$ corresponding to $J$ and $A$ sites in Fig.\ref{fig:RealSpaceForbiddenS5V2}. We see that $A$ sites can be S or  K vertices by comparing with Fig.\ref{fig:PerpendicularSpaceOfPL}.}
    \label{fig:PerpSpaceForbiddenS5V2}
\end{figure}
\begin{figure}[!htb]
    \centering
    \includegraphics[trim=5mm 5mm 5mm 5mm,clip,width=0.4\textwidth]{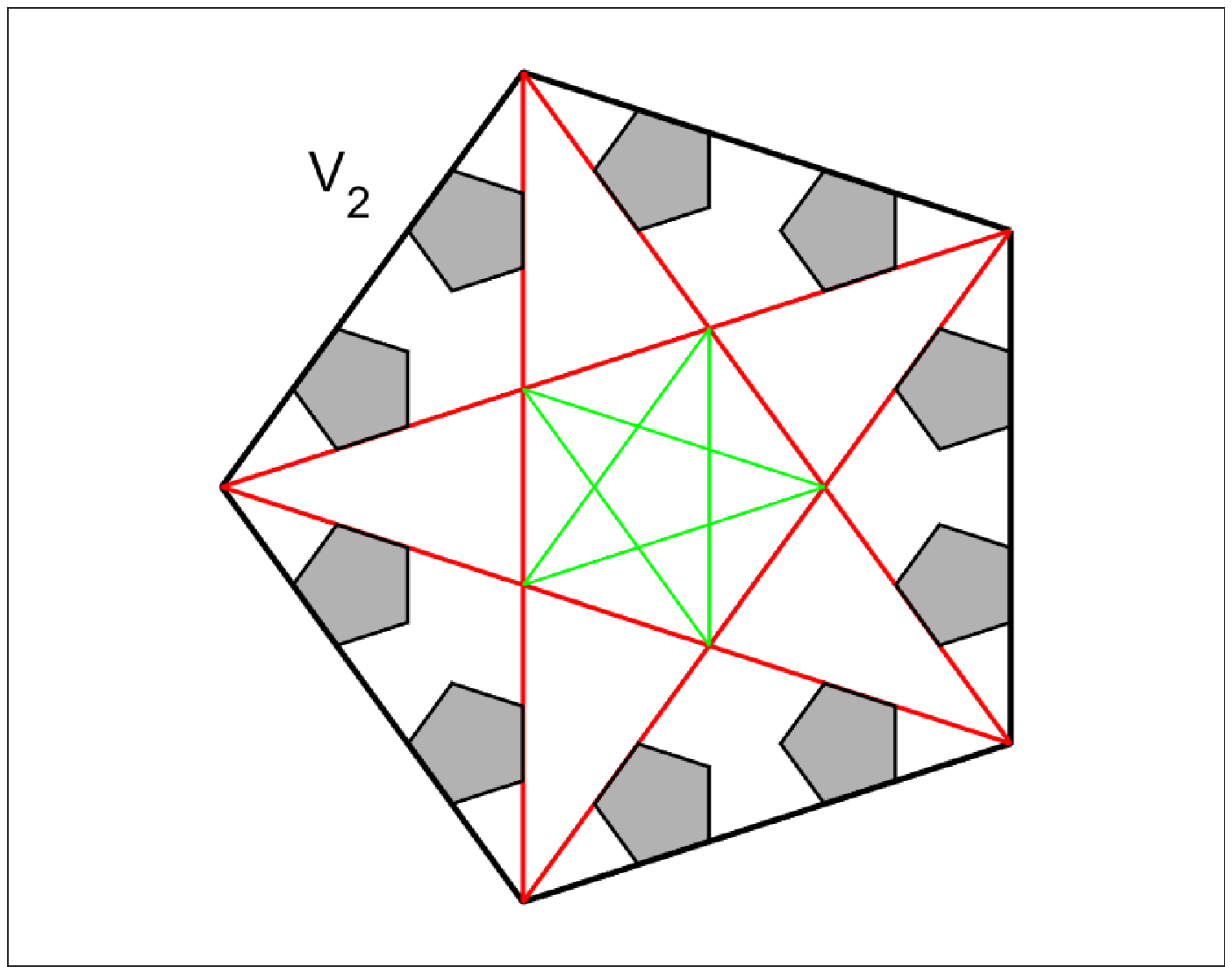}
    \includegraphics[trim=5mm 5mm 5mm 5mm,clip,width=0.4\textwidth]{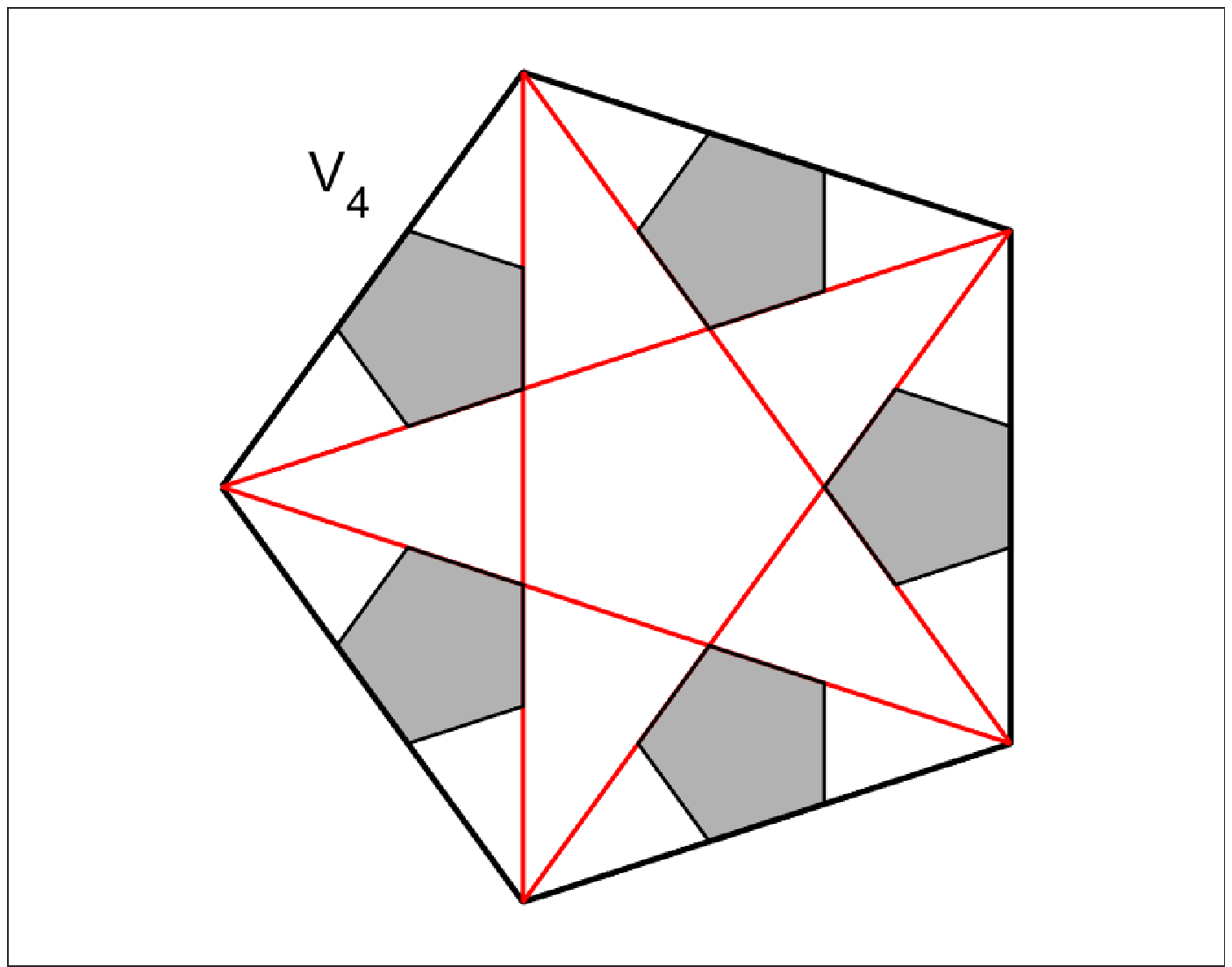}
    \caption{These seed regions in $V_2$ and $V_4$ have two nearest neighbors in regions shown in the previous figure. Their remaining nearest neighbors are consequently forbidden.}
    \label{fig:PerpSpaceSeedRegion}
\end{figure}
\begin{figure}[!htb]
    \centering
    \includegraphics[trim=5mm 5mm 5mm 5mm,clip,width=0.4\textwidth]{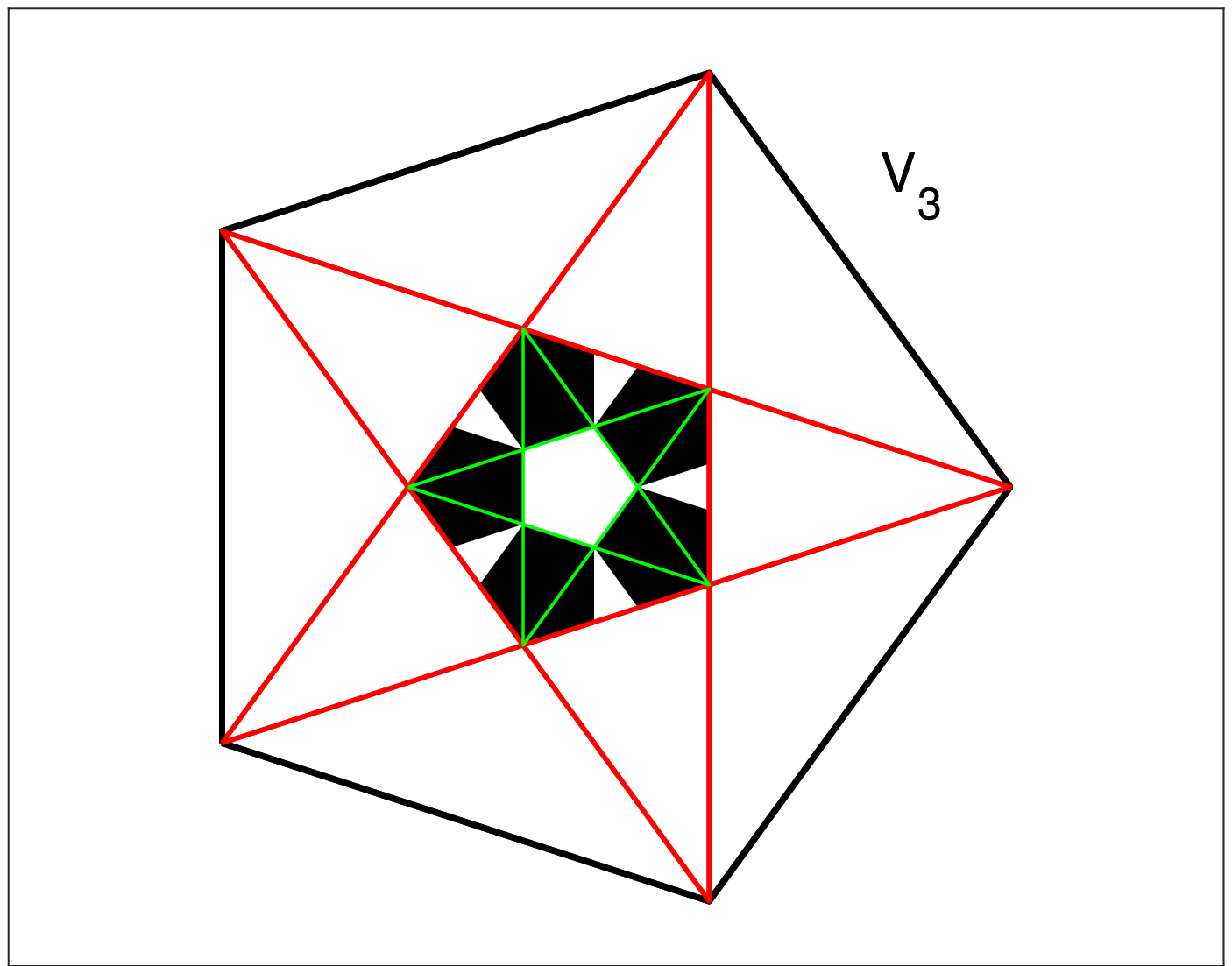}
    \caption{Perpendicular space forbidden regions which are nearest neighbors of seed regions shown in the previous figure. Notice that all areas corresponding to S4 vertices are covered, thus all S4 vertices are forbidden sites. The process displayed in the last three figures can be iterated. }
    \label{fig:PerpSpaceNextForbiddenRegion}
\end{figure}

The perpendicular space images for these J and A sites are shown in Fig.\ref{fig:PerpSpaceForbiddenS5V2}, where all the pentagons have radius $\tau^{-3}$ originating from the S5 region in $V_2$. One can observe that sites labeled with A can be either K or S vertices from $V_1$.
\begin{figure}[!htb]
    \centering
    \includegraphics[trim=5mm 5mm 5mm 5mm,clip,width=0.4\textwidth]{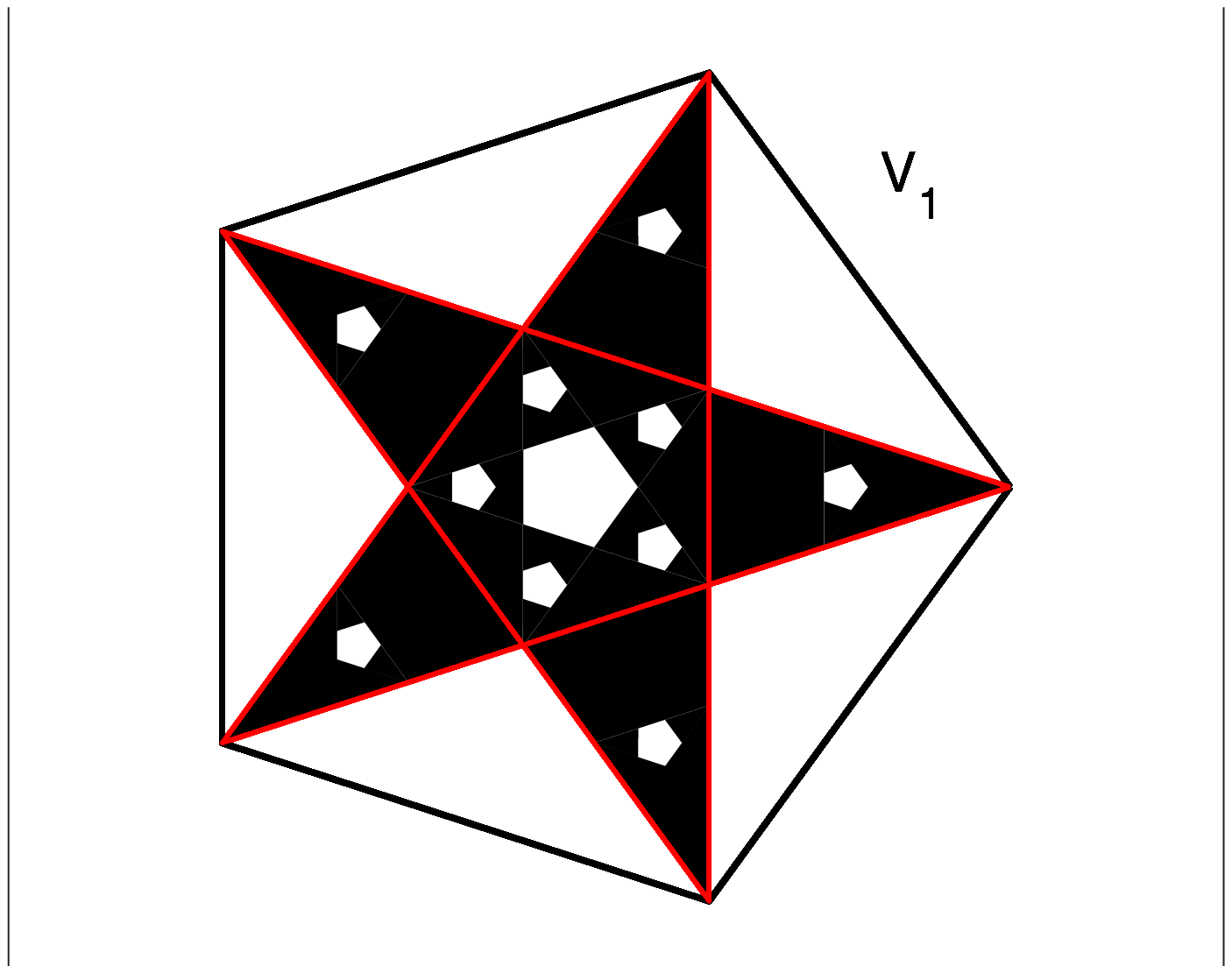}
    \includegraphics[trim=5mm 5mm 5mm 5mm,clip,width=0.4\textwidth]{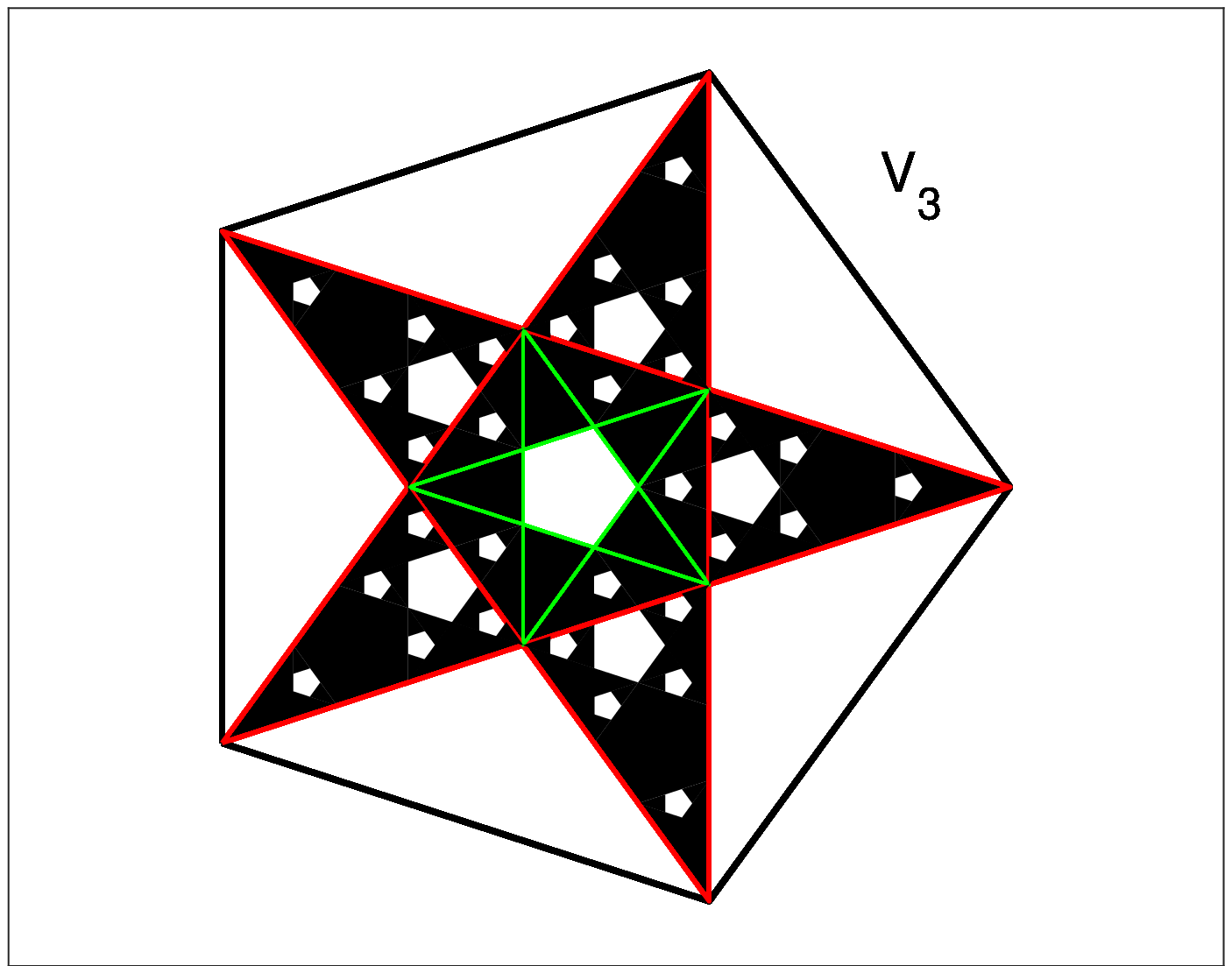}
    \caption{Forbidden regions found as a result of 15 iterations. The iterations terminate at this stage as there are no $N$ edge vertex regions in $V_2$ or $V_4$ which have $N-1$ nearest neighbors in the shown forbidden regions.}
    \label{fig:PerpSpaceFinalForbiddenRegion}
\end{figure}
Now consider the point marked by a red square in Fig.\ref{fig:RealSpaceForbiddenS5V2}. It is a point in the (2-4) sublattice and two of its 3 neighbors are forbidden by the above argument. Consequently the third neighbor, the site labeled by B, must also be a forbidden site. More generally, if a N-edge vertex in (2-4) sublattice has N-1 forbidden neighbors, its last remaining neighbor must also be forbidden. This argument can be easily applied in perpendicular space. Consider the forbidden regions shown in Fig.\ref{fig:PerpSpaceForbiddenS5V2}, the nearest neighbors of the forbidden sites will define regions in $V_2$ and $V_4$. Ten pentagons in D regions of $V_2$ and five pentagons in Q regions of $V_4$ have 2 nearest neighbors in the forbidden regions shown in Fig.\ref{fig:PerpSpaceSeedRegion}. We can calculate the regions corresponding to the third neighbors of these sites, which come out to be five pentagons shown in Fig. \ref{fig:PerpSpaceNextForbiddenRegion}. One should notice that all S4 vertices are shown to be forbidden sites by this argument, as their region are completely covered in Fig.\ref{fig:PerpSpaceNextForbiddenRegion}.

The forbidden regions we have found so far are formed by the union of Fig.\ref{fig:PerpSpaceForbiddenS5V2} and Fig.\ref{fig:PerpSpaceNextForbiddenRegion}. The same algorithm of looking for N-edge neighbors with N-1 neighbors inside forbidden zones can be iterated. Generally, such seed regions get smaller at each iteration, however the forbidden region continues to grow until the fifteenth iteration. At the end of the fifteenth iteration the total amount of forbidden regions is plotted in Fig.\ref{fig:PerpSpaceFinalForbiddenRegion}. There are no N-edge neighbors with N-1 forbidden site neighbors for this region and the iteration process terminates.

\begin{figure}[!htb]
    \centering
        \includegraphics[trim=1mm 1mm 1mm 1mm,clip,width=0.6\textwidth]{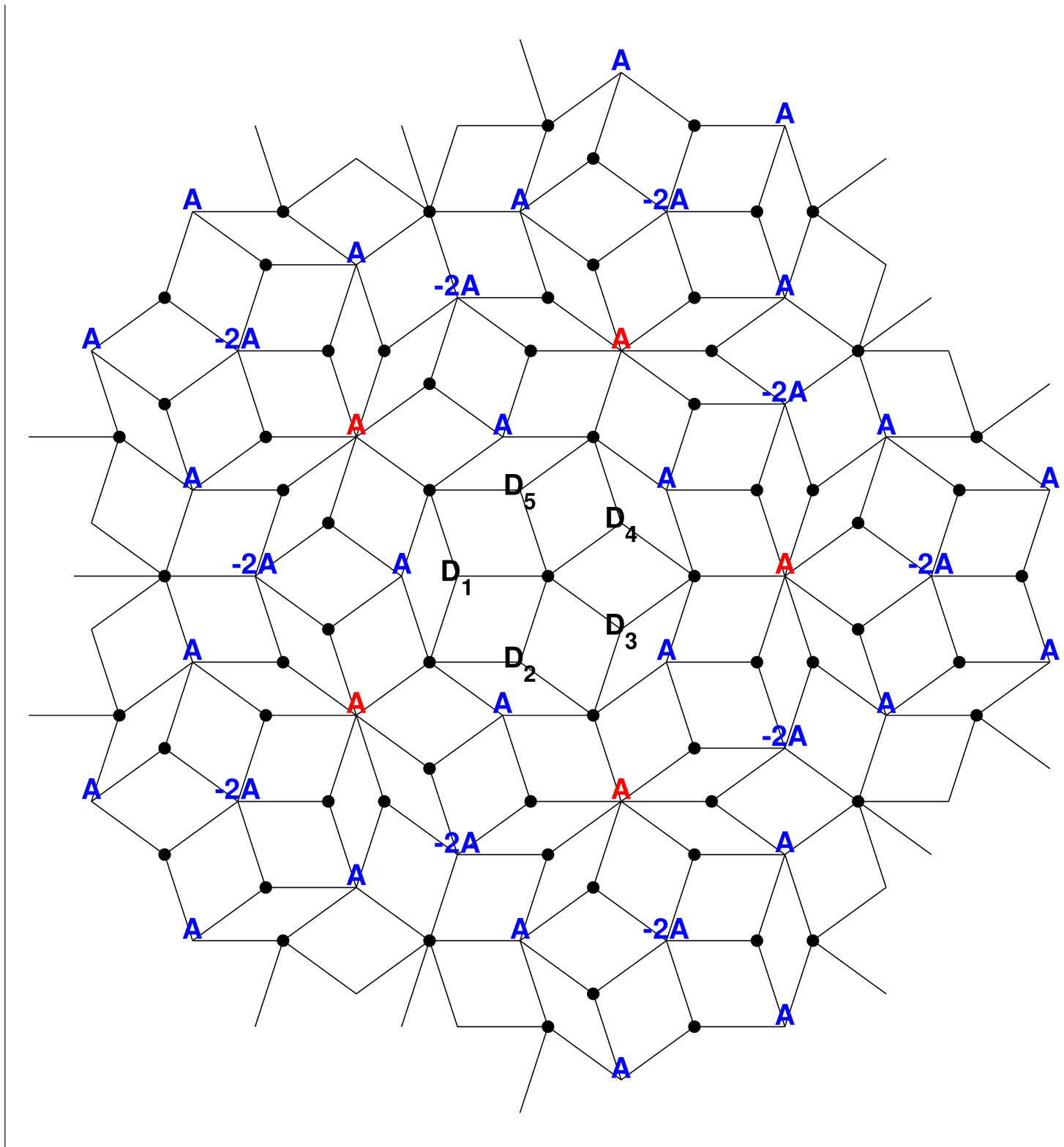}
    \caption{Neighborhood of an S vertex in $V_4$ up to its seventh nearest neighbors. The positions of the sites in this neighborhood are unchanged if S lies in a pentagon of radius $\tau^{-4}$ in perpendicular space. The equations provided by sublattice (2-4) points require all the labeled sites in this figure to be forbidden sites.}
    \label{fig:RealSpaceLargeForbiddenSV4}
\end{figure}

Now, we consider a similar scenario for the neighborhood of an S vertex with index 4. The particular neighborhood we consider is plotted in Fig.\ref{fig:RealSpaceLargeForbiddenSV4}. This neighborhood extending up to the seventh neighbors is unchanged if the central S vertex lies in a pentagon of radius $\tau^{-4}$ at the center of $V_4$. Let's take one of the S3 sites in the figure and call the value of the LS wavefunction on it $A$. By using two D sites sharing the same rhombus, we  see that all the five S3 sites surrounding the central S must have the same wavefunction, which are marked with red A's in the figure. The same logic  is used to mark the five K sites around the central S, and 20 more sites around the 5 outer S vertices with wavefunction $A$, shown as blue A's in the figure. The outer S vertices, as well as J sites connecting them have wavefunction $-2A$.

\begin{figure}[!htb]
    \centering
    \includegraphics[trim=5mm 5mm 5mm 5mm,clip,width=0.4\textwidth]{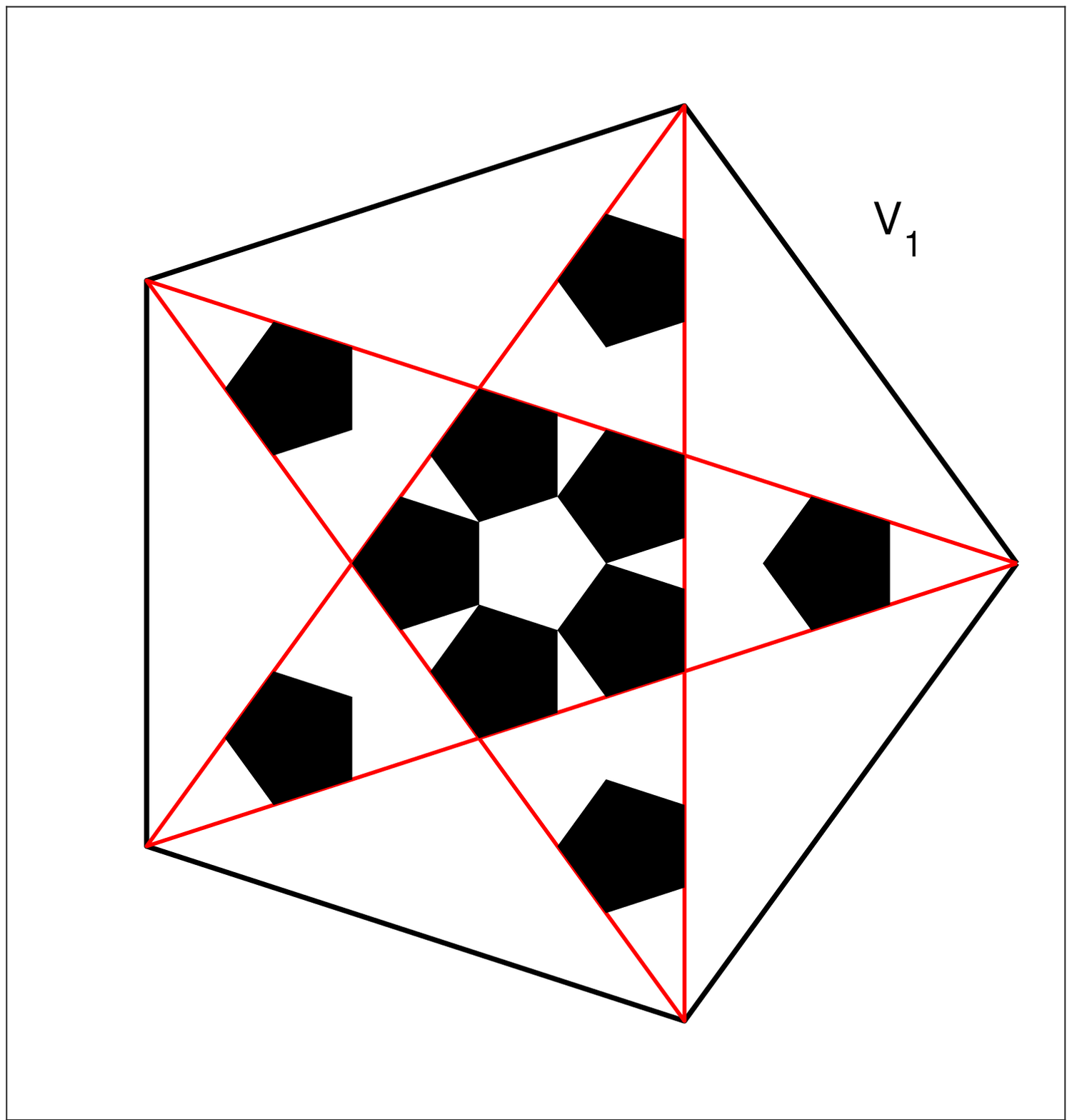}
    \includegraphics[trim=5mm 5mm 5mm 5mm,clip,width=0.4\textwidth]{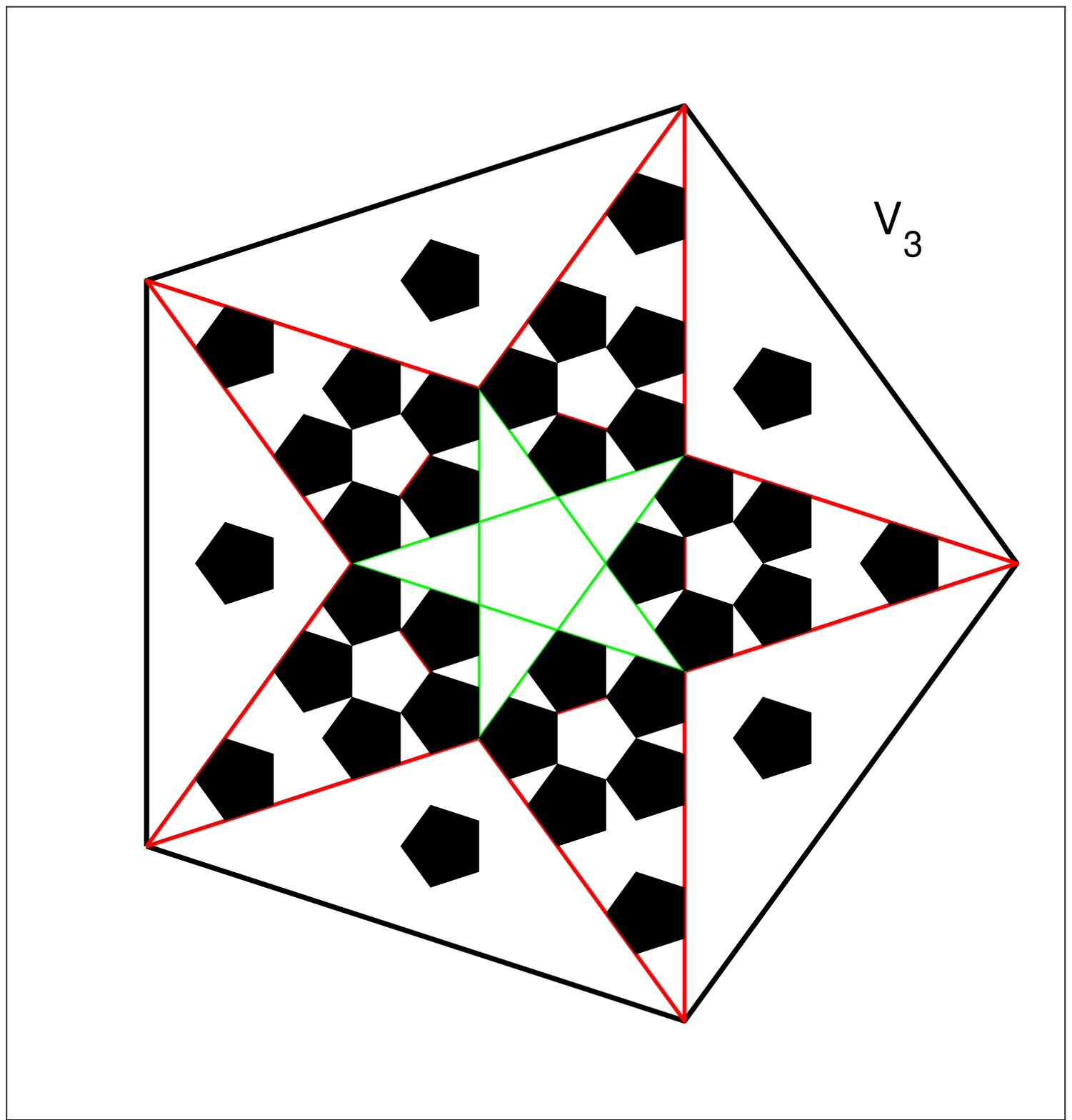}

    \caption{Forbidden regions corresponding to labeled sites in Fig.\ref{fig:RealSpaceLargeForbiddenSV4}. Notice that the pentagons in S3 regions close the holes in Fig.\ref{fig:PerpSpaceFinalForbiddenRegion} making all S3 sites forbidden sites.}
    \label{fig:PerpSpaceForbiddenSV4}
\end{figure}

The five nearest neighbors of the central S are  marked as $D_1,..,D_5$. From the central S we have the equation $D_1+D_2+D_3+D_4+D_5=0$, and the five J sites around the central S give
\begin{equation}
D_1+D_2=D_2+D_3=D_3+D_4=D_4+D_5=D_5+D_1=-3A.    
\end{equation}
The only solution to the above set of equations is $A=D_1=D_2=D_3=D_4=D_5=0$, which shows that all the marked sites in Fig.\ref{fig:RealSpaceLargeForbiddenSV4} are forbidden sites. The corresponding forbidden regions in perpendicular space are given in Fig.\ref{fig:PerpSpaceForbiddenSV4}. We can observe that the new forbidden regions fill in some of the holes in Fig.\ref{fig:PerpSpaceFinalForbiddenRegion}. In particular, all S3 vertices are seen to be forbidden sites. 
If all the allowed regions and the all the forbidden regions above are considered, there are still sites which have not been marked as forbidden or as the support of some LS. We give two more real space neighborhoods to complete the classification. 

\begin{figure}[!htb]
    \centering
     \includegraphics[trim=1mm 1mm 1mm 1mm,clip,width=0.6\textwidth]{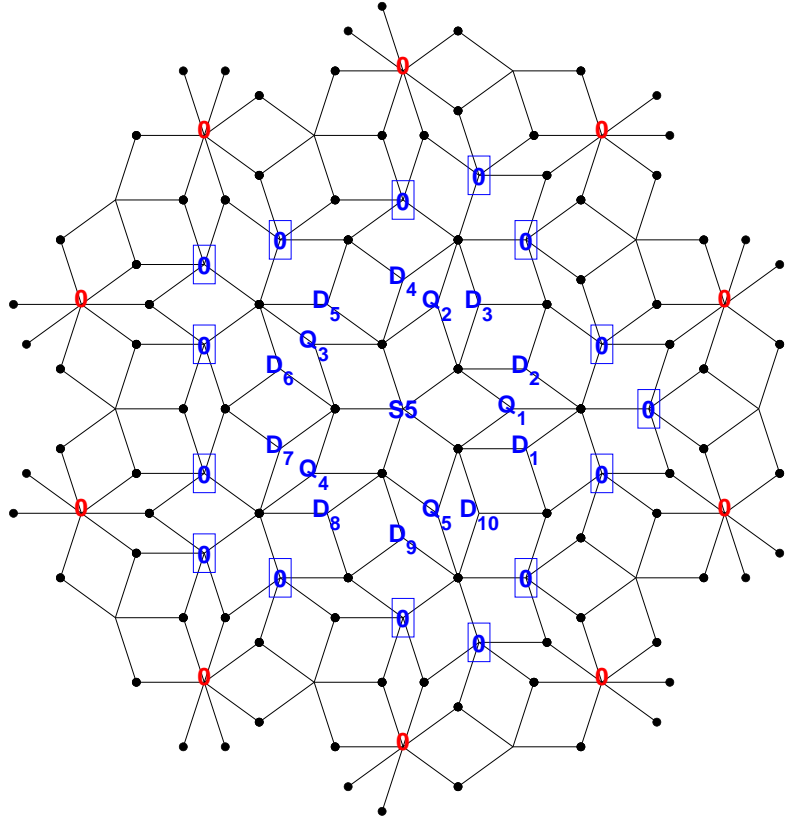}
    \caption{Local neighborhood of an S5 site in $V_3$ up to seventh nearest neighbors. The positions of the sites in the figure are fixed if the S5 vertex lies in a pentagon of radius $\tau^{-5}$ at the center of perpendicular space.}
    \label{fig:RealSpaceForbiddenS5V3}
\end{figure}

We consider the neighborhood of an S5 vertex once again, but choose this vertex to lie in $V_3$, hence have a non zero wavefunction. The particular neighborhood we consider is plotted in Fig.\ref{fig:RealSpaceForbiddenS5V3}. Not all S5 sites in $V_3$ have the same displayed neighborhood (up to the 6$^{th}$ nearest neighbors fixed), but S5 must lie in a pentagon of radius $\tau^{-5}$ at the center of $V_3$.  
\begin{figure}[!htb]
    \centering
    \includegraphics[trim=5mm 5mm 5mm 5mm,clip,width=0.4\textwidth]{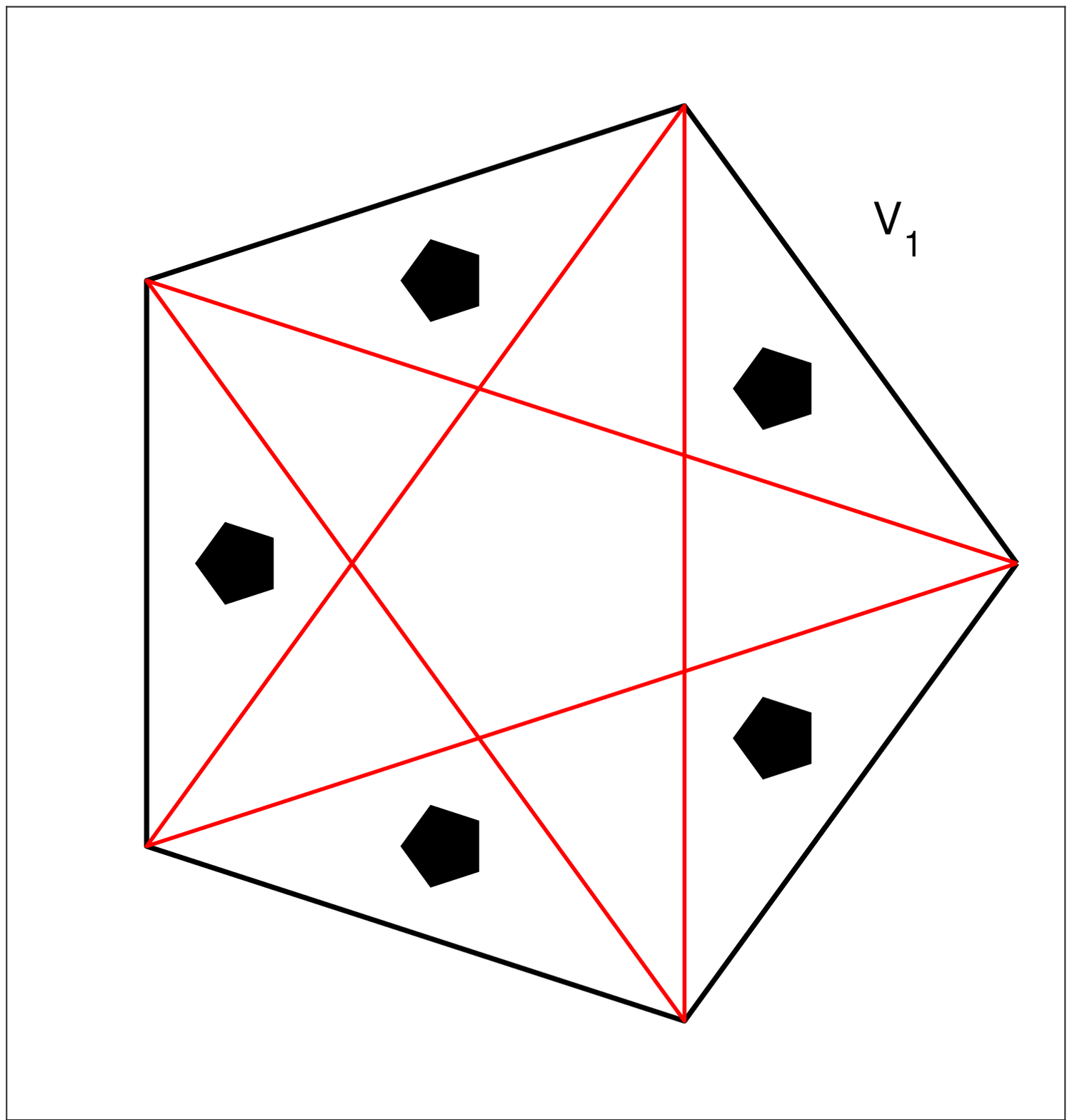}
    \includegraphics[trim=5mm 5mm 5mm 5mm,clip,width=0.4\textwidth]{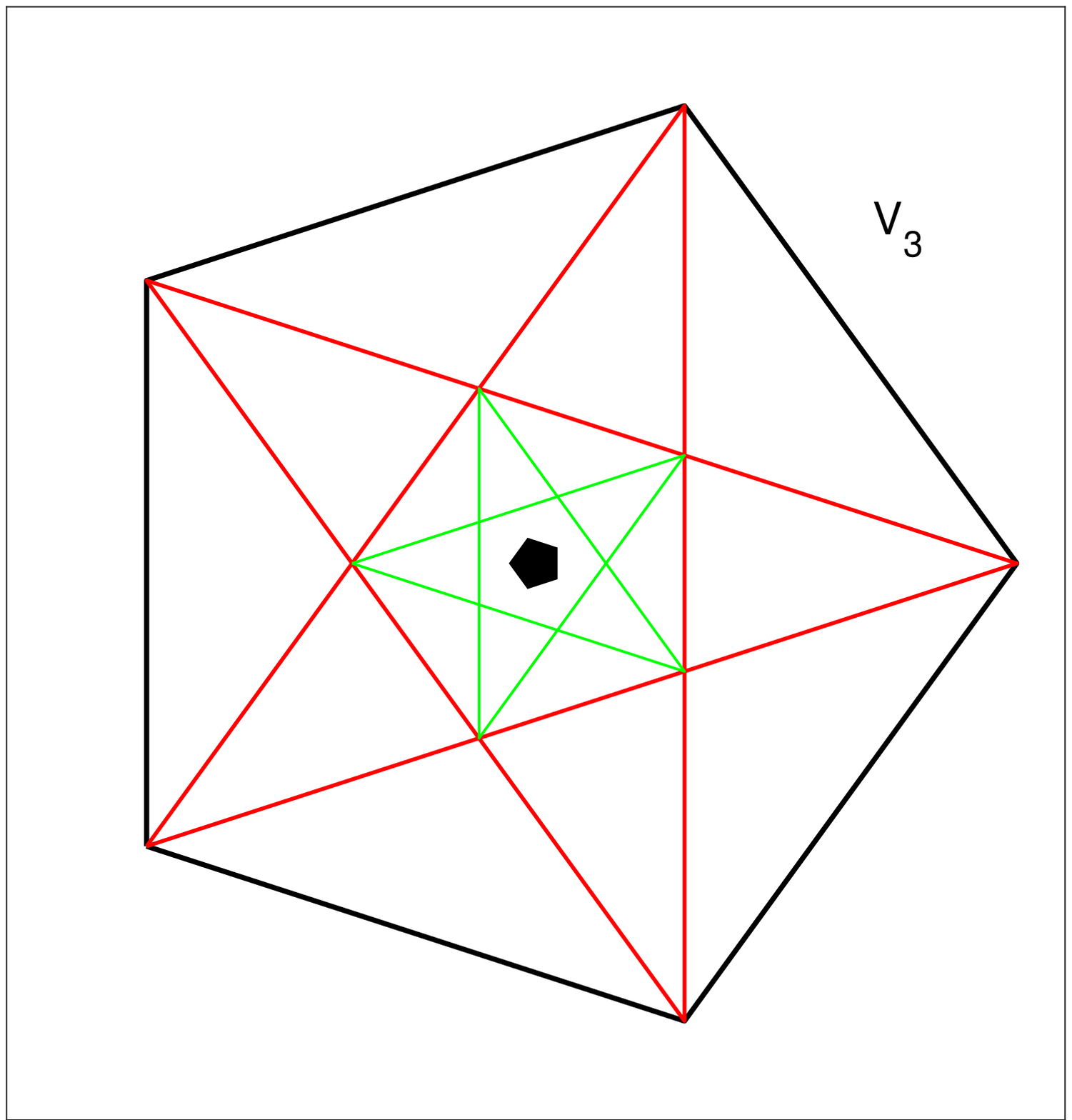}
    \caption{Forbidden regions corresponding to $Q$ and $S5$ labels in Fig.\ref{fig:RealSpaceForbiddenS5V3}.}
    \label{fig:PerpSpaceForbiddenS5V3}
\end{figure}
\begin{figure}[!htb]
    \centering
        \includegraphics[trim=1mm 1mm 1mm 1mm,clip,width=0.6\textwidth]{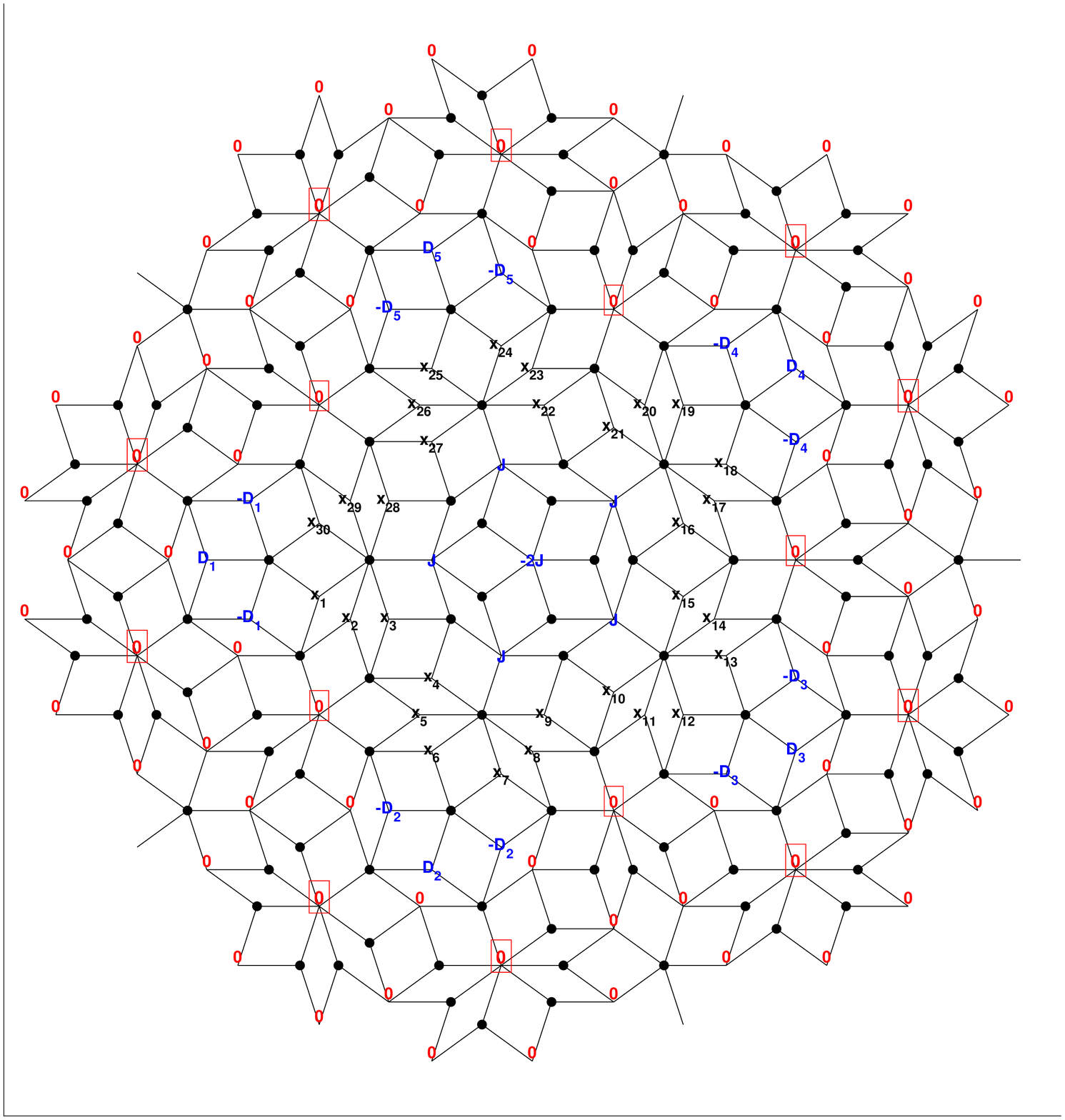}
    \caption{Local neighborhood up to 10$^{th}$ nearest neighbors of an S vertex of index 1. The central S site lies in a pentagon of radius $\tau^{-6}$ at the center of perpendicular space. Central S and five surrounding J sites are forbidden as explained in the text.}
    \label{fig:RealSpaceForbiddenSV1}
\end{figure}
\begin{figure}[!htb]
    \centering
  \centering
    \includegraphics[trim=5mm 5mm 5mm 5mm,clip,width=0.4\textwidth]{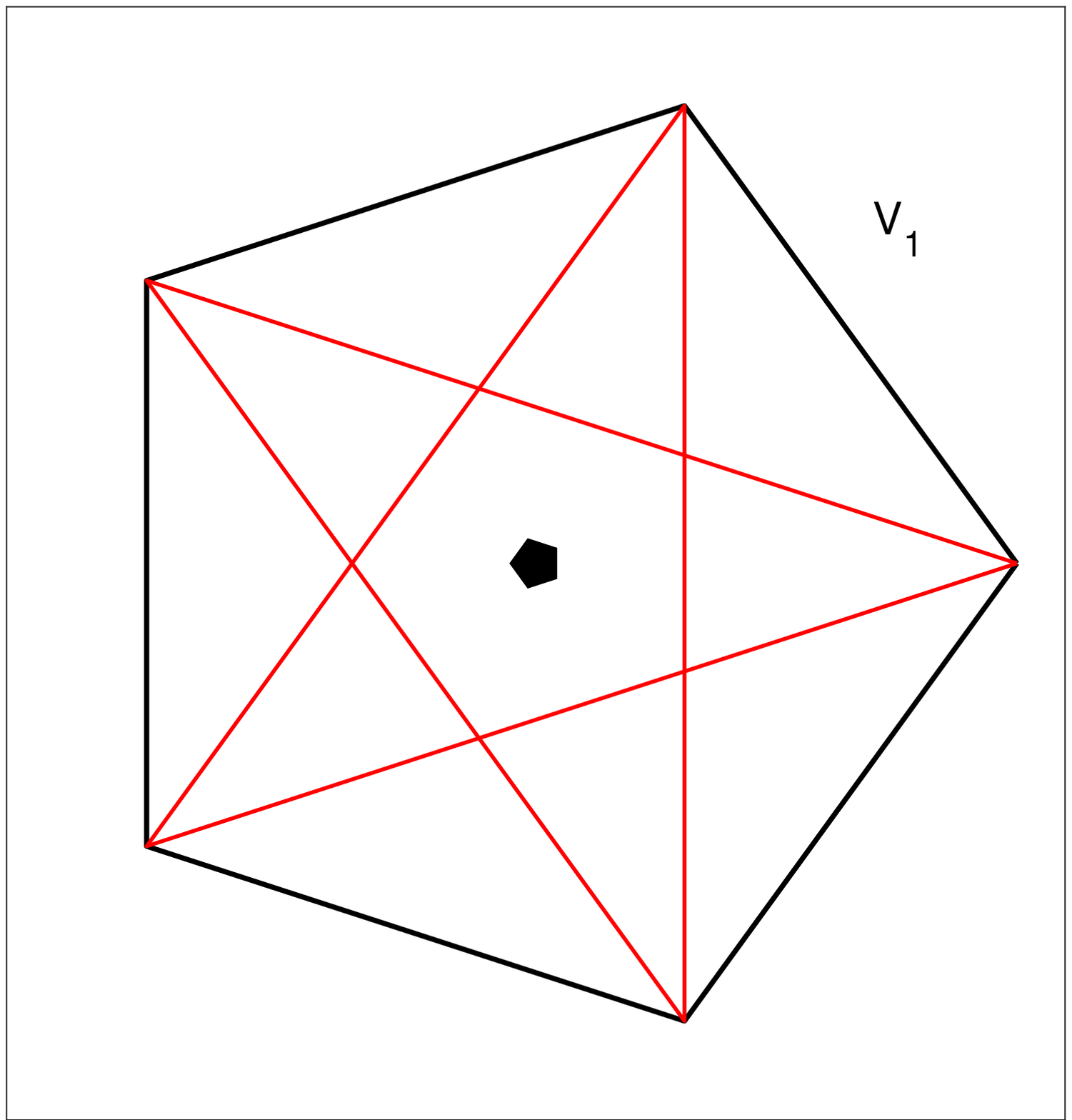}
    \includegraphics[trim=5mm 5mm 5mm 5mm,clip,width=0.4\textwidth]{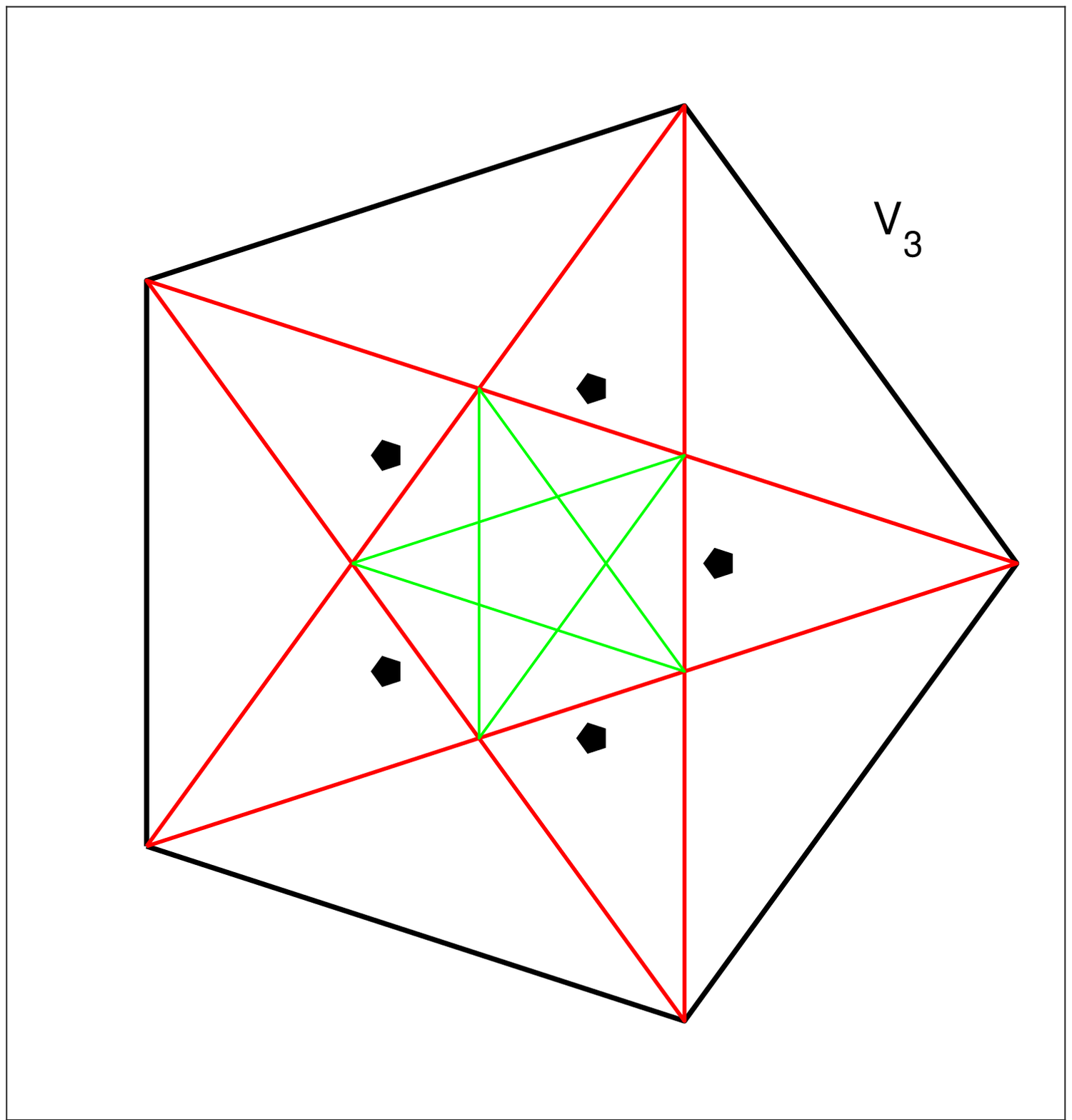}
    \caption{Perpendicular space regions for forbidden sites in Fig.\ref{fig:RealSpaceForbiddenSV1}. }
    \label{fig:PerpSpaceForbiddenSV1}
\end{figure}
The sixth neighbors of central S5 contain ten S3 sites, which were proven above to be forbidden. Hence we can mark them with wavefunction 0, and mark another smaller ring of 0's using 3-edge vertices. The D and Q vertices between this ring and the central S5 are marked with $D_1,..,D_{10}$ and $Q_1,..,Q5$.  The first set of equations come from K sites,
\begin{equation}
    D_2+D_3=D_4+D_5=D_6+D_7=D_8+D_9=D_{10}+D_{1}=0,
\end{equation}
followed by
\begin{eqnarray}
    D_2+D_3+Q_1+Q_2=-S5, \\ \nonumber 
    D_4+D_5+Q_2+Q_3=-S5, \\ \nonumber
    D_6+D_7+Q_3+Q_4=-S5, \\ \nonumber
    D_8+D_9+Q_4+Q_5=-S5, \\ \nonumber
    D_{10}+D_{1}+Q_5+Q_1=-S5.
 \end{eqnarray}
This tells us all Q sites must have the same wavefunction, which is -1/2 of  the central S5. Finally using the S4 site equations
\begin{eqnarray}
    D_1+D_2+Q_1=0, \\ \nonumber
    D_3+D_4+Q_2=0, \\ \nonumber
    D_5+D_6+Q_3=0, \\ \nonumber
    D_7+D_8+Q_4=0, \\ \nonumber
    D_{9}+D_{10}+Q_5=0,
\end{eqnarray}
we conclude both the central S5 and the 5 Q surrounding it are forbidden sites. The perpendicular space regions corresponding to these sites are given in Fig.\ref{fig:PerpSpaceForbiddenS5V3}. 

The final real space configuration we consider is given in Fig, \ref{fig:RealSpaceForbiddenSV1}, consisting of the 10 nearest neighbors of an S site in $V_1$. This neighborhood is fixed if the central S lies in a pentagon of radius $\tau^{-6}$ at the center of $V_1$. 

The five J sites surrounding the center must have the same wavefunction, which is -1/2 of the wavefunction of the central S. Eighth neighbors of S have S3 vertices which we can mark with zero wavefunction as they were proven to be forbidden above. This allows us to mark a number of 6$^{th}$ and 10$^{th}$ neighbors with zero wavefunction, forming an outside ring.  This ring leads us to denote 15 D sites with $\pm D_1,..\pm D_5$. All the remaining sites we mark with $x_1,..x_{30}$.  

The equations coming from S3 sites give us 
\begin{equation}
x_1+x_2+...+x_{30}=-5J,
\label{eq:Xsum}
\end{equation}
But J site equations similar to $x_2+x_3+x_4+x_5=0$, allow us to reduce this sum to $(x_{30}+x_1)+(x_{6}+x_7)+(x_{12}+x_{13})+(x_{18}+x_{19})+(x_{24}+x_{25})=-5J$. We then find
$D_1+D_2+D_3+D_4+D_5=-5J$. The sum in Eq.(\ref{eq:Xsum}) can be reduced in another way using K site equations like $x_3+x_4=-2J$ and J site equations like $x_1+x_2=D_1$, yielding
\begin{eqnarray}
(x_1+x_2)+(x_3+x_4)+...+(x_{29}+x_{30})=-5J,\\ \nonumber
D_1-2J+...+D_1=2(D_1+...+D5)-10J=-5J.
\end{eqnarray}
However as we found $D_1+D_2+D_3+D_4+D_5=-5J$, the equation becomes $20J=5J$, where the only solution is $J=0$. Thus we mark the central S and surrounding J regions in $V_1$ and $V_3$ as forbidden in Fig. \ref{fig:PerpSpaceForbiddenSV1}.

\begin{figure}[!htb]
    \centering
    \includegraphics[trim=5mm 5mm 5mm 5mm,clip,width=0.4\textwidth]{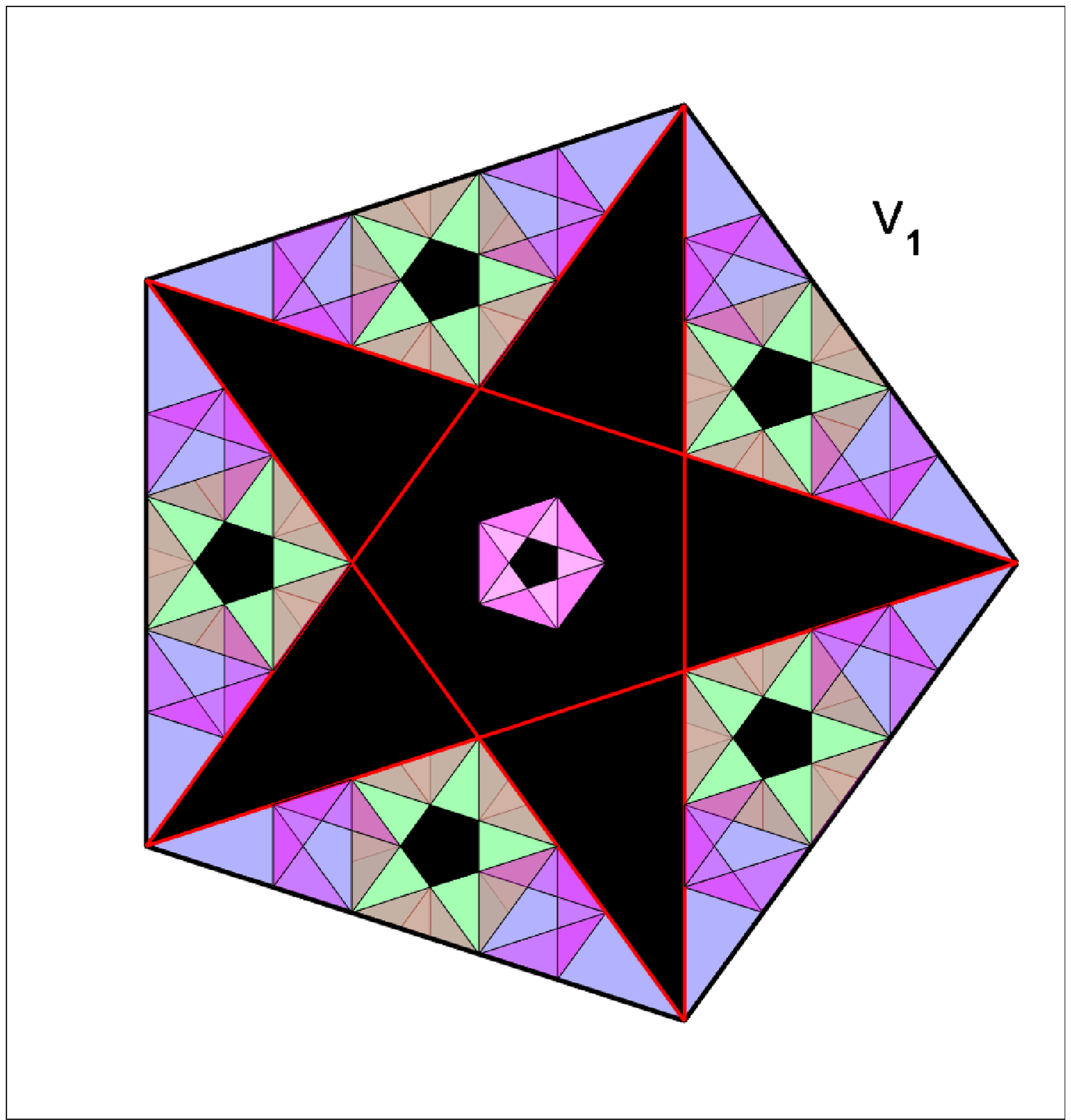}
    \includegraphics[trim=5mm 5mm 5mm 5mm,clip,width=0.4\textwidth]{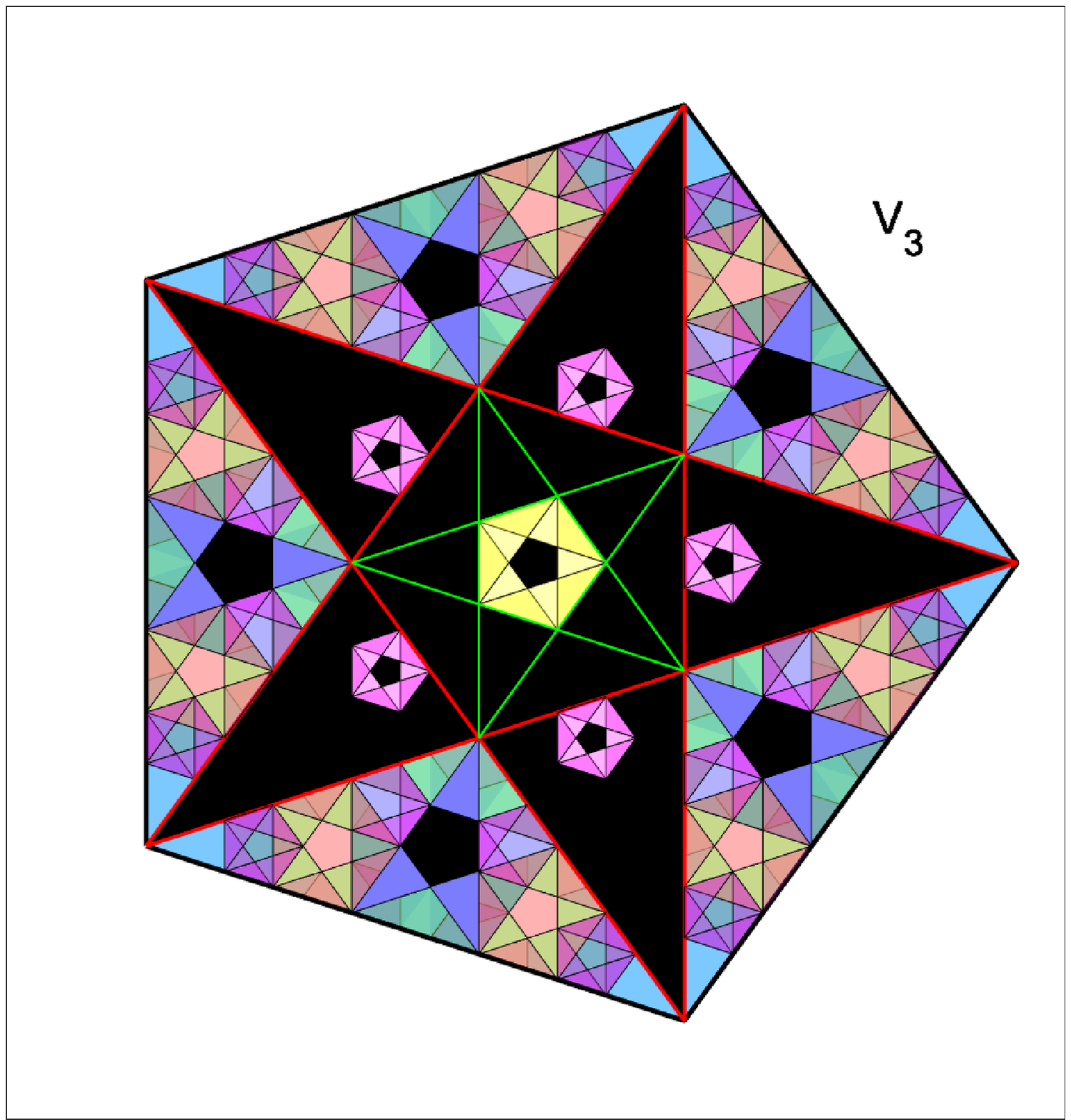}

    \caption{All the forbidden regions (Figs.  \ref{fig:PerpSpaceFinalForbiddenRegion},\ref{fig:RealSpaceForbiddenS5V3},\ref{fig:PerpSpaceForbiddenSV4},\ref{fig:PerpSpaceForbiddenSV1}) and allowed regions (Figs. \ref{fig:Type1PerpSpaceArea},\ref{fig:Type2PerpSpaceArea},\ref{fig:Type3PerpSpaceAreaRotated},\ref{fig:Type4PerpSpaceAreaRotated},\ref{fig:Type5PerpSpaceAreaRotated},\ref{fig:Type6PerpSpaceAreaRotated}) displayed. All areas  inside the four perpendicular space pentagons are filled, thus any site in the PL is either forbidden or in the support of at least one LS.} 
    \label{fig:PerpSpaceAllRegions}
\end{figure}

When we overlay all the allowed regions in Figs. \ref{fig:Type1PerpSpaceArea},\ref{fig:Type2PerpSpaceArea},\ref{fig:Type3PerpSpaceAreaRotated},\ref{fig:Type4PerpSpaceAreaRotated},\ref{fig:Type5PerpSpaceAreaRotated},\ref{fig:Type6PerpSpaceAreaRotated} and all the forbidden regions in Figs.\ref{fig:PerpSpaceFinalForbiddenRegion},\ref{fig:RealSpaceForbiddenS5V3},\ref{fig:PerpSpaceForbiddenSV4},\ref{fig:PerpSpaceForbiddenSV1}    we obtain Fig.\ref{fig:PerpSpaceAllRegions}. There are no areas which are not assigned either as forbidden or as in the support of some LS. This constitutes a geometric proof that a site is either forbidden to host a LS, or is in the support of some LS. Overall $f_{forbidden}=1358-839 \tau\simeq46.95\%$ of PL vertices cannot host a LS.

\section{\label{sec:Conclusion} Conclusion}

Quasicrystal lattices can be defined as projections of  higher dimensional periodic structures. All points in the real space lattice have an image in the projected out dimensions, and this perpendicular space image carries valuable information about the local neighborhood in real space. All the images in perpendicular space lie inside a finite region, and this region is filled densely and uniformly. These properties make it possible to use perpendicular space to label and count local structures in real space.

We applied these perpendicular space accounting methods to label and count the LS in the vertex tight-binding model of the PL. We verified the frequencies of type-1 to type-4 LS, but showed that the number of type-5 and type-6 LS that can be defined on the PL are a factor of $\tau$ larger than previously reported values. We proved the independence of different types of states, and showed how perpendicular space methods can be used to calculate overlaps between them. This allowed us to show that the frequency of linearly independent type-5 and type-6 LS is equal to previously reported values for the total number of these states. We also proved that any site in the PL is either in the support of at least one LS, or forbidden by local connectivity to host a LS. It has been conjectured that the 6 types of LS states span the whole zero energy manifold of the vertex tight-binding model. We have not been able to prove this, but our analysis shows that if another LS state exists it has to be defined on sites which are in the support of type-1 to type-6 LS. Thus, such a LS can not be trivially independent by having a unique site in its support.

The perpendicular space method is independent of the scaling (inflation-deflation) symmetry of the PL, which is generally used to count local structures. It would be interesting to see if this method can be combined with scaling symmetries as done for the Fibonacci chain \cite{mjp16} to investigate eigenstates which are not strictly localized. Similarly other models defined on quasicrystals \cite{fli20,sja08} which have strictly localized excitations may benefit from a perpendicular space approach.

\acknowledgments
M.M. is supported by the Bilkent Comprehensive Undergraduate Scholarship.

\bibliography{PerpendicularSpace}

\end{document}